\documentclass[twocolumn]{aastex61}
\pdfoutput=1 
\usepackage{amsmath,amstext}
\usepackage[T1]{fontenc}
\usepackage{apjfonts} 
\usepackage[figure,figure*]{hypcap}
\usepackage{comment}

\usepackage{graphicx}
\usepackage{natbib}
\usepackage{epsfig}
\usepackage{epstopdf}
\usepackage{enumerate}

\usepackage{tabularx}
\usepackage{multirow}
\usepackage{booktabs}

\newcommand{\push}{\mathrm{push}}
\newcommand{\msun}{M$_{\odot}$}
\newcommand{\kpush}{$k_{\rm push}$}
\newcommand{\trise}{$t_{\rm rise}$}

\shorttitle{Pushing 1D CCSNe to explosions: Explodability and remnant properties}
\shortauthors{Ebinger et al.}

\begin{document}

\title{PUSHing core-collapse supernovae to explosions in spherical symmetry II:\\ 
Explodability and remnant properties}

\author{Kevin Ebinger}
\affiliation{Department of Physics, North Carolina State University, Raleigh NC 27695}
\affiliation{GSI Helmholtzzentrum f\"ur Schwerionenforschung, D-64291 Darmstadt, Germany}
\author{Sanjana Curtis}
\affiliation{Department of Physics, North Carolina State University, Raleigh NC 27695}
\author{Carla Fr\"ohlich}
\affiliation{Department of Physics, North Carolina State University, Raleigh NC 27695}
\author{Matthias Hempel}
\affiliation{Department f\"ur Physik, Universit\"at Basel, CH-4056 Basel, Switzerland}
\author{Albino Perego}
\affiliation{Istituto Nazionale di Fisica Nucleare, Sezione Milano Bicocca, Gruppo Collegato di Parma, I-43124 Parma, Italy}
\affiliation{Dipartimento di Fisica, Universit\`a degli Studi di Milano Bicocca, I-20126 Milano, Italy}
\affiliation{Dipartimento di Scienze Matematiche Fisiche ed Informatiche, Universit\`a di Parma, I-43124 Parma, Italy}
\author{Matthias Liebend\"orfer}
\affiliation{Department f\"ur Physik, Universit\"at Basel, CH-4056 Basel, Switzerland}
\author{Friedrich-Karl Thielemann}
\affiliation{Department f\"ur Physik, Universit\"at Basel, CH-4056 Basel, Switzerland}
\affiliation{GSI Helmholtzzentrum f\"ur Schwerionenforschung, D-64291 Darmstadt, Germany}
\email{kebinge@ncsu.edu}
\email{cfrohli@ncsu.edu}

\begin{abstract}
In a previously presented proof-of-principle study we established a parametrized spherically symmetric explosion method (PUSH) that can reproduce many features of core-collapse supernovae. The present paper goes beyond a specific application that is able to reproduce observational properties of SN1987A and performs a systematic study of an extensive set of non-rotating, solar metallicity stellar progenitor models in the mass range from 10.8 to 120 M$_\odot$. This includes the transition from neutron stars to black holes as the final result of the collapse of massive stars, and the relation of the latter to supernovae\textbf{, possibly faint supernovae, and failed supernovae}. We discuss the explosion properties of all models and predict remnant mass distributions within this approach. The present paper provides the basis for extended nucleosynthesis predictions in a forthcoming paper to be employed in galactic evolution models. 
\end{abstract}


\section{Introduction}
\label{intro:1}

After the hydrostatic burning stages stars more massive than 8~M$_\odot$ undergo core collapse. Up to about 10~M$_\odot$ (the limit is uncertain, see e.g.\ \cite{poelarends2008,woosley2015,Doherty2017}), the collapse occurs due to electron capture (EC) reactions on C-burning products in the O-Ne-Mg core, resulting in fast contraction and the transition of the core matter to NSE followed by the explosion of the star (dubbed EC supernovae). Beyond 10~M$_\odot$ the end of the life of massive stars is initiated by the collapse of the central Fe-core which occurs after central Si-burning when the core exceeds its temperature-dependent Chandrasekhar mass.
This collapse of the stellar core marks the onset of a core-collapse supernova (CCSN),  a violent explosion that disrupts the star, allows for nuclei beyond iron to be formed and ejected into the interstellar medium, and ultimately leaves behind a neutron star (NS) or black hole (BH) as a remnant.

The major open questions are related to the explosion mechanism itself and how the transition from a ``regular'' CCSN with the formation of a neutron star to the formation of a central black hole occurs in nature (if there exists such a clear region of transition, see e.g. \cite{nomoto13}).
This depends strongly on the properties of the stellar progenitor, amongst others the compactness, i.e.\ the central mass concentration.
Detailed reviews and recent findings on the present understanding of the end stages of massive stars are given in e.g.\ \citet[][]{heger10,chieffi13,nomoto06,nomoto13,nomoto17,chieffi17}.
o-called hypernovae / long-duration gamma-ray bursts (GRBs) possibly occur after black hole formation in case of fast rotation and strong magnetic fields \citep[see e.g.][]{macfadyen99,macfadyen01}.
If fast rotation and strong magnetic fields are not present, black hole formation leads to a failed or faint supernova \citep[see e.g.][]{lovegrove13}). For smaller initial stellar masses a neutron star with magnetic fields as high as $10^{15}$~Gauss, known as magnetar, can be formed.
In the present paper we will only address models without rotation and magnetic fields, i.e.\ neutrino-driven CCSNe.

The full solution to the CCSN problem in a self-consistent way is still not converged yet.
Multi-D simulations are a well-suited and necessary tool to investigate the underlying mechanism of CCSNe.
There exists a growing set of 2D and 3D CCSN explosion models \citep[see e.g.][]{janka12,burrows13,nakamura15,janka16,bruenn16,Burrows2018}.
Note that while 2D simulations tend to explode more easily, in 3D the resulting explosion energies may be higher, as is shown e.g.\ in  \cite{takiwaki14,lentz15,melsona15,Melson15,Mueller15,janka16,hix16}.
Multi-dimensional simulations are also computationally intensive.
They are prohibitively expensive if large spatial domains are required and/or if simulations are required to follow the evolution for long timescales beyond the onset of the explosion.
Thus, given the present status, it is still too early to provide complete predictions from multi-D simulations for an extended sample of progenitor stars.

Computationally more affordable models would be a viable tool for studies of large samples of stars but self-consistent state-of-the-art spherically-symmetric simulations of CCSNe do not lead to explosions via the delayed neutrino-driven mechanism \citep[see e.g.][]{liebendoerfer04}. 
The reason is simply the lack of dimensionality which is the basis for many of the proposed explosion mechanisms, e.g.\ the convective or SASI-aided neutrino-driven (see references given above) and the magnetorotational mechanism \citep{Winteler.ea:2012,moesta14,moesta15,nishimura15,nishimura17,halevimoesta18}. 
An exception is the phase-transition mechanism \citep{sagert09,fischer11}, which leads to successful SN explosions in one-dimensional models. Given the complexities and computational costs of multi-D approaches, a number of approximate 1D approaches have been proposed, which try to mimic the net effects of multi-D simulations.
Such parametrized explosions induced in spherically symmetric models are still a pragmatic and valuable approach to study large numbers of stellar progenitors, from the onset of the explosion up to several seconds after core bounce. 

In the past some simplified approaches have been used to predict supernova nucleosynthesis, which artificially induce explosions with estimated typical explosion energies \citep[e.g.][and many more]{woosley95,thielemann96}. 
Interim approaches beyond piston or thermal bomb models 
attempt to mimic multi-D neutrino heating in a spherical approach, in order to obtain more appropriate predictions of the explosion energy, the mass cut between neutron star and ejecta, as well as nucleosynthesis (including the effects of neutrinos on $Y_e$). This includes the ``neutrino light-bulb method'' --- where the proto-neutron star (PNS) is excised and replaced with an inner boundary condition which contains an analytical prescription for the neutrino luminosities 
\citep [e.g.][]{yamasaki05,iwakami08,yamamoto13} --- and the ``absorption methods'' --- where the neutrino energy-deposition is increased by modifying the neutrino opacities in spherically-symmetric models with detailed Boltzmann neutrino transport \citep{froehlich06,froehlich06b,fischer10}. 

More recent spherically-symmetric approaches try to mimic the effect of multi-D neutrino transport in a way that is adapted more consistently to core collapse and PNS accretion. These approaches still need calibrations which can be provided by comparison with a variety of observations of explosion energies, deduced ejected $^{56}$Ni-masses, progenitor properties, as well as the outcome of multi-D studies. \cite{ugliano12} presented a more sophisticated light-bulb method to explode spherically symmetric models, using neutrino energy deposition in post-shock layers. They used an approximate, gray neutrino transport and replaced the innermost 1.1~\msun of the PNS by an inner boundary. The evolution of the neutrino luminosity at the boundary was based on an analytic cooling model of the PNS, which depends on a set of free parameters. These parameters (within the so-called Prometheus-Hot Bubble (P-HOTB) approach) are set by fitting observational properties of SN~1987A for progenitor masses around 20~\msun\  \citep[see also][]{ertl16,sukhbold16}. 

\cite{push1} --- our Paper~I --- utilized the energy in muon and tau neutrinos as an additional energy source that approximately captures the essential effects of multi-D neutrino transport (the PUSH method).
PUSH relies on the so-called delayed neutrino-driven mechanism as the central engine of CCSNe. In particular, it provides an artificially-enhanced neutrino energy deposition inside the gain region in spherically symmetric models by depositing a fraction of the energy carried by heavy flavor neutrinos behind the shock (see detailed discussion in Section~\ref{sec:method}).
This allows to trigger explosions in 1D simulations without modifying $\nu_e$ and $\bar{\nu}_e$ luminosities nor changing charged current reactions. This increases the accuracy of treating the electron fraction for the innermost ejecta, which is a crucial ingredient for nucleosynthesis calculations. 
The PUSH method is also still parametrized, in the sense that the additional energy deposition is calibrated by comparing the obtained explosion energies and nucleosynthesis yields with observations of nearby supernovae (in particular, SN~1987A).

As mentioned above, a major open question is whether core collapse eventually leads to a supernova explosion with a neutron star remnant or whether the final outcome is a central black hole.
Using P-HOTB, \cite{sukhbold16} show that both possible outcomes can occur within the same mass interval, mainly dependent on the pre-collapse stellar model and its compactness \citep{OConnor.Ott:2011}. 
The best observed core-collapse supernova to date is SN~1987A, which is commonly used as the standard against which numerical supernova models are tested. For SN~1987A, observations provide its explosion energy, ejected masses of $^{56,57,58}$Ni and $^{44}$Ti as well as progenitor mass and metallicity. 
Besides the well-observed SN~1987A, many other CCSNe are known.
In addition, observational evidence for failed CCSN explosions has recently become available.
To date, the LIGO-VIRGO collaboration has detected gravitational wave signals from four different BH-BH merger events \citep{ligo_GW150914,ligo_GW151226,ligo_GW170814,ligo-GW170608}. The individual pre-merger BH masses in these events range from $7^{+2}_{-2}\,M_\odot$ (GW170608) to $36^{+5}_{-4} M_\odot$ (GW150914). In two of these events, both pre-merger BHs have masses above 25~M$_{\odot}$. One possible formation mechanism of these stellar mass BHs is through failed SNe of low-metallicity massive stars \citep{ligo_GW150914}.
A complementary observation has been recently made by \cite{kochanek:searchfornothing} where they used \emph{Hubble Space Telescope} imaging to confirm the optical disappearance of a 25~M$_{\odot}$ red supergiant. This suggests that stars with initial masses around 25~M$_{\odot}$ may end their life in a failed explosion, ultimately resulting in a BH.
In a recent review, \cite{smartt.missingRSG:2015} argues, based on a distance-limited sample, that supernovae in the local Universe are, on the whole, not produced by high-mass stars. They interpret their result in such a way that the missing high-mass SN progenitor stars are the high-compactness stars that die as failed SN or directly collapse to a black-hole.
However, \citet{Davis.Beasor:2018} argue that cut-off mass for progenitors of type~II-P and II~L CCSNe may be 25~M$_{\odot}$, which is higher than previously assumed 17~M$_{\odot}$ due to a source of systematic error in converting pre-explosion photometry into an initial ZAMS mass. Full-scale multi-D simulations of core-collapse obtaining resulting BH formation for 40~M$_{\odot}$ and 70~M$_{\odot}$ ZAMS masses \citep{pan18,kuroda18}.

Improved modelling approaches in spherical symmetry like P-HOTB and PUSH represent for the first time treatments that provide supernova explosion energies and resulting neutron star masses, as well as the transition to black hole formation as a function of ZAMS mass. This modeling also permits to investigate the transition from regular CCSNe to faint \textbf{supernovae, if they exist, }or failed supernovae as function of stellar mass.

In Paper~I, the PUSH method was calibrated to SN~1987A, using pre-explosion models representing red giants in the mass range of 18--21~M$_{\odot}$ such that the explosion energy and the yields of $^{56-58}$Ni and $^{44}$Ti are consistent with their corresponding values derived from observations as proof of principle.
Here, in Paper~II, 
we discuss how PUSH is extended to a broad range of progenitor masses. 
We apply the refined PUSH method to large samples of solar-metallicity pre-explosion models from \cite{Woosley.Heger:2002} and \cite{hw07}, with masses between 10.8 and 120~M$_{\odot}$
and we present the explosion properties and the progenitor-remnant connection.
A detailed discussion of the nucleosynthesis obtained with PUSH is presented in a paper by \cite{push3} (Paper~III).

The paper is organized as follows: In Section~\ref{sec:method} our numerical setup and the input models are discussed. The focus is on the changes compared to Paper~I. 
Next, based on general features of CCSN observations, we introduce requirements that the PUSH method should fulfill. These requirements are formulated in the form of constraints on the parametrized neutrino heating in Section~\ref{sec:expls}. In Section~\ref{sec:progenitorscan} we present the resulting systematics of explosion properties across the ZAMS mass range and trends in compactness. Furthermore, we show the predicted neutron star and BH mass distributions resulting from this approach in Section~\ref{sec:remnants} and summarize our results in Section~\ref{sec:discuss}.

\section{Method and Input}
\label{sec:method}

The PUSH method provides a computationally efficient and physically motivated framework to explode massive stars in spherical symmetry. It allows to study multiple aspects related to CCSNe that require modeling of the explosion for several seconds after its onset and for extended sets of progenitors. The method, based on its calibration, is also well-suited to explore the effects of the shock passage through the star
and to predict the neutron-star and black hole mass distribution as function of stellar mass. PUSH relies on the so-called delayed neutrino-driven mechanism as the central engine of CCSNe. In particular, it provides an artificially-enhanced neutrino energy deposition inside the gain region in spherically symmetric models, which do not explode in self-consistent simulations. This more efficient energy deposition is inspired by the increase of the net neutrino heating that a fluid element experiences due to the presence of multi-dimensional effects. Unlike other methods that employ external energy sources or that use modified electron flavor neutrino luminosities to trigger artificial explosions, PUSH deposits a fraction of the energy otherwise carried away by heavy flavor neutrinos ($\nu_x = \nu_{\mu}, \bar{\nu}_{\mu}, \nu_{\tau}, \bar{\nu}_{\tau}$) behind the shock to ultimately provide successful explosion conditions. In self-consistent core-collapse models, the $\nu_x$'s present a marginal dependence on the temporal evolution of the accretion rate \citep[e.g.][]{liebendoerfer04}, 
and their contribution to the energy deposition inside the gain region is negligible.  However, their usage in PUSH presents a number of advantages. The properties of the $\nu_x$ emission, which includes dynamical feedback from accretion history as well as cooling properties of the forming compact object, correlate significantly with the main features of the $\nu_e$ and $\bar{\nu}_e$ emission \citep{oconnor13}. 
Moreover, the accretion luminosity depends not only on the accretion rate but also on the evolution of the mass and radius of the PNS, which is treated accurately and self-consistently in this method. 
This allows to trigger explosions in 1D simulations without modifying $\nu_e$ and $\bar{\nu}_e$ luminosities nor changing charged current reactions. This increases the accuracy of treating the electron fraction for the innermost ejecta, which is a crucial ingredient for nucleosynthesis calculations. 
In addition, unlike the electron (anti-)neutrino luminosities, which decrease suddenly once the shock has been revived in spherically symmetric models, $\nu_x$ luminosities are only marginally affected by the development of an explosion. This allows PUSH to continue injecting energy inside the expanding shock for a few hundreds of milliseconds after the explosion has set in. 
A first implementation of the PUSH method and its calibration strategy was extensively documented in Paper~I \citep{push1}.

\subsection{Simulation Framework of PUSH}
\label{subsec:framework}

\begin{figure}[]  
\begin{center}
	\includegraphics[width=0.48\textwidth]{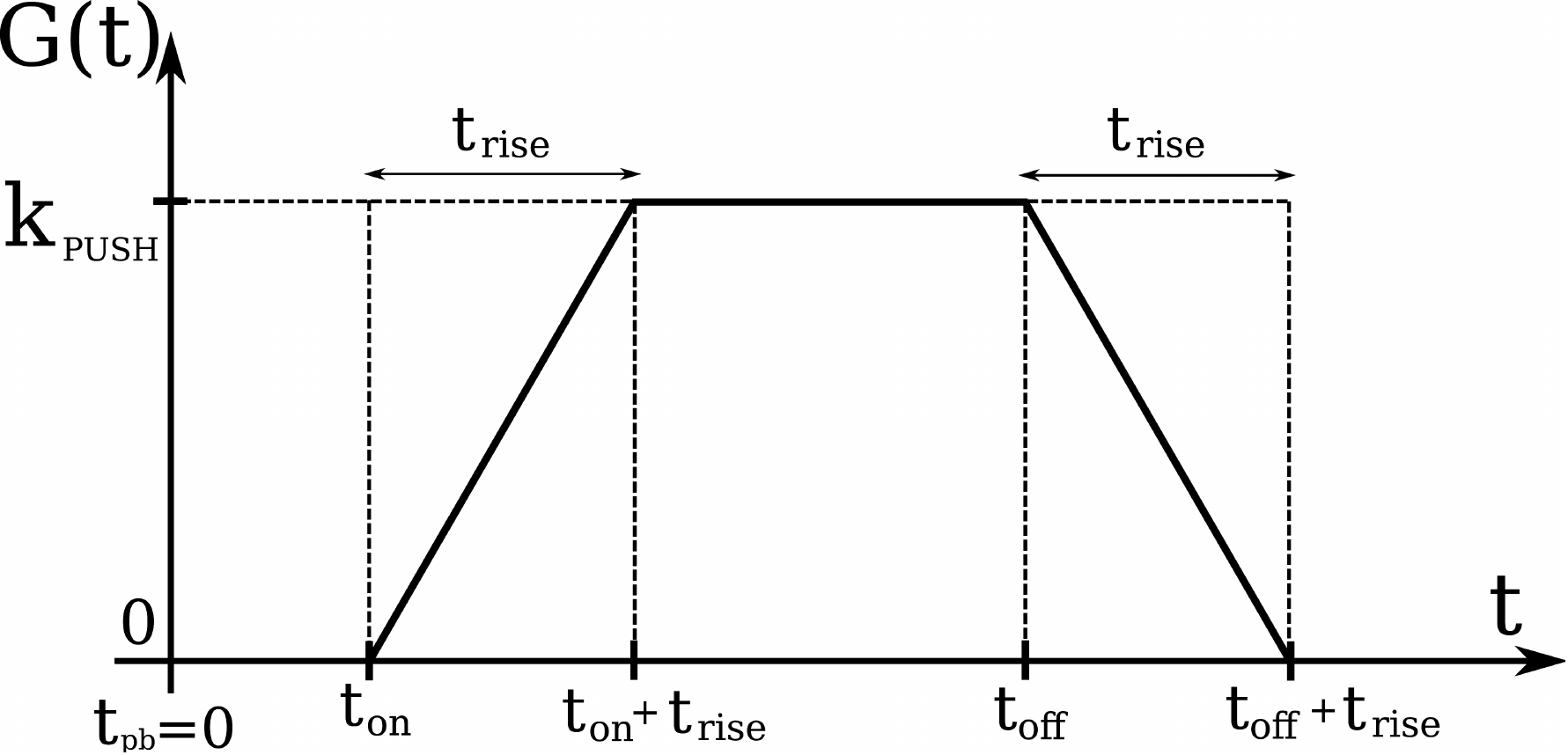}
	\caption{Temporal behavior of the heating due to PUSH. Two parameters ($t_{\text on}$ and $t_{\text{off}}$) are robustly set \citep{push1}, while $k_{\text{PUSH}}$ and $t_{\text{rise}}$ are free parameters.
		\label{fig:push_G-function}
        }
\end{center}
\end{figure}

The simulations presented in this Paper have been performed with the same numerical setup as described in \cite{push1}. Hence, we only provide the essential aspects here and refer the reader to Paper~I for more details. We use the general relativistic hydrodynamics code Agile, which uses an adaptive mesh to achieve a good resolution at the shock front and at the PNS surface \citep{Liebendoerfer.Agile}. For the neutrino-transport we use IDSA for the electron flavor neutrinos \citep{Liebendoerfer.IDSA:2009} and ASL for the heavy flavor neutrinos \citep{perego16}.
Dense and hot nuclear matter in NSE conditions is described by the HS(DD2) nuclear, finite temperature EOS \citep{Hempel.SchaffnerBielich:2010,typel10}. For the non-NSE regime, we use an extension to ideal gas coupled with an approximative alpha-network \citep{Perego2014,Hempel.SchaffnerBielich:2010}.
The PUSH method relies on additional (parametrized) heating from heavy-lepton flavor neutrinos. 
It was introduced to take increased heating efficiencies based on multi-dimensional effects into account and it enables exploding models in spherical symmetry. 
The additional energy deposition is given by the local heating term 
\begin{equation}  
   Q^+_{\push} (t,r) = 4 \, \mathcal{G}(t) \int_0^{\infty} q^+_{\push}(r,E) \, dE ,
   \label{eq:push_integral}
\end{equation}
where
\begin{equation}  
   q^+_{\push}(r,E) \equiv 
   \sigma_0 \;
   \frac{1}{4 \, m_b} 
   \left( \frac{E}{m_e c^2} \right)^2 
   \frac{1}{4 \pi r^2} 
   \left( \frac{dL_{\nu_x}}{dE} \right)
   \mathcal{F}(r,E) .
   \label{eq:push_qdot}
\end{equation}
The term $(dL_{\nu_x}/dE)/(4 \pi r^2)$ is the spectral energy flux for any single $\nu_x$ neutrino species with energy $E$, $\sigma_0$ the typical neutrino cross-section, and $m_b \approx 1.674 \times 10^{-24}{\mathrm g}$ an average baryon mass.
The function $\mathcal{G}(t)$ (which contains the two free parameters $k_{\mathrm{push}}$ and $t_{\mathrm{rise}}$) controls the temporal evolution of PUSH, as shown in Figure~\ref{fig:push_G-function}.

The spatial dependence of PUSH is described by
\begin{equation}   
  \label{eq:fre_push}
  \mathcal{F}(r,E) =
  \left\{
    \begin{array}{ll}
      0 & \mbox{if} \quad r > R_{s} \quad \mbox{or} \quad \dot{e}_{\nu_e,\overline{\nu}_e} < 0 \\
      \exp (- \tau_{\nu_e}(r,E)) & \mbox{otherwise} \\
      \end{array}
  \right. ,
\end{equation}
where $R_s$ is the shock radius, $\dot{e}_{\nu_e,\bar{\nu}_e}$ the net specific energy deposition rate due to electron neutrinos and anti-neutrinos, and $\tau_{\nu_e}$ is the optical depth for electron neutrinos.
Note that the entropy-gradient criterion from Paper~I (where PUSH only provides extra heating in regions where $ds/dr<0$) is no longer included in the spatial heating term.
The effect of lifting the entropy-gradient criterion on the simulations and on our calibration procedure will be discussed in details in Section~\ref{subsec:dsdr}.

We stress that we do not modify the electron neutrino and antineutrino transport (hence, preserving a consistent $Y_e$-evolution) and that the mass cut emerges naturally from the simulations. Moreover, we want to emphasize that we can self-consistently follow the simulation through to the formation of a black hole ($\rho_{\mathrm{c}} > 10^{15}$g~cm$^{-3}$) since we include the full PNS in our computational domain.

\subsection{Initial Models}
\label{subsec:progenitors}

We use two sets of non-rotating pre-explosion models with solar-metallicity from the stellar evolution code {\tt KEPLER} presented in \cite{Woosley.Heger:2002} and \cite{hw07}. From here on, we will refer to these sets of pre-explosions models as WHW02 and WH07, respectively. The two sets of models span zero-age main sequence (ZAMS) masses between 10.8~M$_{\odot}$ and 75~M$_{\odot}$ (WHW02) and between 12~M$_{\odot}$ and 120~M$_{\odot}$ (WH07).
Table~\ref{tab:all_prog_series} summarizes all the pre-explosion models used, including their ZAMS mass and a unique label for each model.

\begin{table}  
\begin{center}
	\caption{Pre-explosion models used in this study.
    	\label{tab:all_prog_series}
	}
	\begin{tabular}{lllllc}
	\tableline \tableline 
Series & Label & Min Mass & Max Mass & $\Delta m$ & Ref. \\
 &  & (M$_{\odot}$) & (M$_{\odot}$) & (M$_{\odot}$) &  \\
	\tableline
WHW02 & s & $10.8$ & $28.2$ & $0.2$ & 1 \\
      &   & $29.0$ & $40.0$ & $1.0$ & 1 \\
      &   & $75.0$ &        &       & 1 \\
WH07 & w & $12.0$ & $33.0$ & $1.0$ & 2 \\
     &   & $35.0$ & $60.0$ & $5.0$ & 2 \\
     &   & $70.0$ &        &       & 2 \\
     &   & $80.0$ & $120.0$& $20.0$& 2 \\
	\tableline
	\end{tabular}
\end{center}
\tablecomments{All models have solar metallicity. Note that the pre-explosion models are irregularly spaced in mass. The $\Delta m$ indicates the mass spacings of available pre-explosion models for each mass interval Min Mass to Max Mass}
\tablerefs{(1)~\citet{Woosley.Heger:2002}; (2)~\citet{hw07}}
\end{table}

\begin{figure}[]  
	\includegraphics[width=0.48\textwidth]{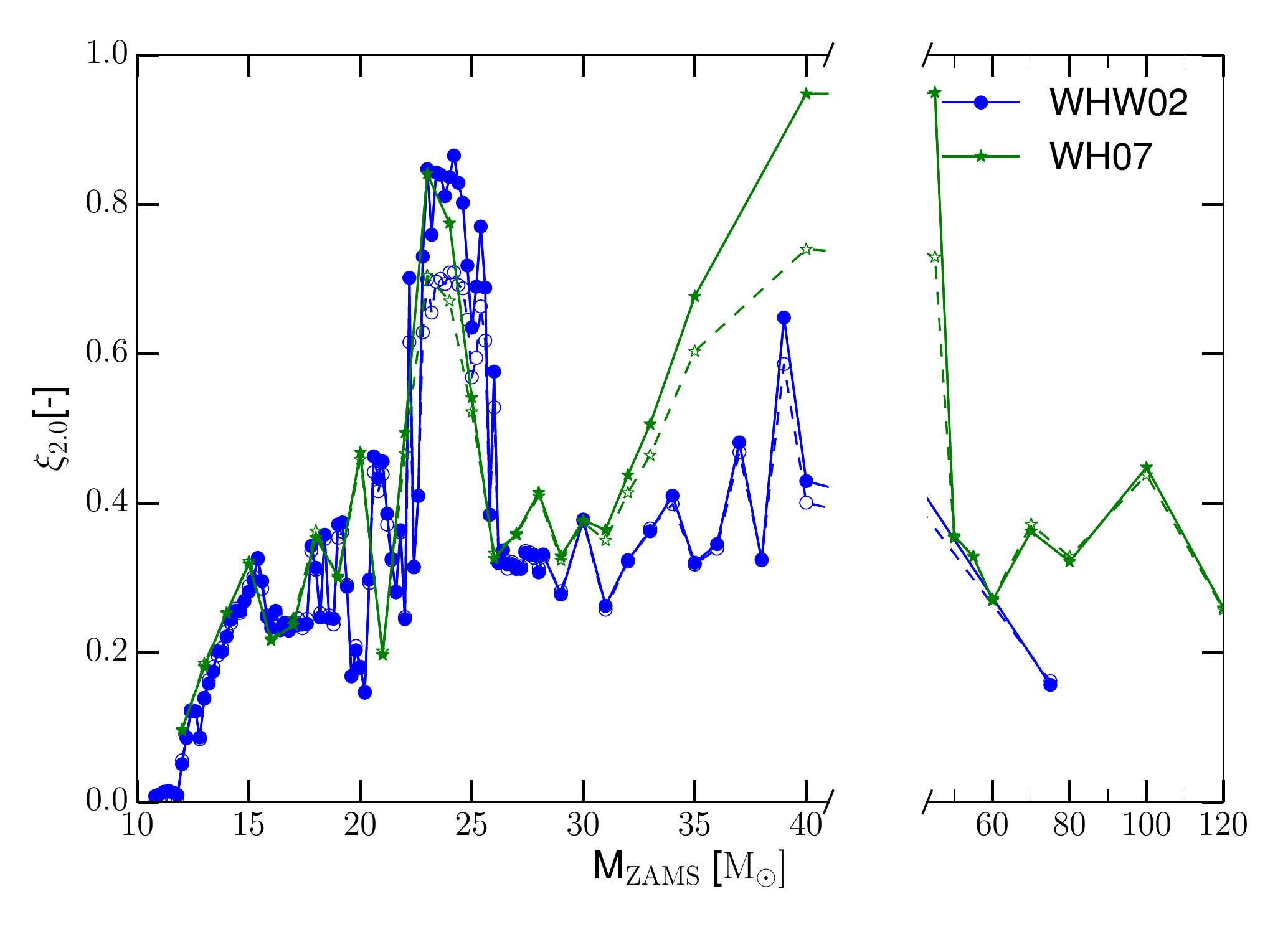}
	\caption{Compactness $\xi_{2.0}$ as function of ZAMS mass for the WHW02 (blue) and the WH07 (green) progenitors. The solid lines denote the compactness at bounce; dashed lines indicates the compactness of the pre-explosion model (typically at the onset of collapse).
		\label{fig:prog_compactness}
        }
\end{figure}

\begin{figure}[]  
	\includegraphics[width=0.48\textwidth]{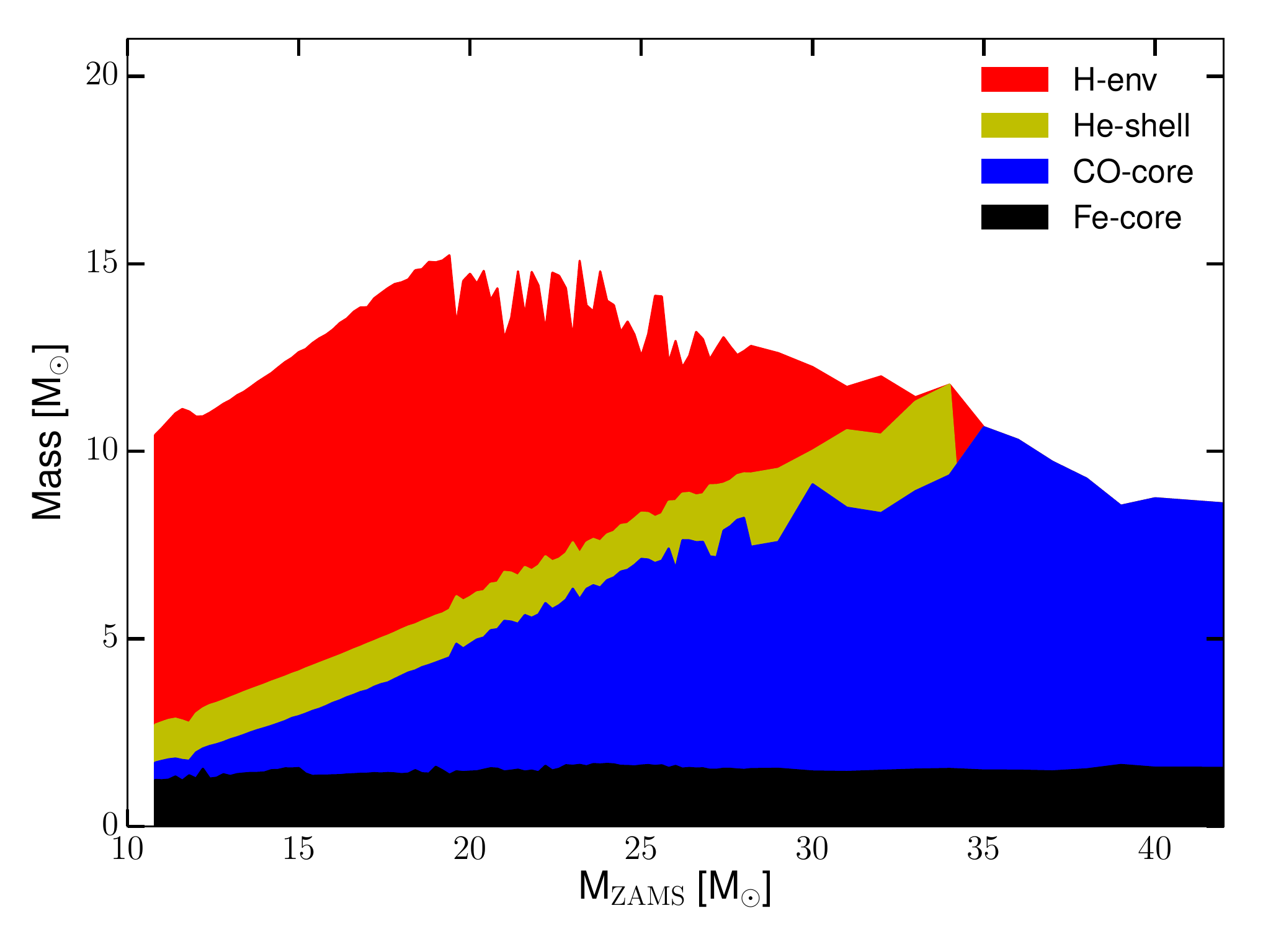}
	\caption{Fe-core mass (black), CO-core mass (blue), He-shell mass (yellow), and total mass including the H-envelope (red) as function of ZAMS mass at the onset of collapse for the WHW02 progenitors.
		\label{fig:prog_mass_structure}
        }
\end{figure}

\begin{figure}[]  
	\includegraphics[width=0.48\textwidth]{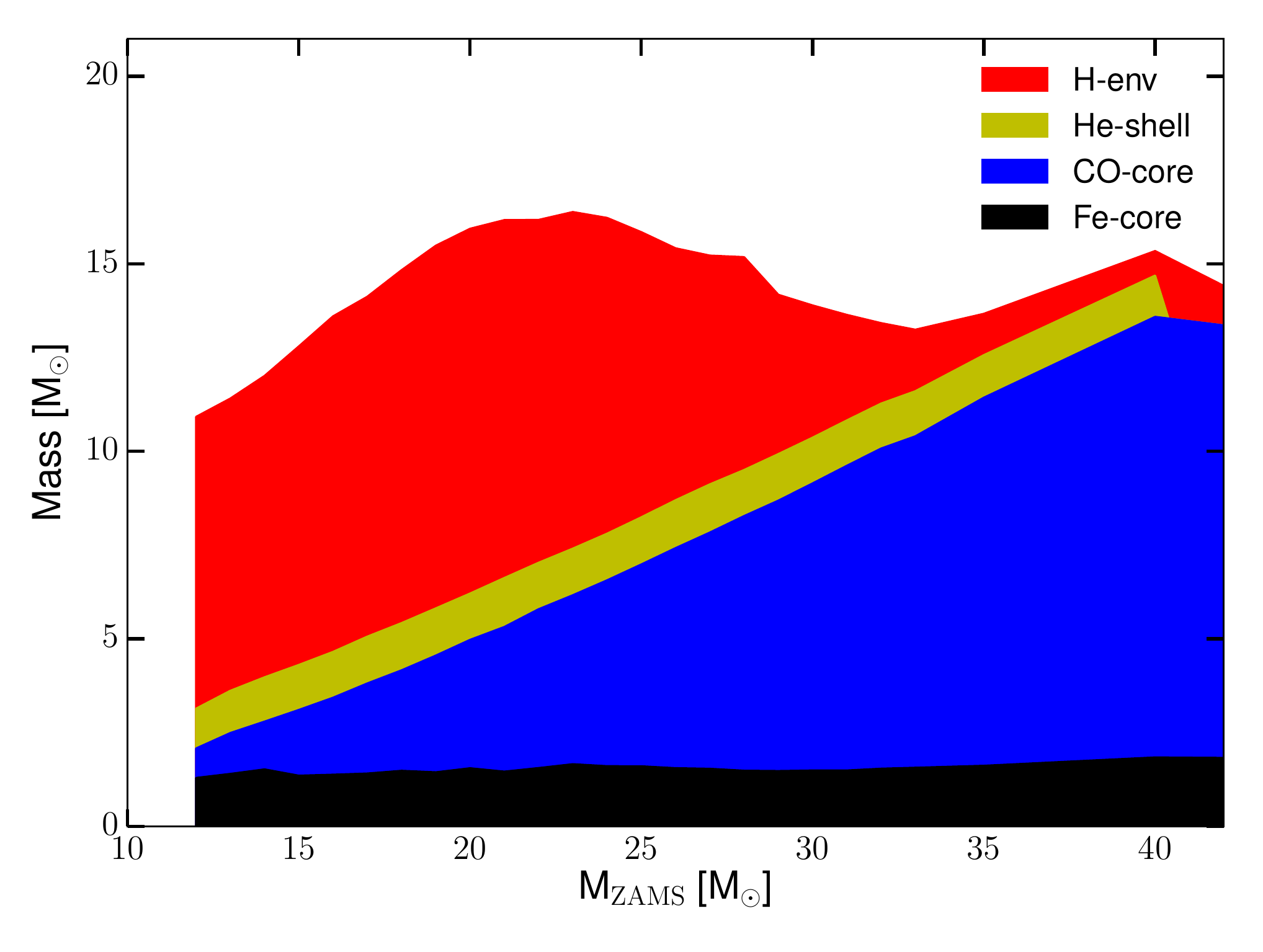}
	\caption{Fe-core mass (black), CO-core mass (blue), He-shell mass (yellow), and total mass including the H-envelope (red) as function of ZAMS mass at the onset of collapse for the WH07 progenitors.
		\label{fig:prog_mass_structure2}
        }
\end{figure}

It is well-known that the initial stellar mass is a poor predictor of the final fate (successful or failed explosion) of massive stars. Hence, we make use of the compactness parameter \citep{OConnor.Ott:2011}
\begin{equation} 
\xi_{M} \equiv \frac{M/M_{\odot}}{R(M)/1000\mathrm{km}}.
\label{eq:compactness}
\end{equation}
Compactness values evaluated for $M$ between 1.5 and 3.0~M$_{\odot}$ are commonly used, see e.g. \cite{OConnor.Ott:2011,oconnor13,nakamura15,Burrows2018}.
In our Paper I, where we investigated exploding models that were aimed to reproduce SN1987A, we used $M=1.75$~M$_{\odot}$. In this work, we investigate a much broader ZAMS mass range of progenitors, which involves the formation of BHs and more massive neutron stars than in the previous study. To account for the larger possible mass included in the layers of the star that are crucial for the explosion mechanism as well as the larger possible neutron star masses and BH formation, we chose compactness values (at bounce) evaluated at 2~M$_{\odot}$.
The compactness parameter for all WHW02 and WH07 models is shown in Figure~\ref{fig:prog_compactness} where solid lines denote the compactness at bounce and dashed lines at the onset of collapse. The difference in compactness between the two sets illustrates the uncertainties involved in the pre-explosion models themselves.
The compactness curves reflect the structure of the pre-explosion models, as illustrated in Figures~\ref{fig:prog_mass_structure} and \ref{fig:prog_mass_structure2} which show the Fe-core mass (defined as the layers with $Y_e < 0.495$), the carbon-oxygen (CO) core mass (enclosed mass with $X_{\mathrm{He}} \leq 0.2$, i.e. up to the beginning of the He-shell), He-core mass (mass regions with $X_{\rm H}\leq 0.2$, i.e.\ up to the beginning of the H-shell), and the total mass including the H-envelope at the end of the stellar evolution simulation for the WHW02 and WH07 models.
For models with ZAMS masses up to $\sim$ 35~M$_{\odot}$ (WHW02) and $\sim$40~M$_{\odot}$ (WH07), respectively, the CO-core mass grows continuously with increasing initial mass, while the envelope exhibits a different behavior: The helium-envelope remains approximately constant up to 20--25~M$_{\odot}$ ZAMS mass. For higher ZAMS masses, mass loss decreases the envelope mass (hydrogen and helium) until at approximately 30--40~M$_{\odot}$ ZAMS mass, the models are almost completely stripped of their envelope. The decrease of the H/He-envelope mass is mirrored in the compactness.
Beyond 30~M$_{\odot}$ the rise in compactness reflects the increasing CO-core mass, as these models have lost virtually all of their H- and He-envelope.
For initial masses above 40~M$_{\odot}$, the evolution is quite uncertain in particular because mass loss is poorly understood. We include these models for completeness, however we caution the reader that they may be quite uncertain.
Moreover, successful explosions of models beyond 30~M$_{\odot}$ result in Type Ib or Ic SNe due to the loss of the H- (and He-) envelope.
More generally, the efficiency of semiconvection and overshooting, together with the metallicity, leads to a rather complex interplay of different burning shells \citep{rauscher2002} which contributes to the rapid, non-monotonic variations in compactness with ZAMS mass \citep{sukhbold.woosley:2014,sukhbold17}.

\subsection{Hydrodynamic simulations}
\label{subsec:hydro}

Using the framework described in Section~\ref{subsec:framework} and the initial models from Section~\ref{subsec:progenitors}, we perform hydrodynamic simulations of the collapse, bounce, and post-bounce phases. We include the pre-explosion model up to the He-layer in the hydrodynamic simulations with the exception of the low mass model s11.2 where the computational domain ends just below the He-layer and a few models with ZAMS masses above 30~M$_{\odot}$ with large pre-explosion mass loss. 
The total run time of the simulation depends on its outcome which we categorize based on the explosion energy (representing the kinetic energy of the ejected matter) and the central density (indicating black hole formation if in excess of $\sim 10^{15}$gm~cm$^{-3}$ and ultimately diverging). We compute the explosion energy as in Paper~I: 
we assume that the total energy of the ejecta with rest-masses subtracted eventually converts into kinetic energy of the expanding supernova remnant at late times and we include the gravitational binding energy of the progenitor (see also Equations (12)--(17) in Paper~I).
For exploding models, we define the explosion time $t_{\rm expl}$ as the time when the shock reaches 500 km, measured with respect to core bounce.
We distinguish between successful explosions, black hole formation, and failed explosions, defined as follows:
(i) If a simulation completes the set total simulation time of $t_{\mathrm{final}}=5$~s (approximately 4.6~s -- 4.8~s post bounce) and/or has a saturated positive explosion energy it corresponds to an explosion. 
(ii) If a simulation forms a black hole in the 5~s run time it corresponds to a non-exploding model (black hole formation).
(iii) If a simulation has a negative explosion energy at times when PUSH is no longer active (or at $t_{\mathrm{final}}$) it corresponds to a failed explosion and hence a non-exploding model which eventually will form a black hole.

\subsubsection{Entropy-gradient criterion} \label{subsec:dsdr}

\begin{figure}  
	\includegraphics[width=0.48\textwidth]{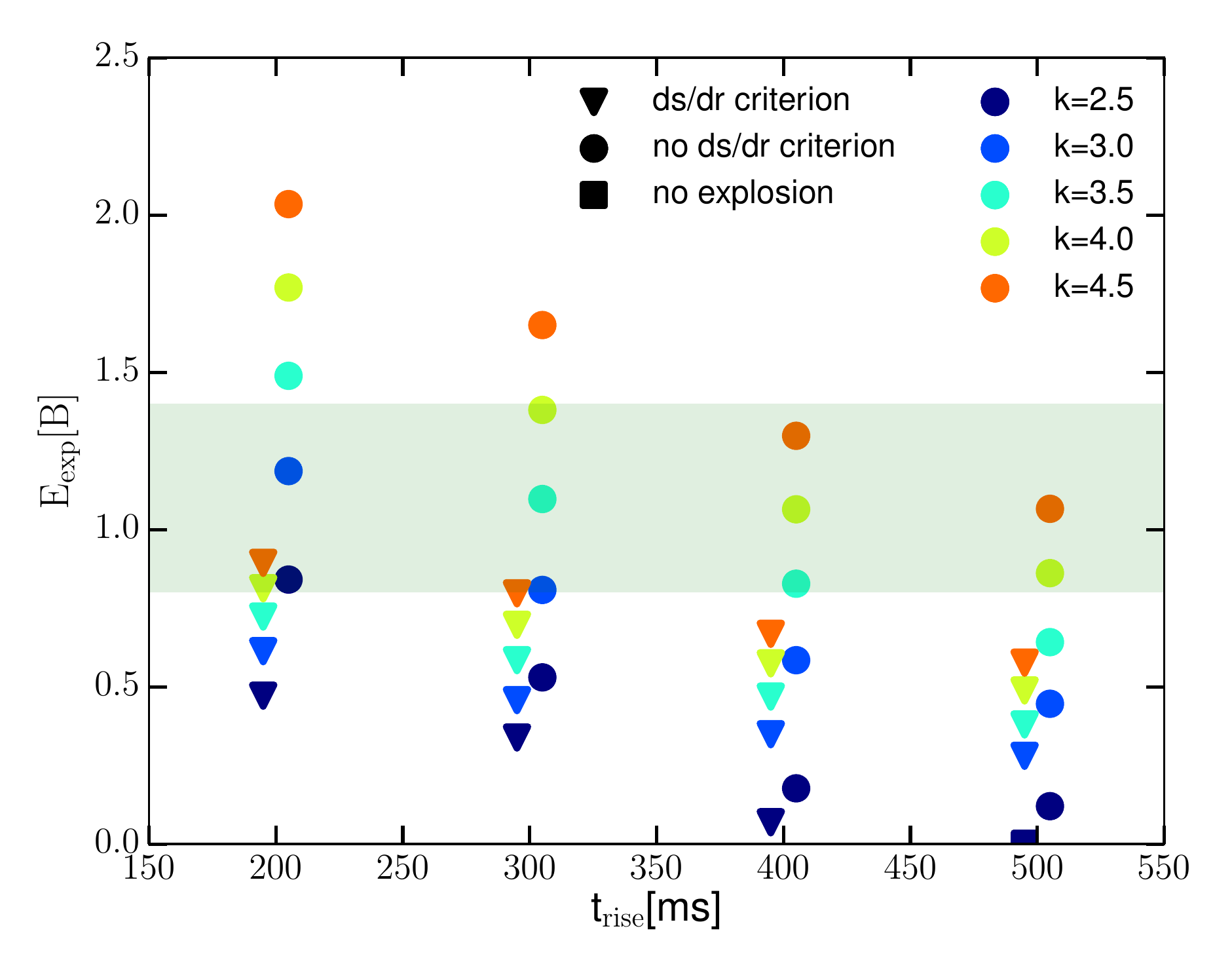}
	\caption{Explosion energy as a function of the PUSH parameter $t_{\mathrm{rise}}$ for different values of $k_{\rm{push}}$. The circles are without the entropy-gradient criterion (this work); the triangles are with the entropy-gradient criterion (Paper~I). The black square indicates a simulation that did not explode. The shaded band indicates the observational explosion energies for SN~1987A. Note that the circles (triangles) are slightly offset to the right (left) of the true value of $t_{\mathrm{rise}}$ for visualization purposes.
		\label{fig:dsdr_effects}
     }
\end{figure}

As indicated in Section~\ref{subsec:framework}, the current version of the PUSH heating term does not include the entropy-gradient criterion anymore.
Unlike simulations in spherical symmetry, multi-dimensional simulations exhibit simultaneous downflow and outflow of matter. These regions are at similar distances from the PNS, but have quite different entropies (see for example \cite{Pan.Flash.Agile:2016}). This is a fundamentally multi-dimensional behavior and difficult to reproduce in spherically symmetric simulations.
The spatial heating term $\mathcal{F}(r,E)$ in Paper~I included a criterion (the entropy-gradient criterion) which restricted heating to regions where the one-dimensional system fulfills the condition for convective instability according to the Schwarzschild criterion (i.e.\ to zones with $ds/dr<0$).
This proved to be too restrictive in spherically symmetric simulations and not allow for the extended heating seen in multi-dimensional systems. It introduces a self-canceling effect of the increased heating efficiency on a relatively short timescale. This severely limits the heating and considerably confines the allowed parameter space.
This self-canceling effect of the heating due to the entropy-gradient criterion required relatively small values of \trise in Paper~I and hence led to a rapid evolution of the entropy profiles.
When combined with the observational constraints from SN~1987A (explosion energy and Ni ejecta yields), this resulted in relatively early explosions. 
The explosion time impacts the location of the mass cut and hence the ejecta mass. The mass cut determines the total amount of Ni ejecta and also how much material with $Y_e<0.5$ is ejected (and hence the relative amounts of $^{57}$Ni and $^{58}$Ni).
For early explosions, the mass cut resides deeper in the pre-explosion structure while for late explosions the mass cut is further out in the star. 
With the entropy criterion on (Paper~I), it was only possible to reproduce SN~1987A if we imposed 0.1~M$_{\odot}$ of fallback by hand.
Generally, fallback consists of two components: the early fallback which can be only determined in multi-dimensional simulations and the late fallback which requires simulation times far longer than what is feasible with the presented PUSH setup. The late fallback, due to shock reflection at density jumps of outer shell boundaries is not included in this study as its effect is considered to be small \citep{sukhbold16}. 
In this work, we relaxed the entropy-gradient criterion. In this case, a broader range of values for \trise allows for explosion with energies around 1~Bethe.
This can be seen in Figure~\ref{fig:dsdr_effects} which shows explosion energy as function of \trise and \kpush without (triangles) and with (circles) the entropy-gradient criterion for the s18.8 model. Points within the shaded region represent simulations that are consistent with observational constraints of explosion energy. Also note that larger values for \trise result in lower explosion energies and in later explosions. For the largest value of \trise considered here, only simulations without the entropy-criterion result in explosion energies of around 1~Bethe.
This leaves the choice between early explosions with the necessity to impose fallback or later explosions without the need for additional fallback. 
This choice only plays a very minor role for the outcome of the simulations (successful explosion with neutron star or failed explosion with black hole), as in both cases the final ejecta mass is very similar (the difference in mass cut is similar to the amount of imposed fallback which is required to reproduce the observed nucleosynthesis yields).
However, it is not obvious how much fallback should be imposed for models with progenitor masses outside of the range of SN~1987A.
We find that the new setup of this Paper (without the entropy-gradient criterion) enables explosions that are in better agreement with the 
temporal evolution of the shock radius, neutrino heating rates and entropy profiles of multi-dimensional simulations by \citet{Pan.Flash.Agile:2016,bruenn16,ebinger.phd}, 
and that have nucleosynthesis yields and explosion energy consistent with SN~1987A.
In the following, all the results presented are for simulations using this new setup.

\subsection{Nucleosynthesis Postprocessing}
\label{subsec:nucpost}

We use a post-processing approach to predict the detailed composition of the ejecta. The ejecta are divided into tracer shells of $10^{-3}$~M$_{\odot}$ each. We follow the trajectory of each mass element that reaches a peak temperature $\geq$ 1.75~GK using a nuclear reaction network which includes isotopes from free nucleons to $^{211}$At. For details we refer to Paper~III.

Within each tracer, we record the evolution of fundamental hydrodynamical and neutrino quantities, including radius, density, temperature, electron fraction, neutrino luminosities and mean energies.
For each mass element of the ejecta we start the nucleosynthesis post-processing when the temperature drops below 10~GK ($\sim$0.86~MeV), using the NSE abundances (determined by the current electron fraction) as the initial composition. For the tracer particles that never reach 10~GK we start the post-processing at the beginning of the hydrodynamic simulation and use the abundances from the approximate $\alpha$-network at this point (WHW02) or from the progenitor model (WH07) as the initial composition.
For the times t$<$t$_{\rm final}$, we use the hydrodynamic evolution. 
If the conditions at $t_{\mathrm{final}}$ are still sufficiently hot for nucleosynthesis to occur, we extrapolate radius, temperature, and density as follows:
\begin{eqnarray}
r(t) &=& r_{\rm final}  + t v_{\rm final} \label{eq:extrapol_rad}, \\
\rho(t) &=& \rho_{\rm final} \left( \frac{t}{t_{\rm final}} \right)^{-3}, \\
T(t) &=& T[s_{\rm final},\rho(t),Y_e(t)] \label{eq:extrapol_t9},
\end{eqnarray}
where $r$ is the radial position, $v$ the radial velocity, $\rho$ the density, $T$ the temperature, $s$ the entropy per baryon, and $Y_e$ the electron fraction of the tracer.  The subscript ``final'' indicates the end time of the hydrodynamical simulation. 
Equations~(\ref{eq:extrapol_rad})~--~(\ref{eq:extrapol_t9}) correspond to a free expansion for the density and an adiabatic expansion for the temperature \citep[e.g.][]{korobkin2012}. 
For the extrapolation we calculate the temperature at each time-step using the equation of state of \citet{Timmes.Swesty:2000}.

\section{Explosion Modeling with PUSH}
\label{sec:expls}

In this Section, we develop PUSH into a method that can be applied to a broad set of pre-explosion models to predict remnant properties (this paper) and detailed nucleosynthesis (Paper~III).
The direct application of the best fit parameters from Paper~I to the whole progenitor range may sound straight forward but we found that a more variable approach is necessary to reproduce the available observational properties of core-collapse supernovae. This complication does not come as a big surprise, since, after all, the understanding of the nature of the explosion mechanism of CCSNe has been elusive for decades.

\subsection{Calibration against SN~1987A}
\label{subsec:87Afit}

In Paper~I, the calibration of PUSH was performed using solar-metallicity models between 18 and 21~M$_{\odot}$ from \cite{Woosley.Heger:2002}. This resulted in two possible candidates, s18.0 and s19.4, both of which were able to reproduce the observed Ni yields when we imposed 0.1~M$_{\odot}$ of fallback by hand.
In this Paper, we use an updated spatial heating term that does not include the entropy criterion anymore (see Section~\ref{subsec:dsdr}).
Therefore, we repeat the calibration procedure as described in Paper~I using the pre-explosion models from the WHW02 set with ZAMS masses between 18.0 and 21.0~M$_{\odot}$ in order to find a suitable candidate able to reproduce the observed properties of SN~1987A. 
We find that the s18.8 model is in good agreement with the observed explosion energy and yields of Ni and Ti for $k_{\mathrm{push}}=4.3$ and $t_{\mathrm{rise}}=400$~ms  (Table~\ref{tab:fitparams}) without the need for any additional fallback.
The observational estimates for SN~1987A are summarized in Table~\ref{tab:bestfit}. These values are the same as used in Paper~I except for $^{44}$Ti where we now use the newer results from \cite{Boggs2015} instead of \cite{Seitenzahl2014}. For comparison, Table~\ref{tab:bestfit} also gives the corresponding values from our calibrated model for SN~1987A.

\begin{table}  
\begin{center}
  \caption{Parameter values from calibration.
  			\label{tab:fitparams}}
  \begin{tabular}{cccc}
  \tableline \tableline 
  	$k_{\mathrm{push}}$ [-] & $t_{\mathrm{rise}}$ [ms] & $t_{\mathrm{on}}$ [ms] & $t_{\mathrm{off}}$ [s] \\
  \tableline
  	4.3 & 400 & 80 & 1 \\
  \tableline 
\end{tabular}
\end{center}
\end{table}

\begin{table}  
\begin{center}
  \caption{Observed and calculated properties of SN~1987A
         \label{tab:bestfit}}
  \begin{tabular}{ccc}
  \tableline \tableline
            Quantity  & SN~1987A & PUSH \\
                      & (observed) & (s18.8) \\
            \tableline
			$E_{\rm expl}$ $(10^{51}~\rm erg)$ & 1.1 $\pm$ 0.3  & 1.2 \\
			$M_{\rm prog}$ (\msun) & 18-21  & 18.8 \\
			$^{56}{\rm Ni}$~(\msun) & $(0.071 \pm 0.003)$ & 0.069 \\ 
			$^{57}{\rm Ni}$~(\msun) & $(0.0041 \pm 0.0018)$ & 0.0027 \\ 
			$^{58}{\rm Ni}$~(\msun) & 0.006 & 0.0066 \\ 
			$^{44}{\rm Ti}$~(\msun) & $(1.5 \pm 0.3) \times 10^{-4}$ &  $3.05 \times 10^{-5}$ \\
			\tableline
\end{tabular}
\end{center}
\tablecomments{The s18.8 model was identified as the progenitor which reproduces SN~1987A for  $k_{\mathrm{push}}=4.3$ and $t_{\mathrm{rise}}=400$~ms. The nucleosynthesis yields for SN~1987A are taken from \cite{Seitenzahl2014} except for $^{58}$Ni which is taken from \cite{Fransson.Kozma:2002} and $^{44}$Ti which reflects the value from \cite{Boggs2015}. The explosion energy is adapted from \cite{Blinnikov2000}. The corresponding values from our calibration model are shown for comparison.
}
\end{table}

\subsection{Black Hole Formation}
\label{subsec:bh-formation}

An important question in the investigation of CCSNe across the whole progenitor mass range is whether a collapsing star ultimately leads to a successful supernova explosion or whether it fails to explode and forms a BH. In this Section we turn our attention to failed SNe and the formation of BHs.
To do this, we investigate the behavior of spherically-symmetric CCSN models without the application of the extra heating from PUSH (i.e.\ setting $k_{\mathrm{push}}=0$). These simulations are not expected to explode. Rather, they will collapse to BHs.  The timescale on which the different models undergo collapse can depend on the progenitor structure and on the choice of the EOS. 
We use a subset of models which samples the mass range up to 40~M$_{\odot}$ of both sets of pre-explosion models (WHW02 and WH07) and perform simulations for two equations of state (SFHO and HS(DD2); \cite{Hempel.SchaffnerBielich:2010,fischer14,sfho}) until the formation of a central black hole which starts to occur for central densities around  $\sim10^{15}$~g/cm$^3$.

\begin{figure}  
	\includegraphics[width=0.48\textwidth]{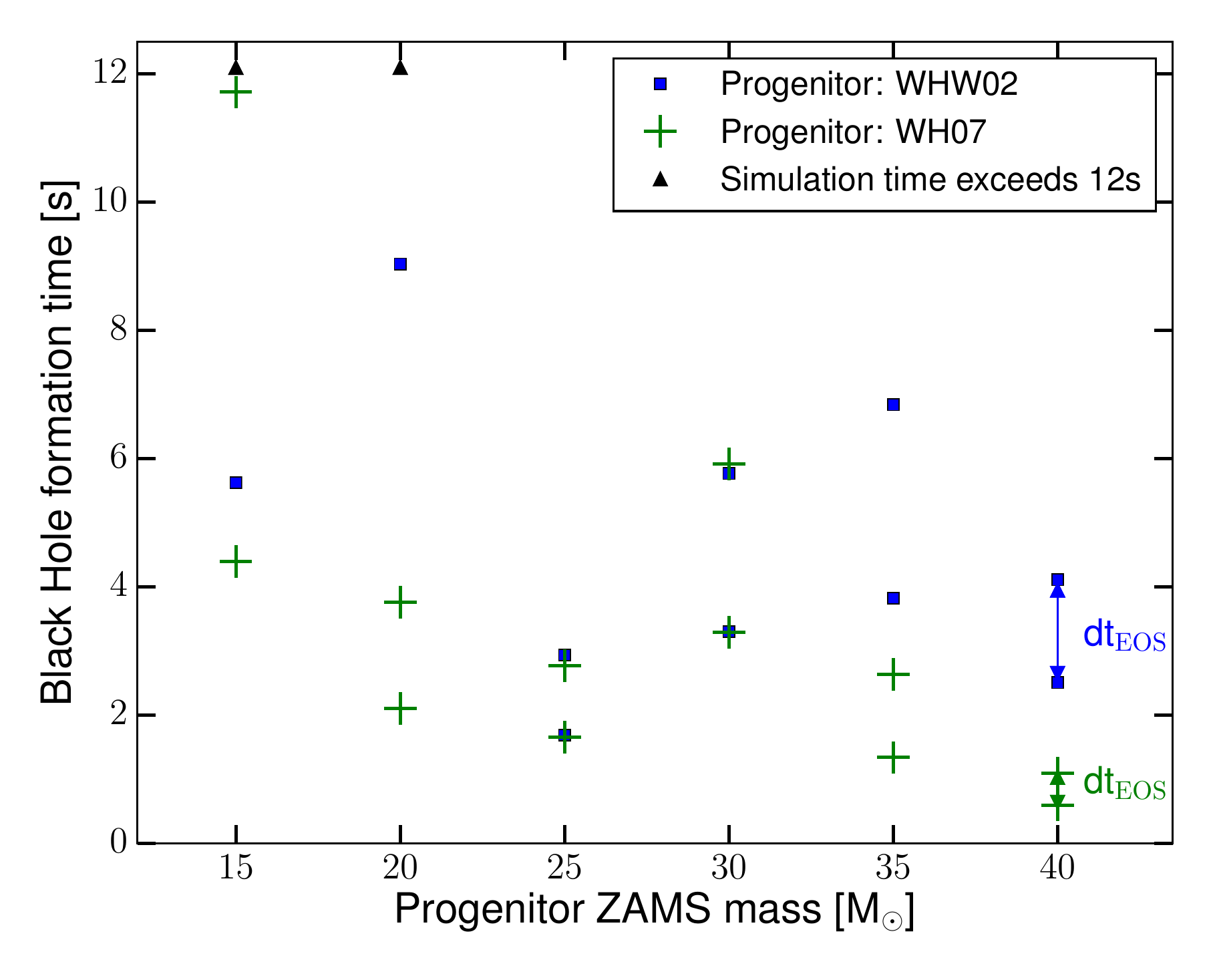}
	\caption{Black hole formation times for a collection of different progenitor ZAMS masses from two different progenitor sets (WH07 in green and WHW02 in blue) and two equations of state (HS(DD2), SFHO). The shorter BH formation time for each model corresponds to the SFHO EOS.
		\label{fig:overviewcollapse}
     }
\end{figure}

\begin{figure}  
	\includegraphics[width=0.48\textwidth]{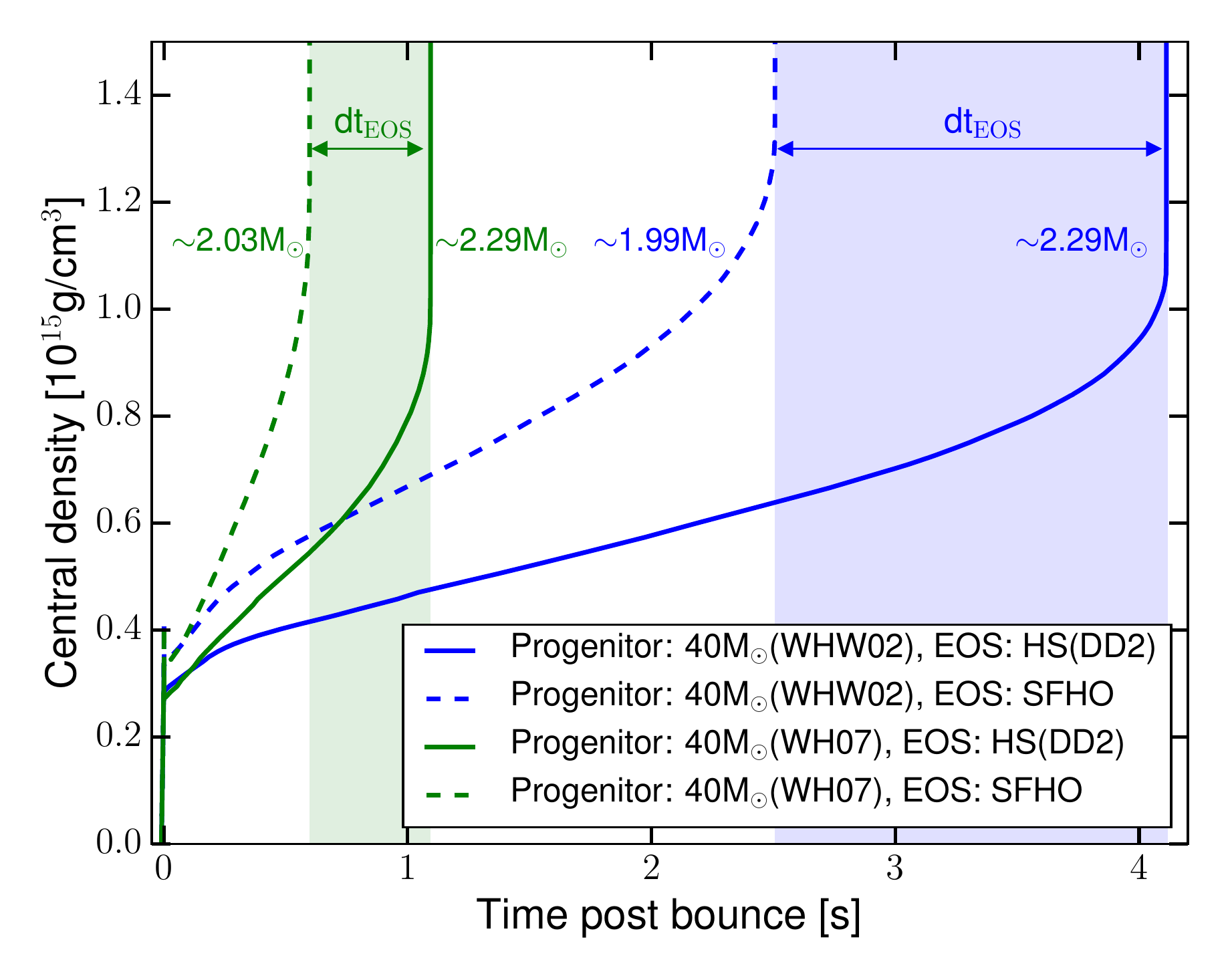}
	\caption{Temporal evolution of the central density of a 40 M$_{\odot}$ solar metallicity star for two progenitor models (WHW02 in blue and WH07 in green) and two equations of state (HS(DD2) solid lines, SFHO dashed lines, \cite{Hempel.SchaffnerBielich:2010},\cite{fischer14}). Gravitational PNS masses at collapse are given next to the corresponding central density curves.
		\label{fig:detailcollapse}
     }
\end{figure}

In Figure~\ref{fig:overviewcollapse} the BH formation times for several pre-explosion models from WHW02 and WH07 are given. The differences in BH formation time between the progenitors can be related to different accretion rates, which are correlated to compactness (see Equation~\ref{eq:compactness}).
Figure~\ref{fig:detailcollapse} shows the temporal evolution of the central density of the two 40~M$_{\odot}$ models from the two pre-explosion model sets (WHW02 in blue; WH07 in green) and for two different equations of state (HS(DD2) solid lines, SFHO dashed lines). It is evident that the BH formation time strongly depends on the EOS used (indicated by the colored areas; models using the SFHO EOS collapse faster than the ones using the HS(DD2) EOS) and even stronger on the progenitor model (difference between green and blue lines). Thus, both aspects can have an impact on the explodability of numerical models since they set the estimated upper limit for the timescale on which a delayed mechanism should revive the stalled shock.

From the collapse timescales (see Figure \ref{fig:overviewcollapse}) we can draw first conclusions on the explodability of the different progenitors. As expected, progenitors with higher compactness have a shorter collapse time. In addition, the WH07 models collapse to BHs on shorter timescales overall. This relation between collapse time, or BH formation timescale, and compactness is also discussed in other works (see e.g. \cite{OConnor.Ott:2011}). In our simulations, the BH candidates have compactness $\xi_{2.0}>0.5$. The BH formation timescales from simulations with $k_{\mathrm{push}}=0$ can be interpreted as upper bound for the timescale on which the neutrino-driven mechanism can yield successful explosions. 
From this, we expect that (at least some) pre-explosion models with compactness $\xi_{2.0}>0.5$ do not explode (i.e.\ form black holes), which can later be summarized as a constraint on the PUSH method when applied to an entire set of pre-explosion models.

\subsection{Standard neutrino-driven CCSNe}
\label{subsec:parabola-k}

The PUSH method represents a parametrization of the neutrino-driven mechanism intended for the investigation of CCSNe in spherically symmetric simulations.
Hence, we focus on standard neutrino-driven CCSNe (between $\sim~10$ and 21~M$_{\odot}$ where 21M$_{\odot}$ represents the upper mass limit of SN~1987A which we assume to be a standard neutrino-driven explosion) and the \textbf{putative} faint/failed SN branch (above 21~M$_{\odot}$).
\textbf{Note that a definitive consensus on the existence of such a ``faint supernova branch'' is still debated based on the explosion observations and progenitor mass determination (see Appendix~\ref{subsec:obs-constraints}).}

Based on general features of CCSN observations, we require that the PUSH method should
\begin{enumerate}[(i)]
\item reproduce the observed properties of SN~1987A for a suitable pre-explosion model, 
\item allow for the possibility of black-hole (BH) formation,
and 
\item result in lower explosion energies for stars with masses of $\lesssim$ 13 M$_{\odot}$ (``Crab-like SNe'') 
\end{enumerate}
when applied across the entire mass range.

In the following, we quantify the general requirements and discuss how they are included in setting the parameters of the PUSH method such that it ultimately can be applied to an entire mass range of pre-explosion models and is in agreement with general observational properties of CCSNe across the ZAMS mass range.

As a first step, we applied the parameters from the calibration (see Table~\ref{tab:fitparams}) to all WHW02 pre-explosion models. This approach leads to robust explosions for all considered stars if reasonable explosion energies of the order of 1~Bethe are supposed to be achieved for suitable candidates of SN1987A.
The resulting explosion energies cannot explain the observations \textbf{and t}hey even partially behave opposite to the expectations (see Section~\ref{subsec:bh-formation} and also Appendix~\ref{subsec:obs-constraints}).
With a constant \kpush for the entire mass range ruled out, a compactness-dependent \kpush factor is the next obvious choice for the calibration of our effective model such that it can fulfill all three requirements we imposed.

\begin{figure}  
	\includegraphics[width=0.48\textwidth]{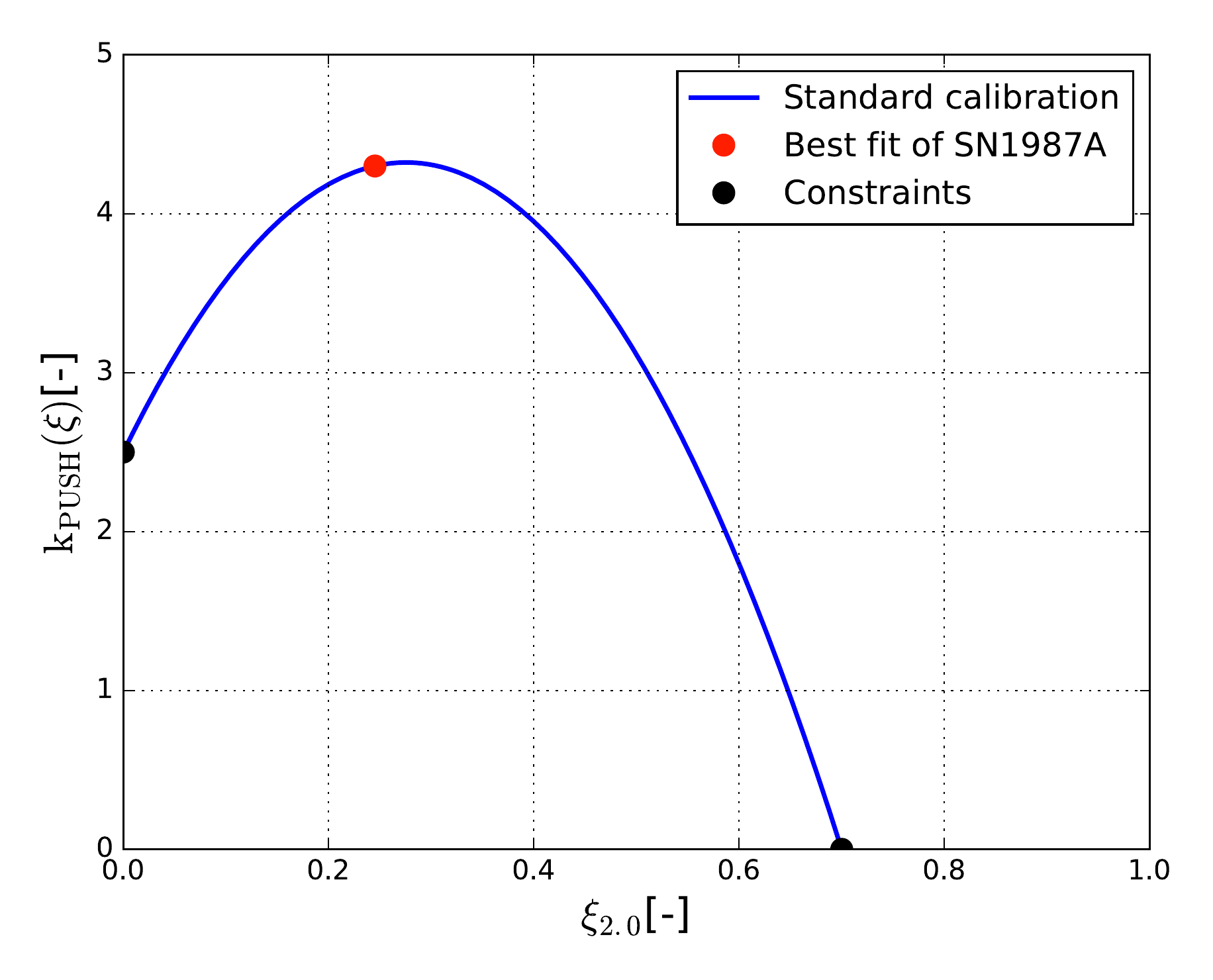}
	\caption{Compactness-dependence of parameter \kpush (solid line) together with the three constraints for the standard calibration of PUSH: ``Crab-like SNe'' (left black dot; $k_{\mathrm{push}}=2.5$ at $\xi_{2.0}=0.0$), SN~1987A (orange dot; $k_{\mathrm{push}}=4.3$ at $\xi_{2.0}=0.245$ (best fit model s18.8)), and BH formation (right black dot; $k_{\mathrm{push}}=0.0$ at $\xi_{2.0}\geq 0.7$).
		\label{fig:parabola}}
\end{figure}

Any calibration of PUSH has to include the best fit for SN~1987A, i.e.\ have $k_{\rm{push}}=4.3$ for $\xi_{2.0}=0.245$ (model s18.8).  
The compactness values of progenitors for ZAMS masses below 21~M$_{\odot}$ are $\xi_{2.0} \leq 0.4$--$0.5$. To emulate a transition from the standard convective neutrino-driven mechanism to a regime of less efficient convective neutrino-driven mechanism and eventually to BH formation we tune down the \kpush parameter  above $\xi_{2.0}=0.4$--$0.5$.
This leads to the constraint for \kpush to approach zero for compactness values above $\xi_{2.0} \sim 0.5$ and to be set to zero for compactness values $\xi_{2.0} \geq 0.7$.
The observed explosion energies of progenitors at the lower end of the mass range indicate weaker explosions (and hence lower values of \kpush).
These lower ZAMS mass progenitors coincide with lower compactness values. Thus, we obtain the constraint for \kpush to be smaller for lower compactness values.
We choose $k_{\rm{push}}=2.5$ for $\xi_{2.0}=0.0$  as the third constraint which the compactness-dependent PUSH parameter function \kpush($\xi$) ultimately has to fulfill. 
In summary, the three points defining the dependence of \kpush on the pre-explosion model compactness are:
\begin{enumerate}[(i)]
\item $k_{\mathrm{push}}=2.5$ at $\xi_{2.0}=0.0$ (``Crab-like SNe''), 
\item $k_{\mathrm{push}}=4.3$ at $\xi_{2.0}=0.245$ (SN1987A calibration model s18.8),  
and
\item $k_{\mathrm{push}}=0.0$ at $\xi_{2.0}\geq 0.7$ (BH formation).
\end{enumerate}
We assume a polynomial dependence of $k_{\mathrm{push}}$ on the compactness and, given that we have three fixed points, we consider a parabolic dependence: $k_{\rm{push}}(\xi)=a\xi^2+b\xi+c$ (see also Figure~\ref{fig:parabola}). 
The resulting parabola through the three points given above is described by $a=-23.99$, $b=13.22$, and $c=2.5$.

In addition to the parabola, we also tested piece-wise linear dependences of \kpush on the compactness (using the same three fixed points as for the parabola) and comparable values for the calibration to SN~1987A. The results are very similar, confirming that our results are not sensitive to the functional dependence of \kpush on the compactness.

\begin{figure*}
\begin{center}
	\begin{tabular}{cc}
	\includegraphics[width=0.48\textwidth]{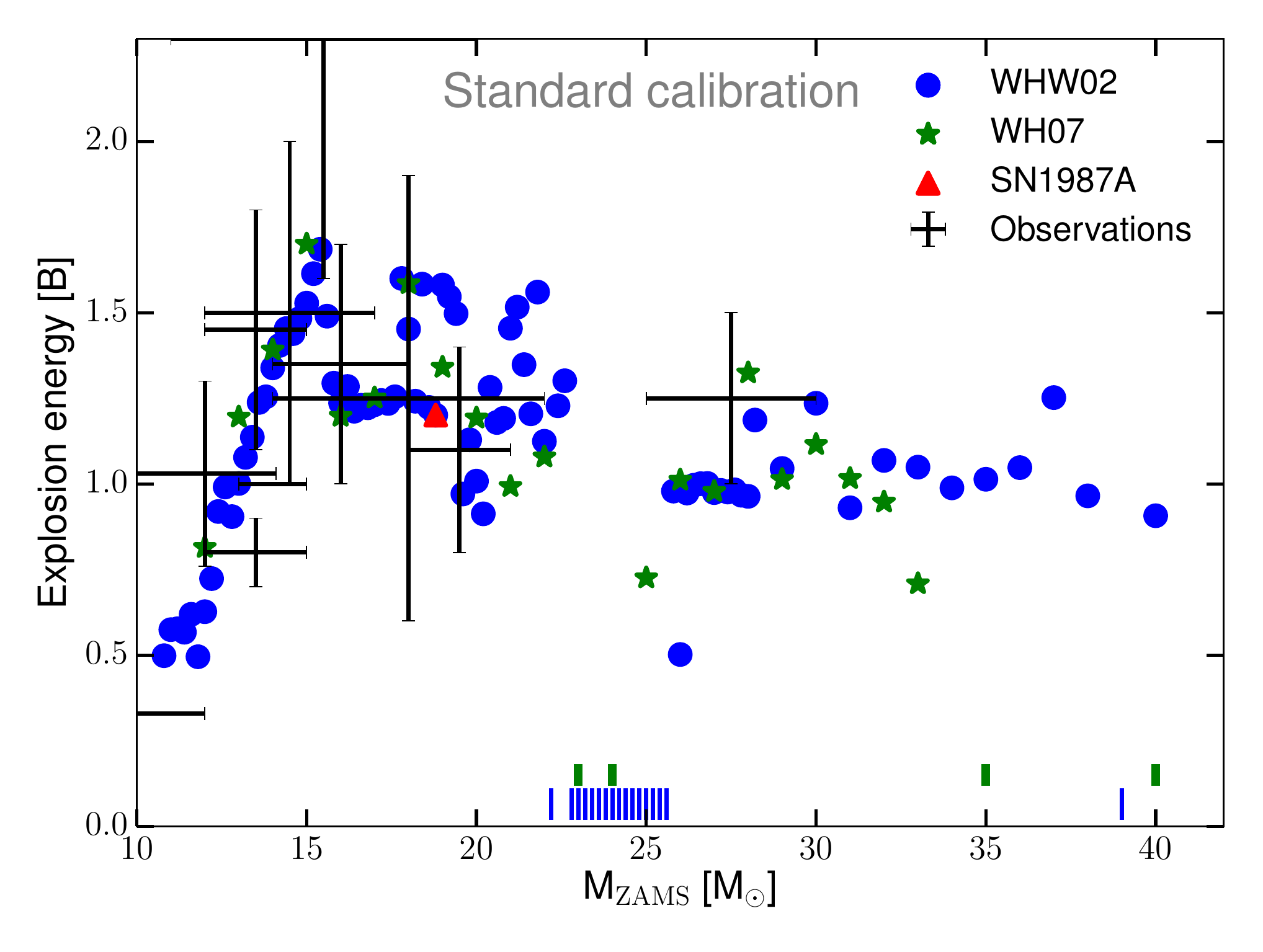} &
    \includegraphics[width=0.48\textwidth]{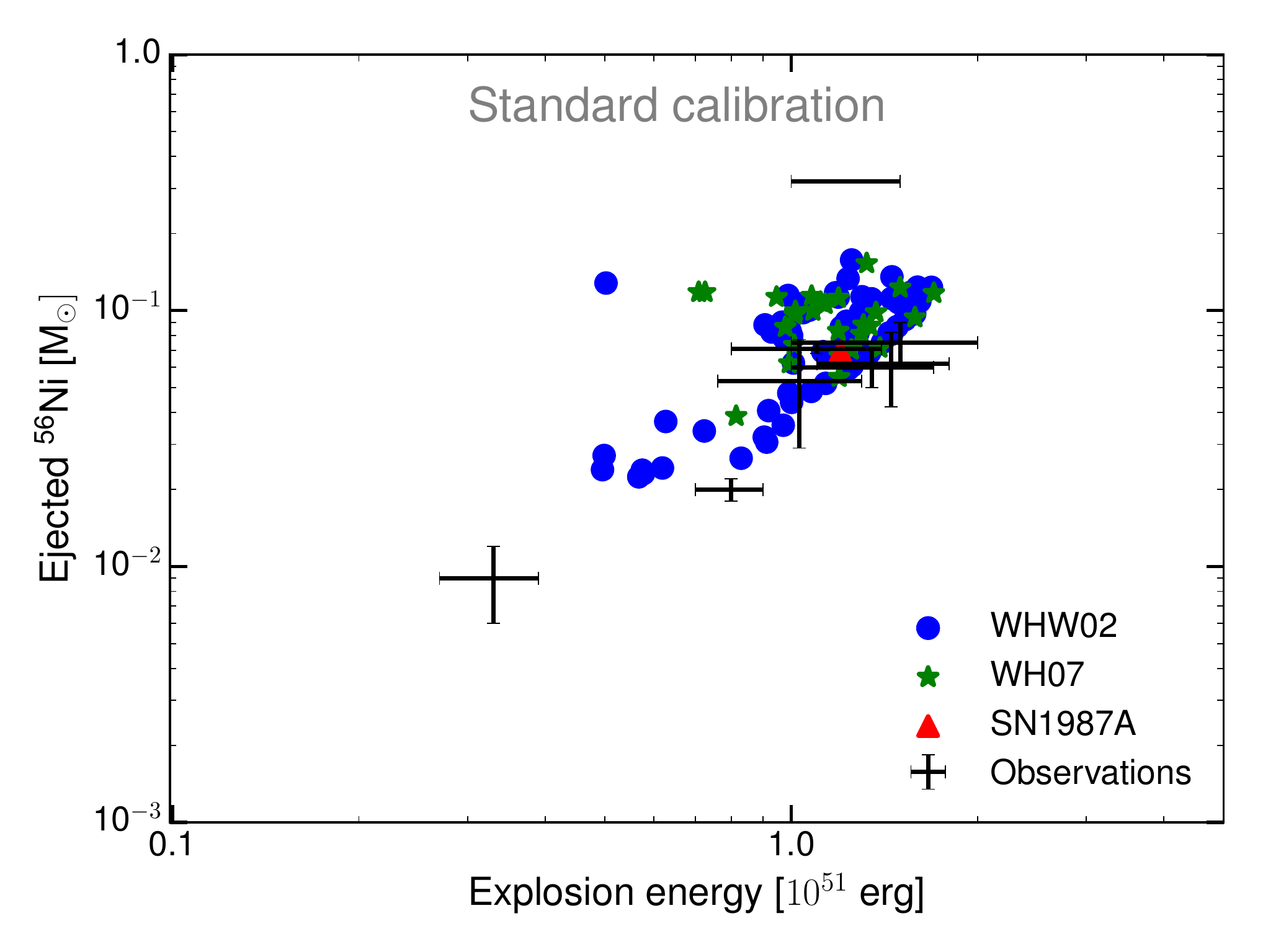} \\
	\end{tabular}    
	\caption{\emph{Left: } Explosion energies as function of ZAMS mass for observed supernovae (black crosses with error bars), for pre-explosion models from WHW02 (blue circles) and WH07 (green stars) from the standard calibration. The vertical dashes at the bottom of the Figure indicate masses for which a BH was formed.
    \emph{Right: } Ejected $^{56}$Ni masses as function of explosion energy for the WHW02 progenitors (blue circles) and the WH07 progenitors (green stars) for the same models. In both panels, the black crosses with error bars represent the observational data and the red triangle indicates the SN~1987A model s18.8. The lower left cross in the right panel represents SN~2005cs with ZAMS mass of 9~M$_{\odot}$ which is below the mass range of our models (see Table~\ref{tab:all_prog_series}).
        \label{fig:parabolaI_s02_w07}
		\label{fig:scan_Eexpl-56ni}
    }
\end{center}
\end{figure*}

\begin{figure*}   
\begin{center}
\begin{tabular}{cc}
	\includegraphics[width=0.48\textwidth]{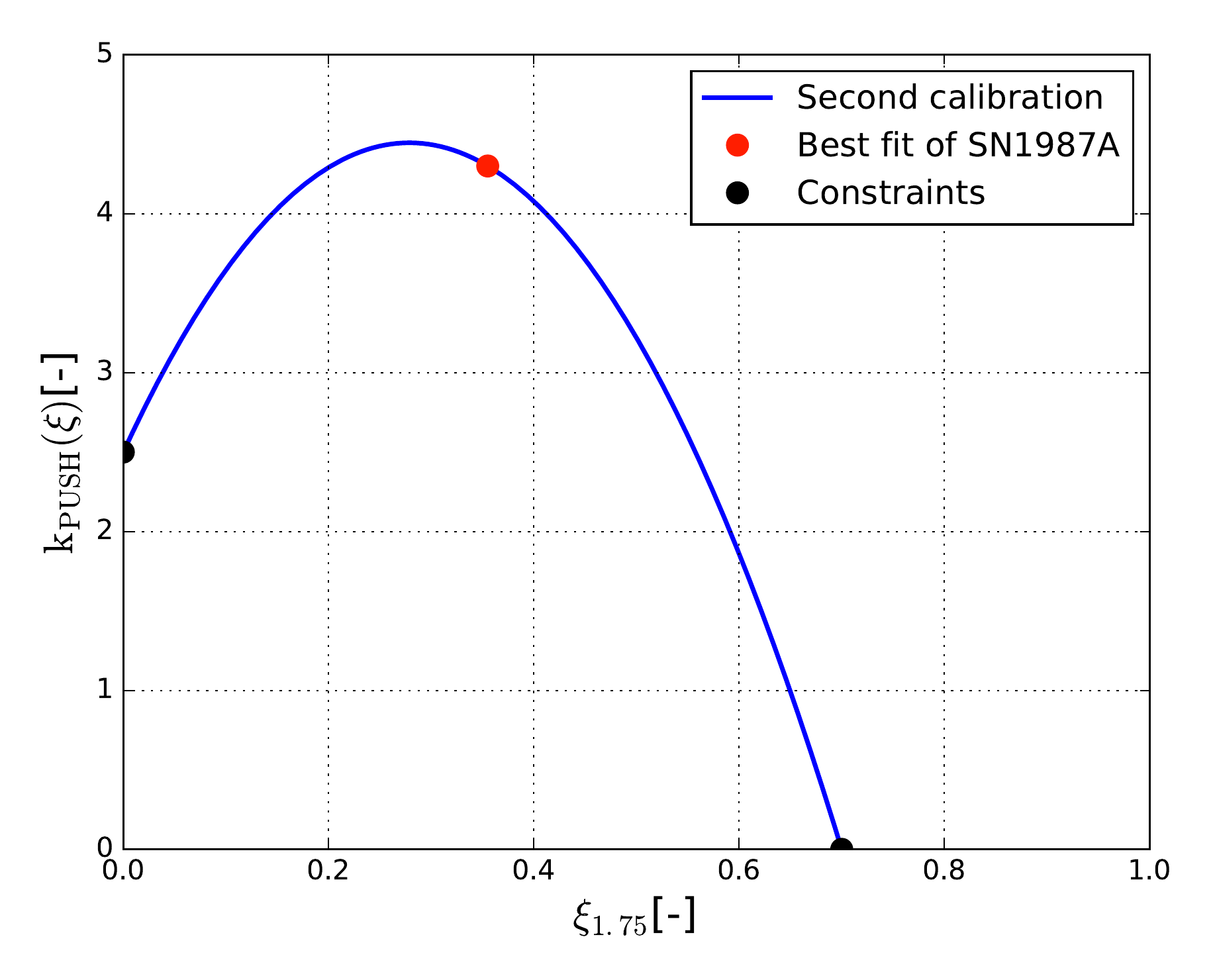} &
	\includegraphics[width=0.48\textwidth]{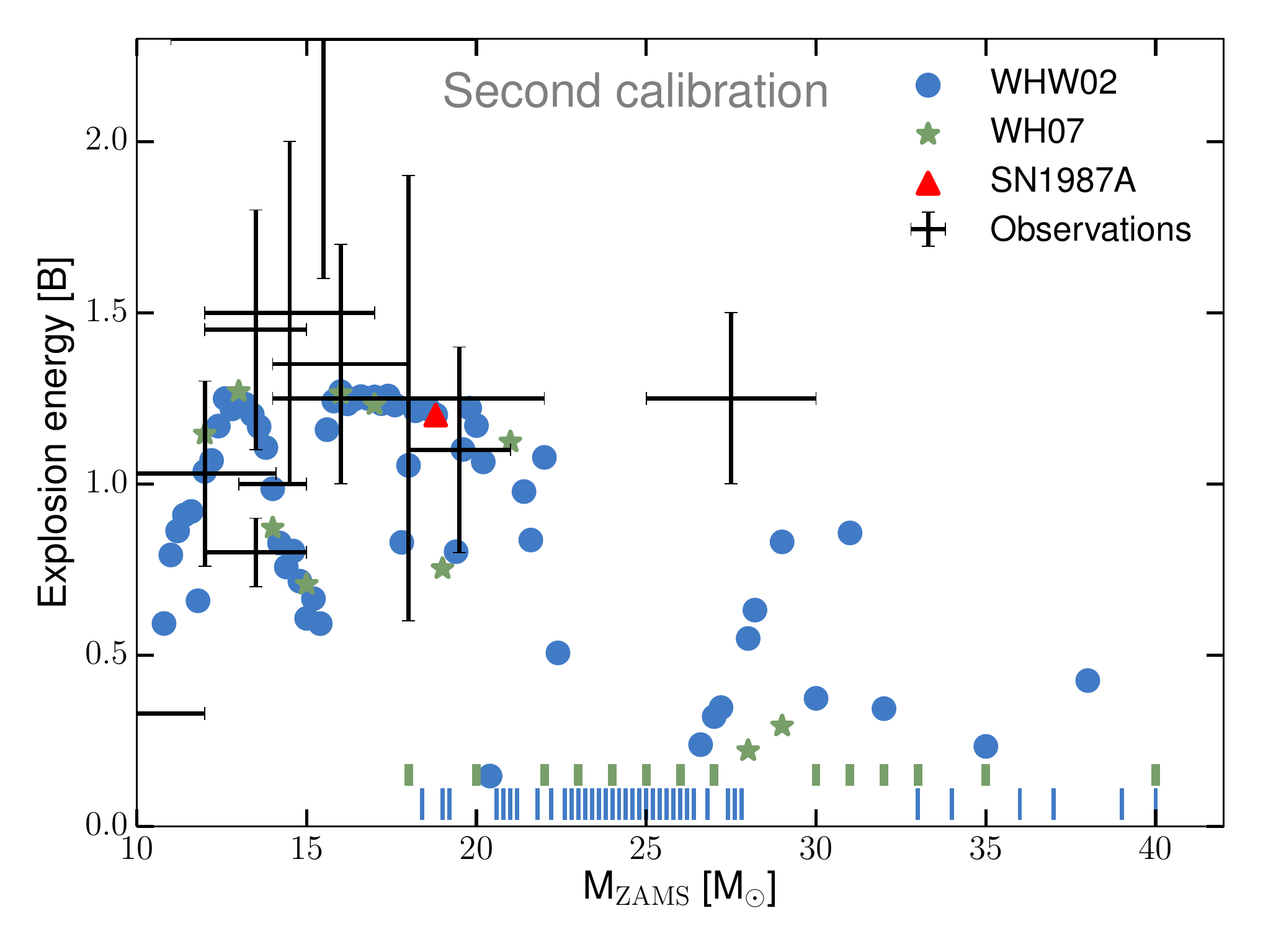} \\
\end{tabular}
\caption{
\emph{Left: } 
Compactness-dependence of parameter \kpush (solid line) together with the three constraints for the second calibration. ``Crab-like SNe'' (left black dot; $k_{\mathrm{push}}=2.5$ at $\xi_{1.75}=0.0$), SN~1987A (orange dot; $k_{\mathrm{push}}=4.3$ at $\xi_{1.75}=0.3551$ (best fit model s18.8)), and BH formation (right black dot; $k_{\mathrm{push}}=0.0$ at $\xi_{1.75}\geq 0.7$). 
\emph{Right: }
Explosion energies as function of ZAMS mass for observed supernovae (black crosses with error bars), for pre-explosion models from WHW02 (blue circles) and WH07 (green stars) for the second calibration which is more prone to lower explosion energies and BH formation. The red triangle indicates the SN~1987A model s18.8.
\label{fig:parabola2}
\label{fig:parabolaII_s02_w07}
}
\end{center}
\end{figure*}

Now, we apply this compactness-dependent \kpush to both series of pre-explosion models. The left panel of Figure~\ref{fig:parabolaI_s02_w07} shows the resulting landscapes of explodability for all the progenitors in the WHW02 and WH07 sets in comparison with observed explosion energies (see Appendix~\ref{subsec:obs-constraints} and \cite{bruenn16} for the selection of observed SNe and a discussion of the associated uncertainties).
For both sets, the explosion energy increases from ``Crab-like SNe'' at the lowest mass-end to robust explosions (with $E_{\mathrm{expl}} \approx 0.8$ -- $1.6$~Bethe) between 15~M$_{\odot}$ and 21~M$_{\odot}$. Above about 25~M$_{\odot}$, no observational data are available. We find that models between 22~M$_{\odot}$ and 26~M$_{\odot}$ form black holes (see the vertical dashes at the bottom of the Figure in this mass range). Above 26~M$_{\odot}$, we find a mixture of successful explosions (models that have lost most of their hydrogen and helium envelopes) and BHs (models with higher CO-core and Fe-core masses). 
\textbf{Due to the relative sparseness of pre-explosion models at higher ZAMS masses, it is not clear whether we obtain many faint explosions, or if we instead go back and forth between regular explosions and BH formation (cf.\ Figures \ref{fig:scan_Eexpl-56ni} and \ref{fig:parabola2}). }
The models with ZAMS mass above 50~M$_{\odot}$ explode for both sets. 
When looking at the Ni ejecta masses for the same models together with the explosion energy (Figure~\ref{fig:scan_Eexpl-56ni}, right panel), our models lie within the same range of explosion energies and Ni ejecta masses as the observations.

\begin{figure}  
\begin{center}
	\includegraphics[width=0.48\textwidth]{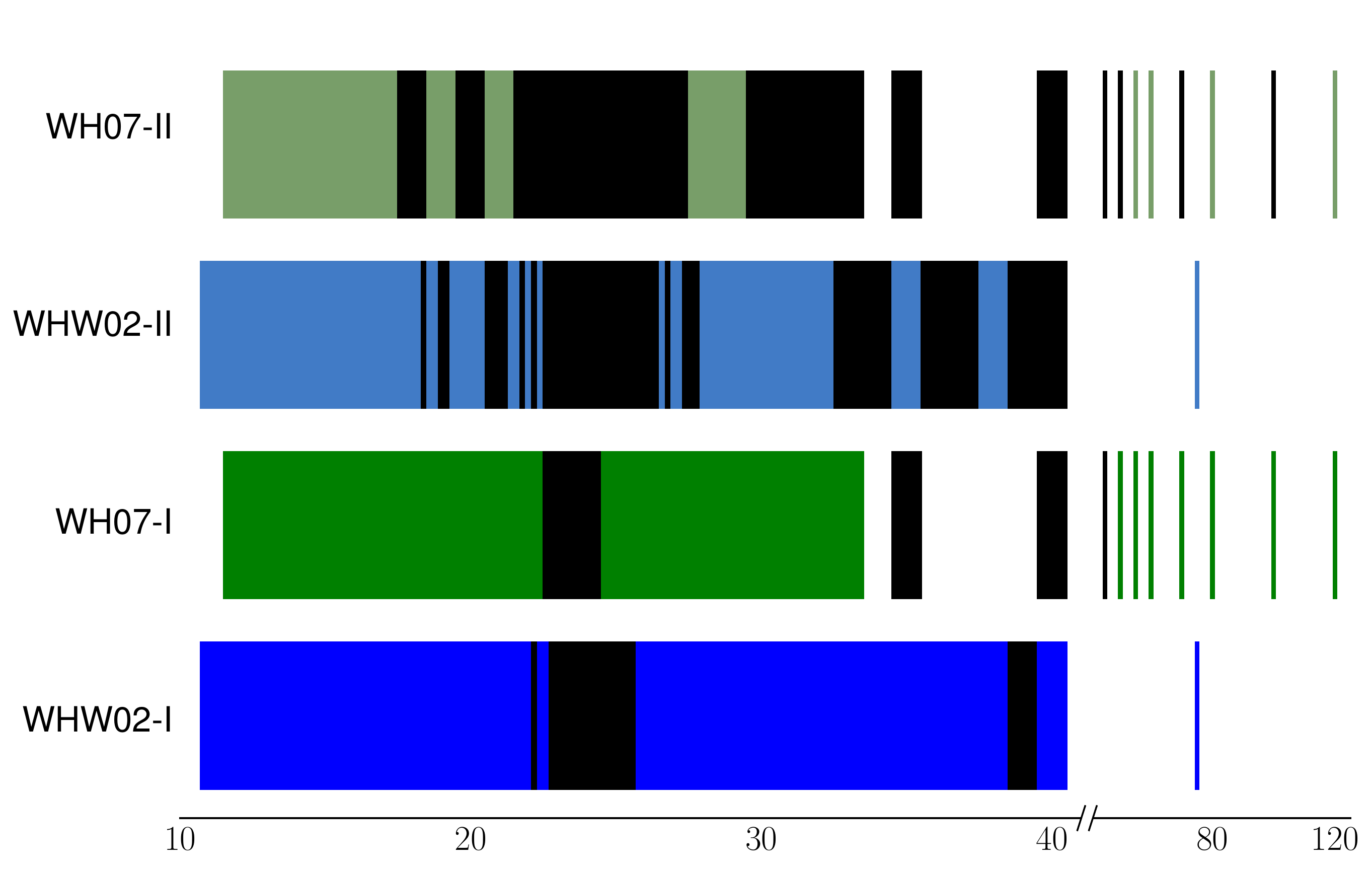}
	\caption{Explosion outcomes for the two progenitor sets and the two different calibrations. The colored areas indicate exploding models which leave behind a neutron star as a remnant and black areas indicate failed explosions which lead to BH formation. Dark colors are used for the standard calibration, lighter colors for the second calibration.
	\label{fig:engine_s02_w07}
    	}
\end{center}
\end{figure}

We also investigated a somewhat different interpretation of the observational constraints by using the compactness $\xi_{1.75}$ instead of $\xi_{2.0}$\footnote[1]{The parameters for the second calibration parabola function of \kpush are:$a=-25.05$, $b=-13.96$, $c=2.5$.}.
The compactness $\xi_{1.75}$ is more strongly dependent on the iron core mass and shows a slightly different behavior than the compactness $\xi_{2.0}$.
The left panel of Figure~\ref{fig:parabola2} shows the parabolic dependence of \kpush for the second calibration and the right panel of Figure~\ref{fig:parabolaII_s02_w07} shows the resulting explosion energies compared to observational data for this second calibration.
In Figure~\ref{fig:engine_s02_w07}, we summarize the outcomes of both calibrations for both progenitor samples. It is evident, that the second calibration results in a lower overall explodability and leads to a larger fraction of faint explosions and black holes. The first calibration which represents our standard calibration is in good agreement with observations (SN~1987A, \textbf{putative} faint SN branch, and BH formation) and is well suited to investigate the full range of CCSN progenitors.
The relation between compactness and explodability (i.e.\ explosion energies and black hole formation times; first investigated in \cite{OConnor.Ott:2011}), is expected to be one of the relevant aspects of the neutrino-driven mechanism. We showed that with a relatively simple calibration in compactness we can fulfill the constraints we formulated and obtain an effective model to investigate the properties of CCSNe.   
From here on, we discuss and use the standard calibration of PUSH introduced in this Section unless otherwise indicated.

\section{Systematics across the mass range}
\label{sec:progenitorscan}

In this Section, we present and discuss the explosion properties of our simulations for the entire WHW02 and WH07 samples (combined including ZAMS masses of 10.8~--~120~M$_{\odot}$) based on the standard calibration presented in the previous Section. 
These same models are used in the detailed discussion of the corresponding nucleosynthesis yields and their possible impact an the galactic chemical evolution in Paper~III.

\subsection{Explosion properties}
\label{subsec:scan_expls}

\begin{figure*}  
\begin{center}
	\begin{tabular}{cc}
	\includegraphics[width=0.48\textwidth]{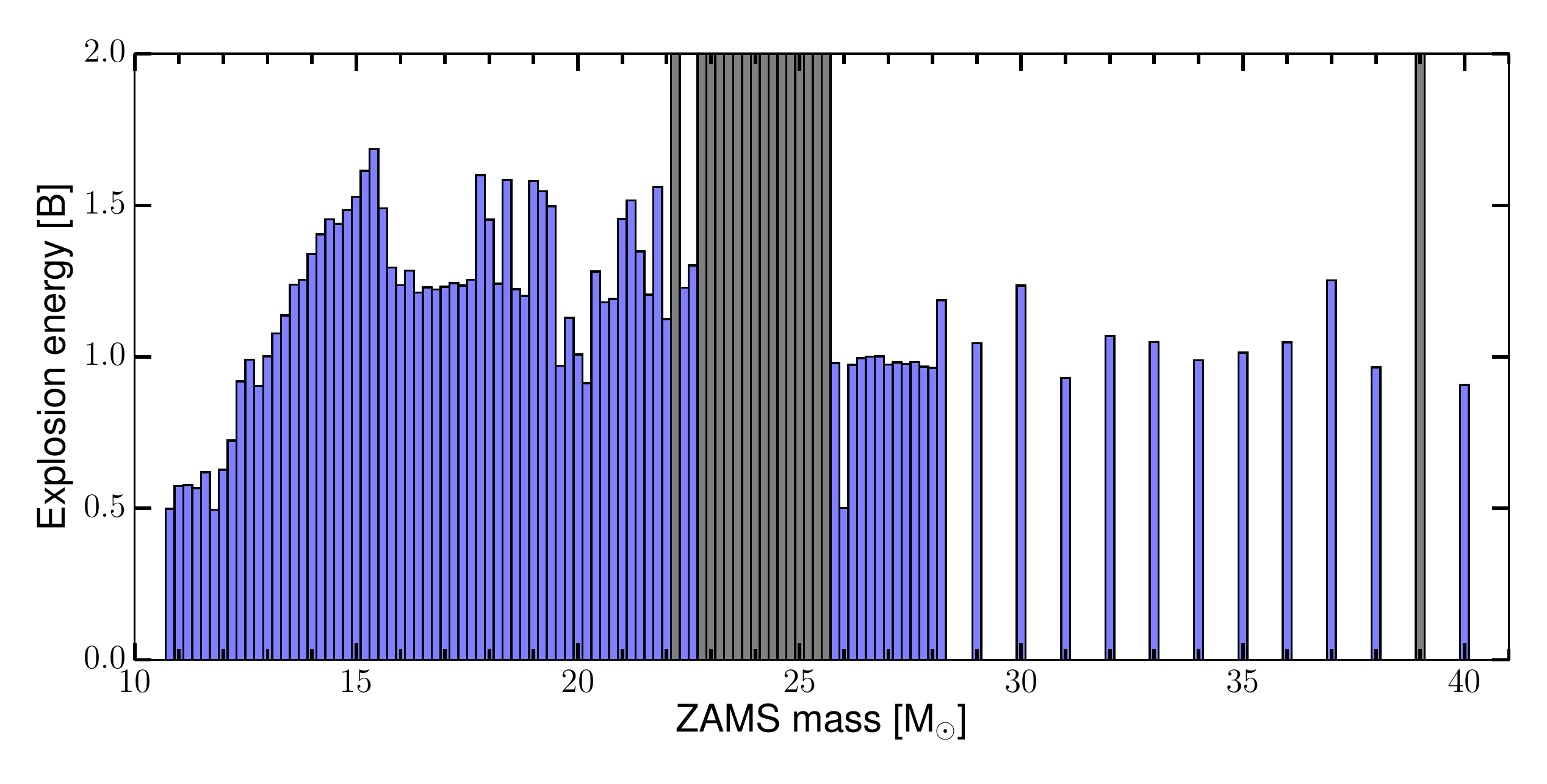} & 
    \includegraphics[width=0.48\textwidth]{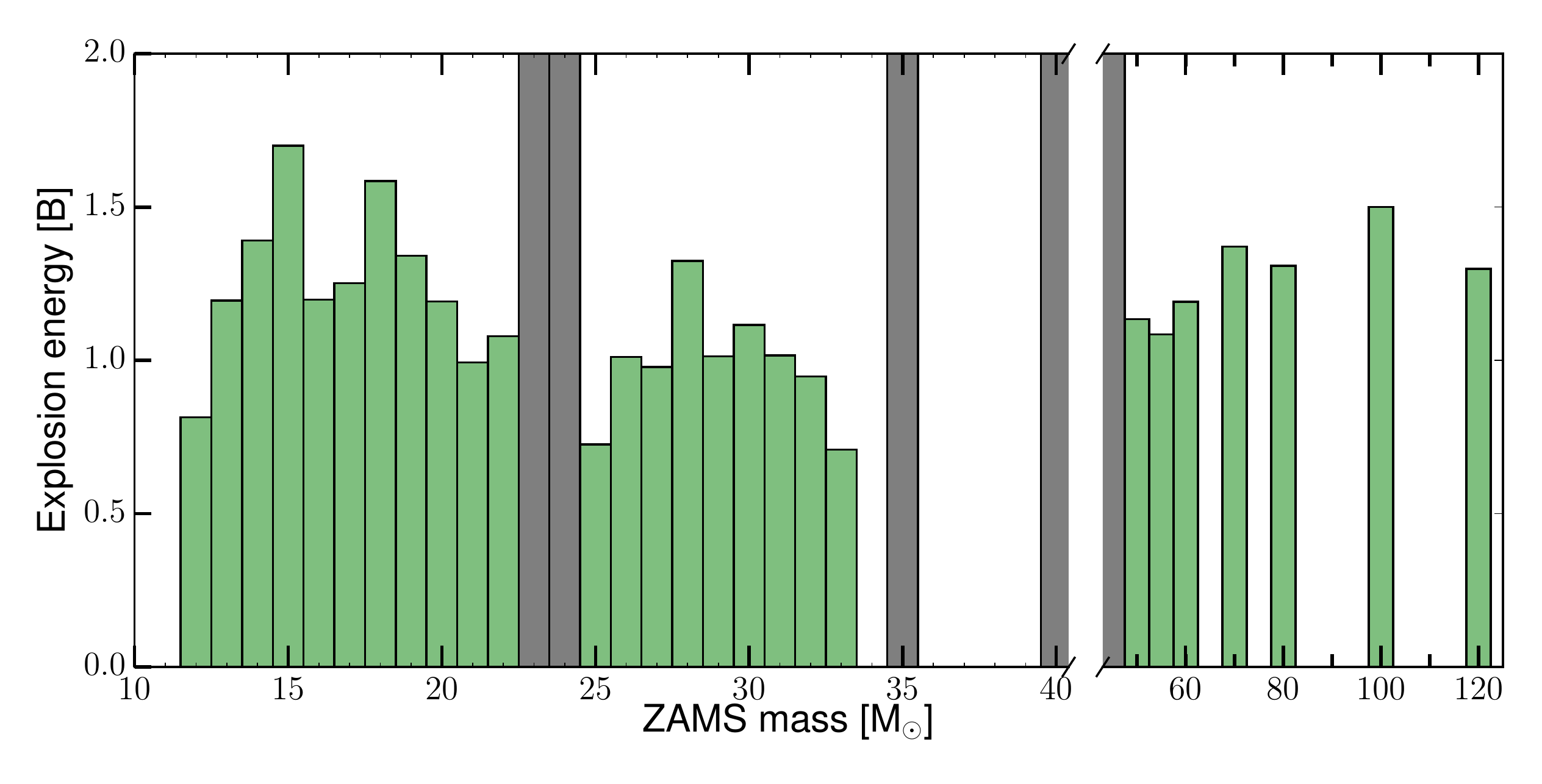} \\ 	
    \includegraphics[width=0.48\textwidth]{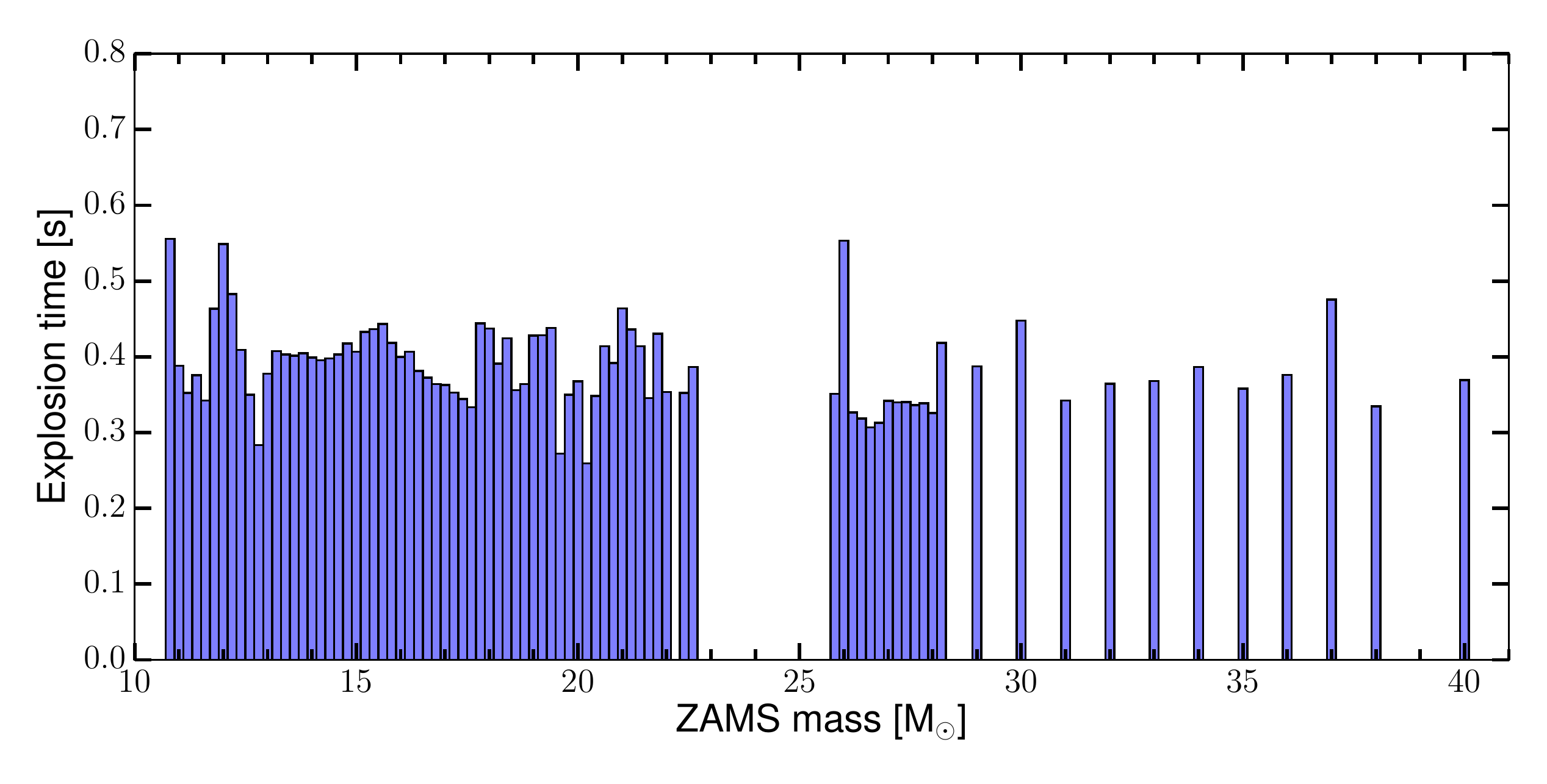} & 
    \includegraphics[width=0.48\textwidth]{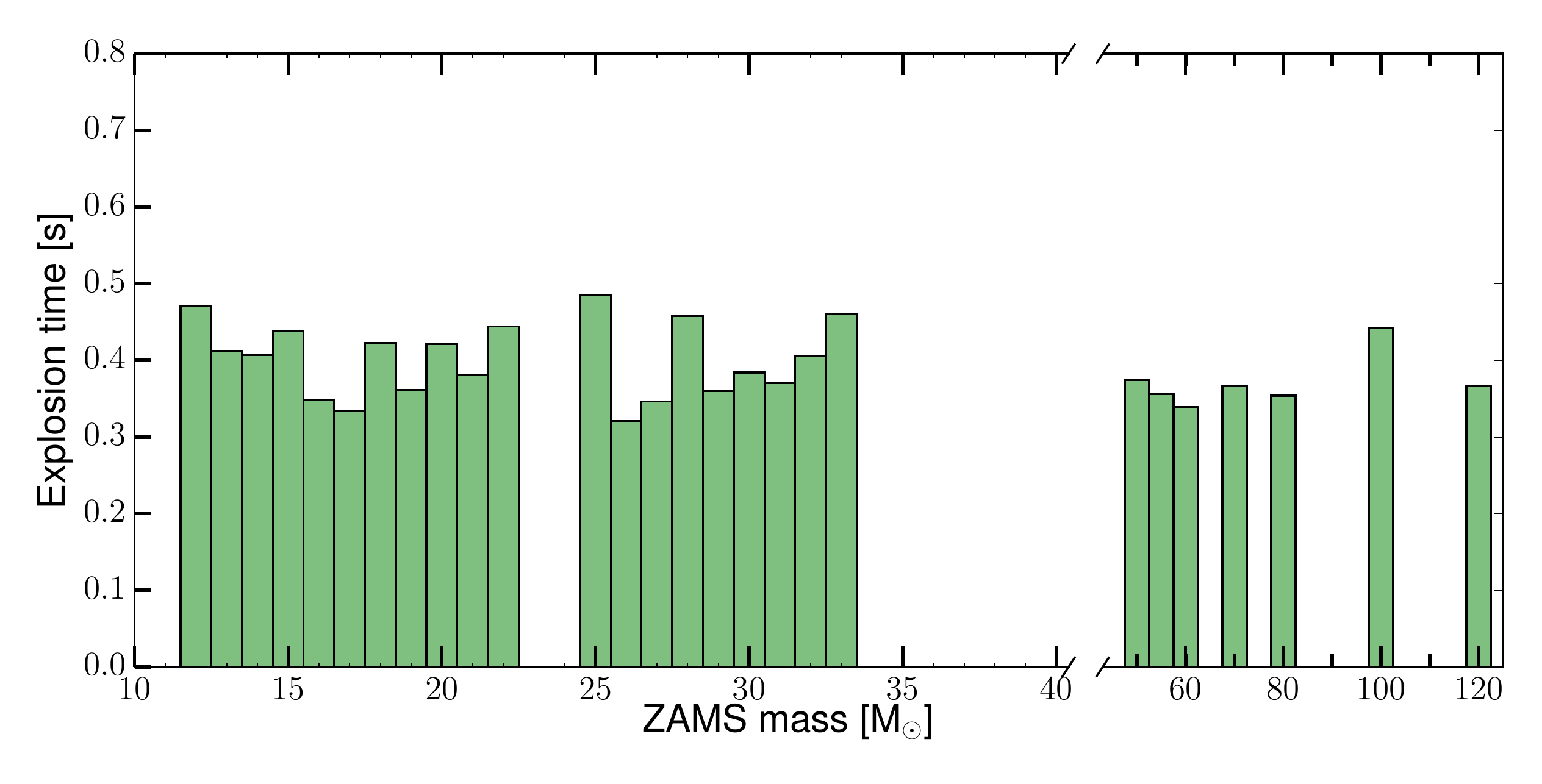} \\
    \includegraphics[width=0.48\textwidth]{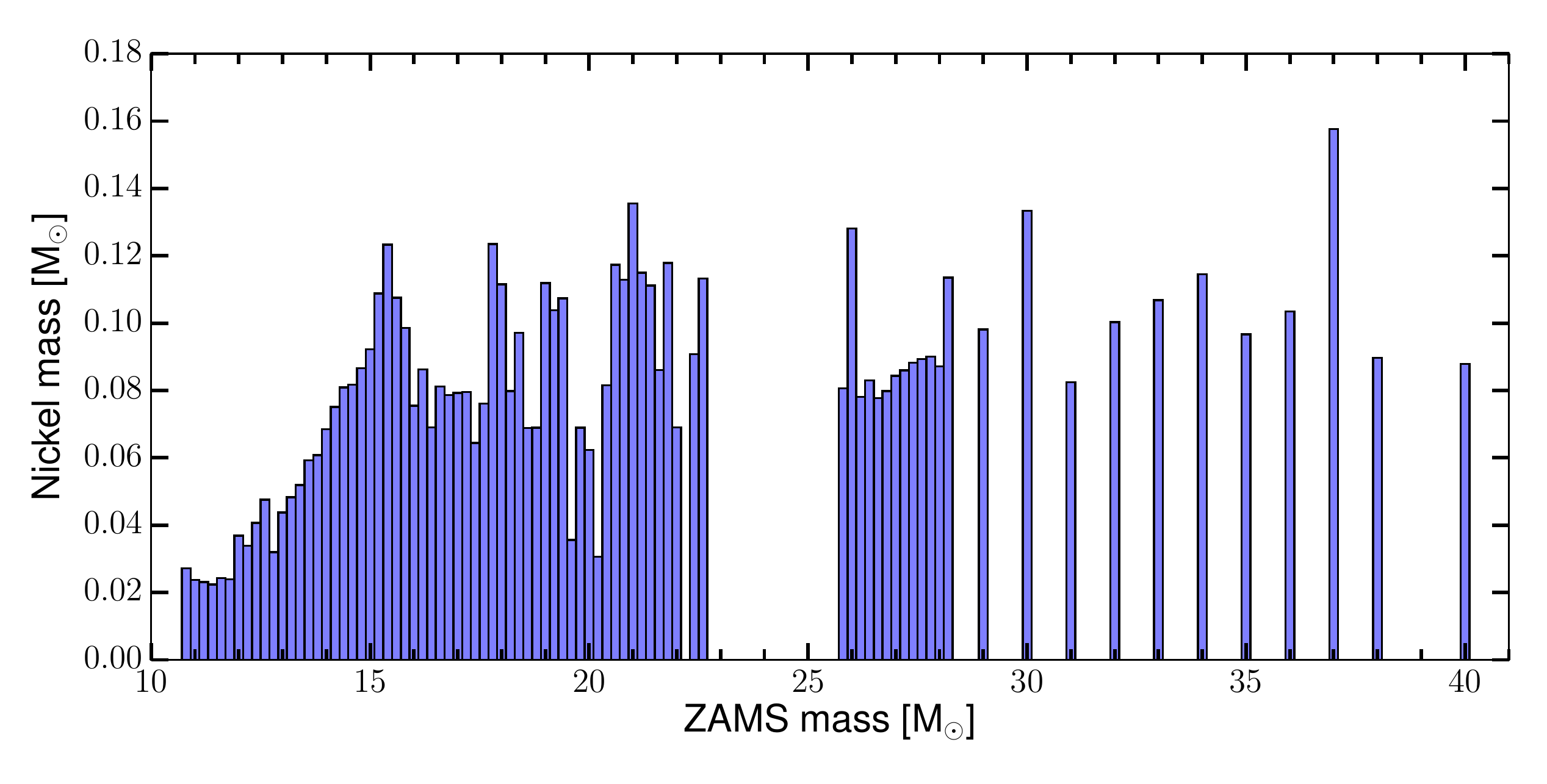} & 
    \includegraphics[width=0.48\textwidth]{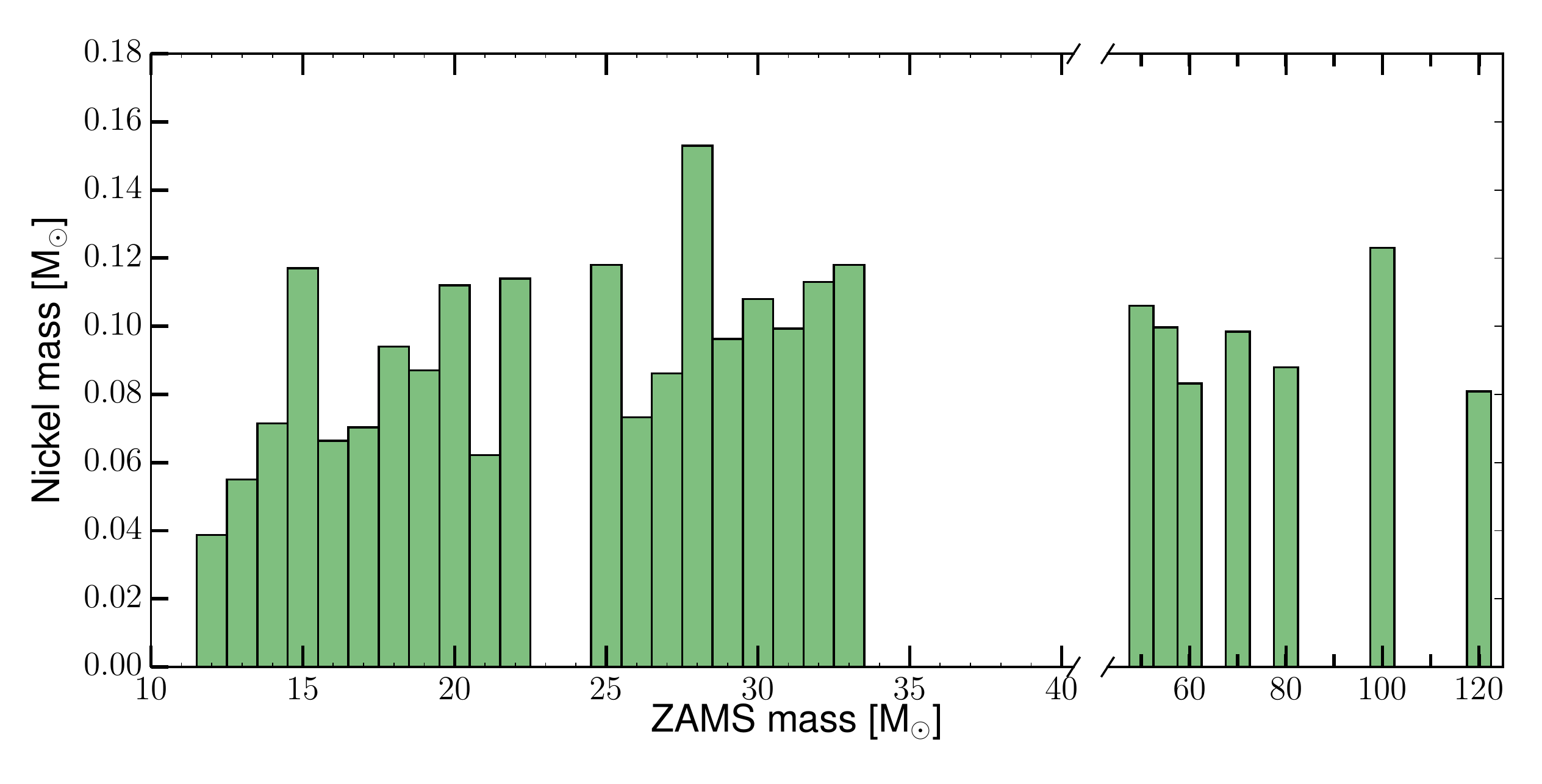} \\
    \includegraphics[width=0.48\textwidth]{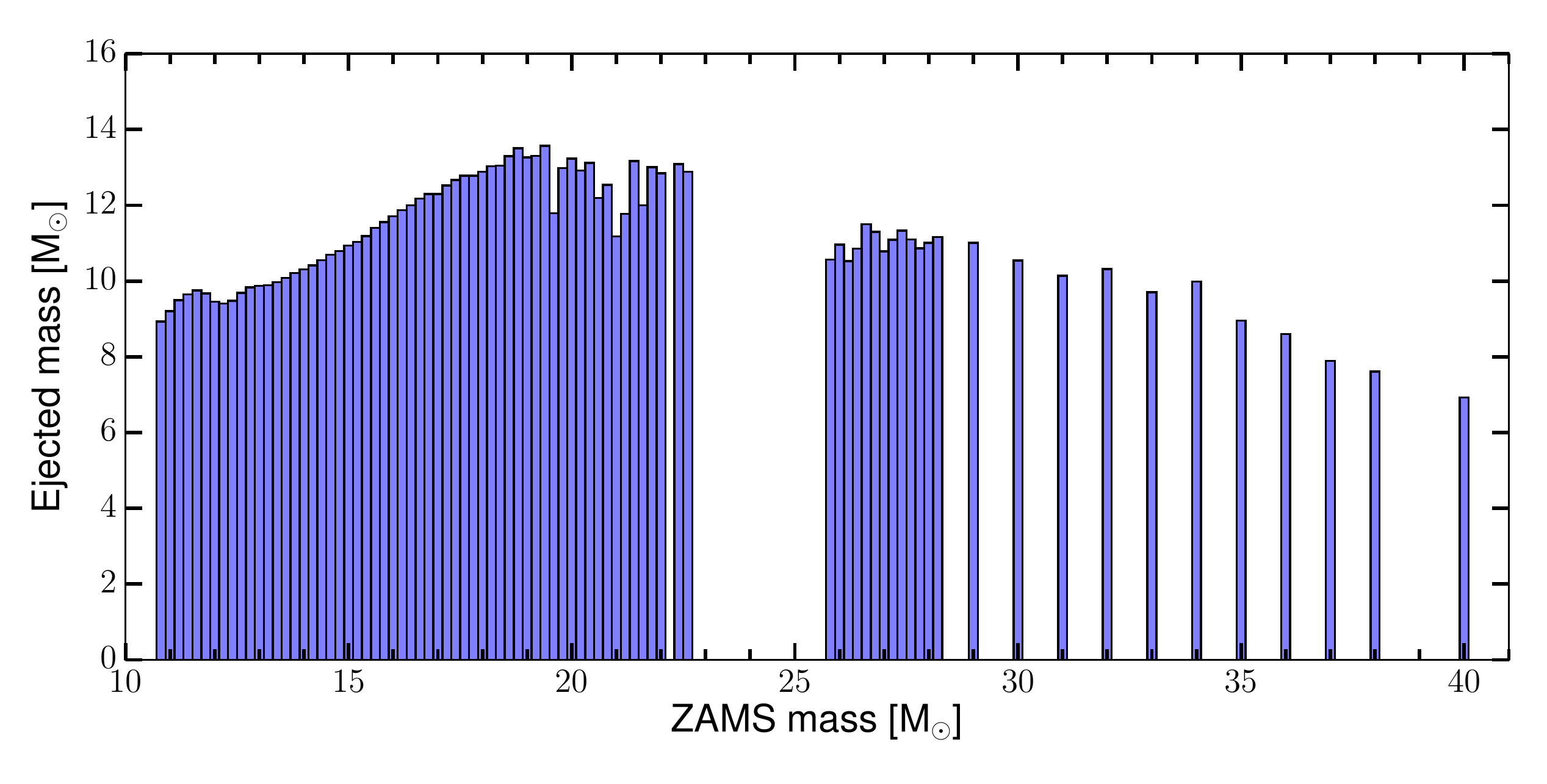} & 
    \includegraphics[width=0.48\textwidth]{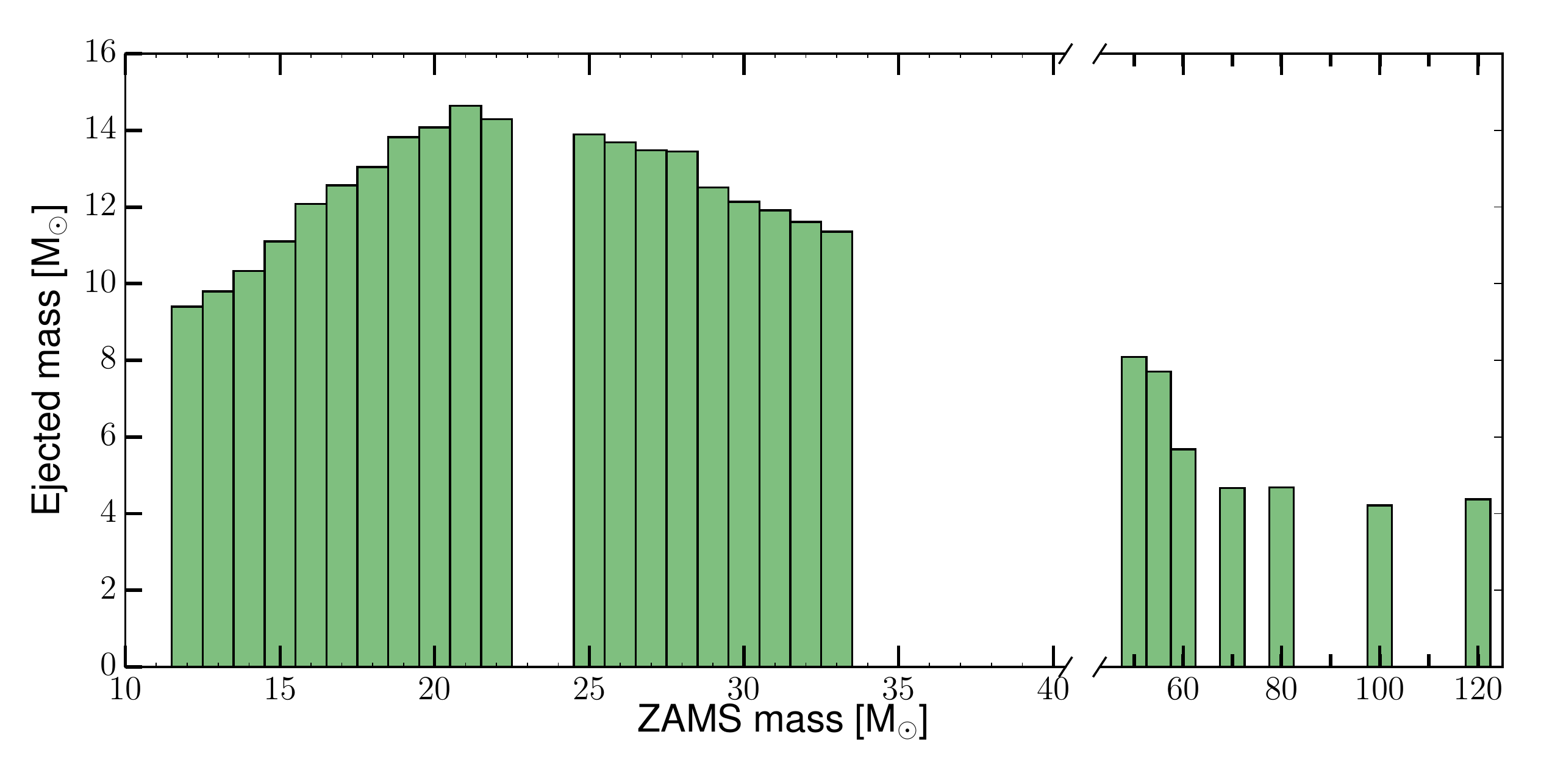} \\      
    \includegraphics[width=0.48\textwidth]{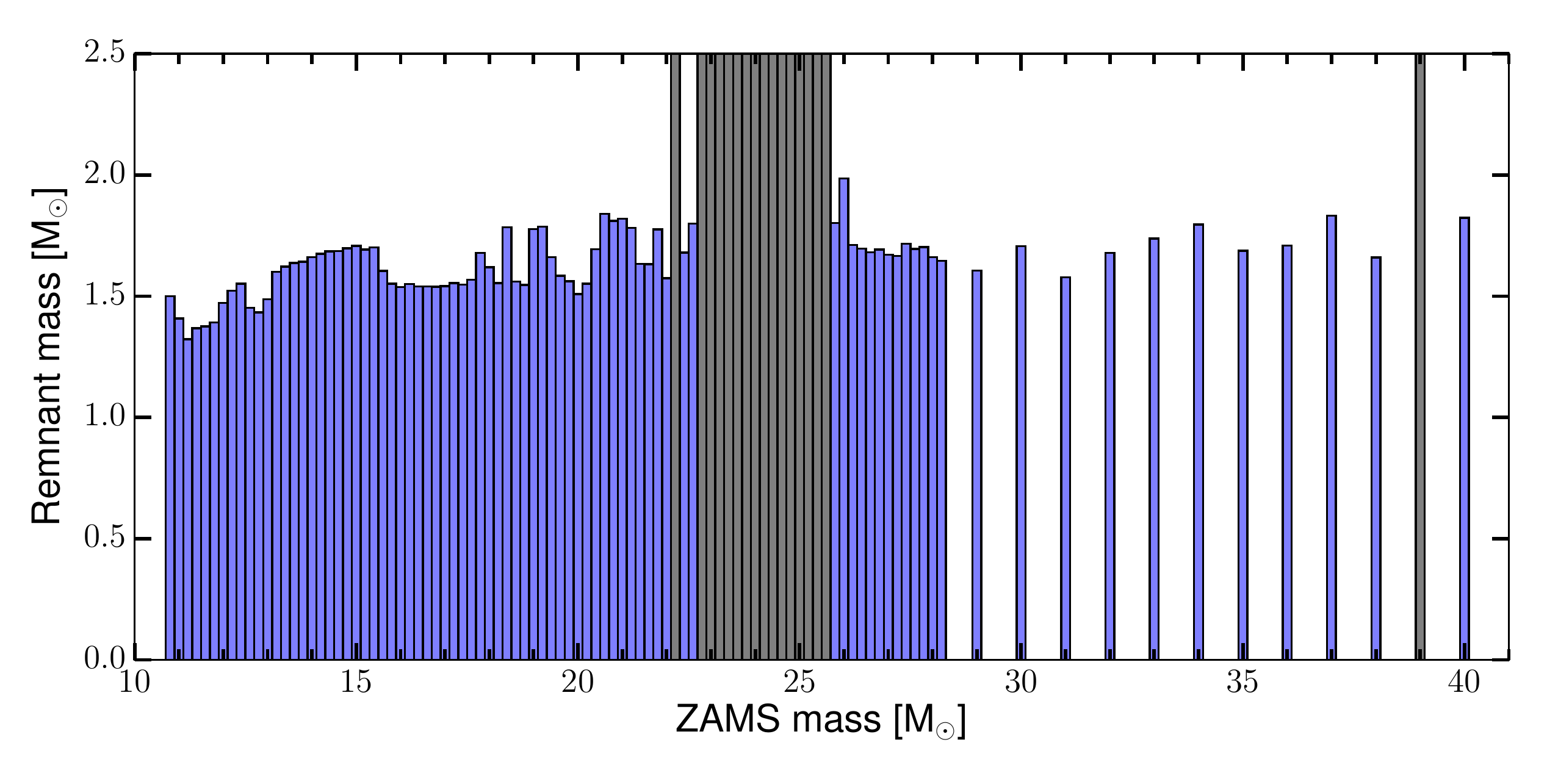} & 
    \includegraphics[width=0.48\textwidth]{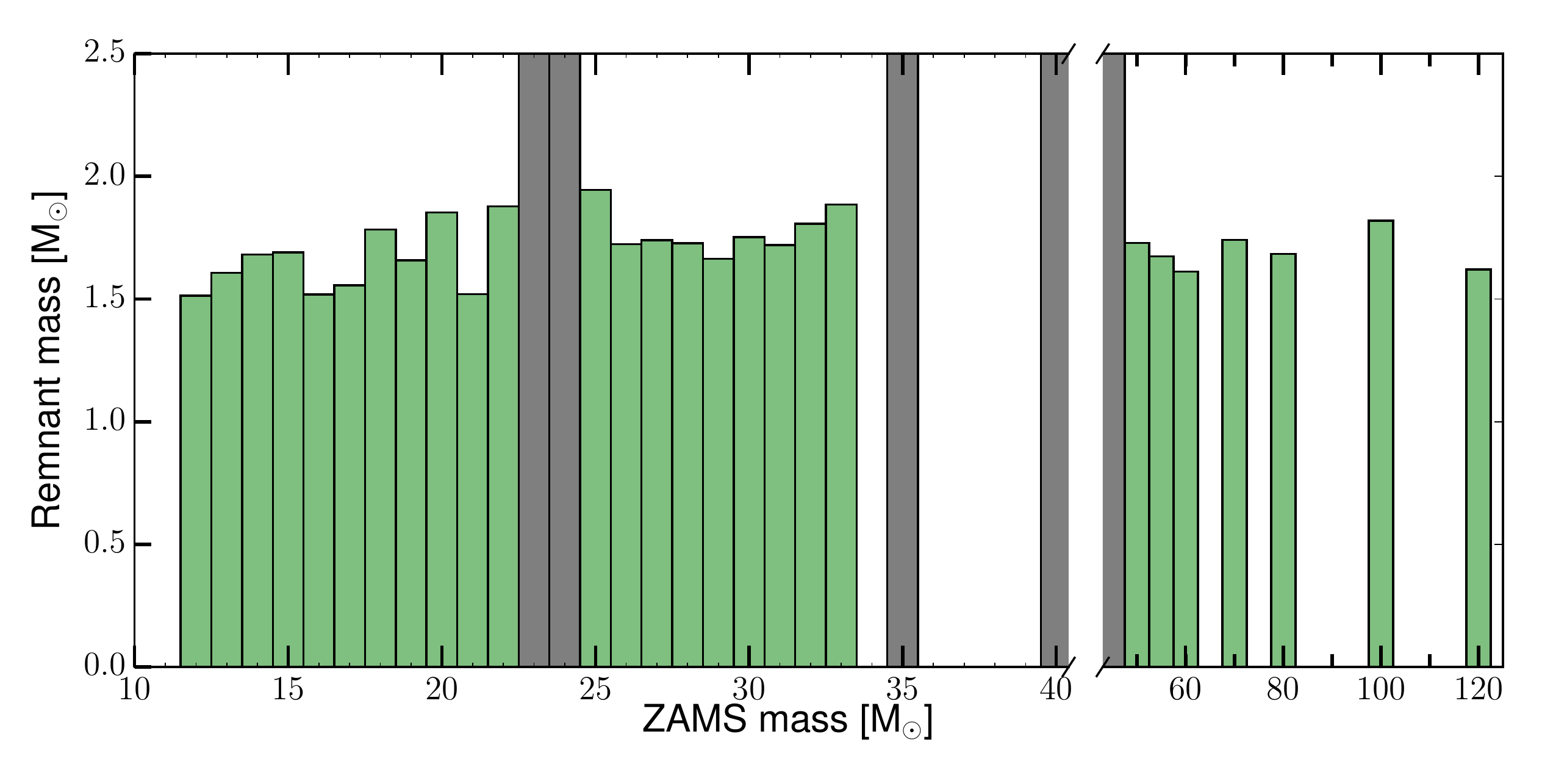} \\  
	\end{tabular}
 	\caption{From top to bottom: explosion energy, explosion time, ejected Ni mass, total ejecta mass, and remnant mass (baryonic mass) for the WHW02 series (left column) and WH07 series (right column) as function of the ZAMS mass using the standard calibration. Dark bars in the explosion energy and remnant mass panels indicate models that did not explode, i.e.\ ultimately formed black holes. The presented data is available as machine-readable Table. A sample Table can be found in Appendix~\ref{appx:tables}.
		\label{fig:scan_properties-mass}
    }
\end{center}
\end{figure*}

Figure~\ref{fig:scan_properties-mass} gives an overview of the resulting explosion and the predicted remnant properties for all pre-explosion models considered in this study. 
From top to bottom, we show the explosion energy, the explosion time, the ejected Ni mass, 
the total ejecta mass, the remnant mass (baryonic mass) as a function of ZAMS mass for the WHW02 (left column) and WH07 (right column) pre-explosion models. Note that the WH07 series consists of a smaller number of models than the WHW02 series.
The general features are similar for both series. 
With the standard calibration, we obtain explosion energies from less than 0.5 to 1.7~Bethe, gravitational NS masses from 1.2 to 1.8~M$_{\odot}$ (see also Section~\ref{sec:remnants}) and ejected nickel masses from 0.02 to 0.16~M$_{\odot}$. 
The lowest explosion energies and Ni masses are obtained for the lowest-mass progenitors (around 11--12~M$_{\odot}$), while the highest explosion energies and Ni masses result from pre-explosion models around 15~M$_{\odot}$, 18~M$_{\odot}$, and 21~M$_{\odot}$ ZAMS mass.

For the most massive pre-explosion models ($\gtrsim 30$~M$_{\odot}$), there is more variation both between the two series used in this work as well as in comparison with other works. The pre-SN evolution of such massive stars is much less certain \citep[e.g.][]{hirschi2015} and the outcome of core collapse is much more sensitive to the details of the prescription (see for example Figures~8 and 9 in \cite{sukhbold16}). 
If these stars explode, they are of Type Ib or Ic. 
Overall, the WH07 series has lower explosion energies and is more prone to BH formation (as discussed in Section~\ref{subsec:bh-formation}). Coupled to that, this series results in somewhat more massive neutron stars. The WH07 series also has ejecta masses that are slightly larger for the same ZAMS mass when compared to the WHW02 series. 
Note that for lower metallicity, massive stars experience less mass-loss during their evolution and hence their mass at collapse is closer to their initial ZAMS mass. Above $\sim$30~M$_{\odot}$ they also have a higher compactness than their counterparts with solar metallicity. From this, we expect pre-explosion models at low metallicity to not have any successful explosion above $\sim$30-35~M$_{\odot}$, constituting a potential origin of the BHs seen by LIGO-VIRGO.

Our results for the solar metallicity pre-explosion models show a moderate trend of the explosion properties with ZAMS mass up to about 15~M$_{\odot}$ as expected from our calibration. This is the same mass range where the compactness exhibits an increasing trend with ZAMS mass (and where the CO-core mass grows with increasing ZAMS mass, see Figures~\ref{fig:prog_compactness}, \ref{fig:prog_mass_structure}, and \ref{fig:prog_mass_structure2}). Beyond about 15M$_{\odot}$, there is no obvious trend with ZAMS mass. In this mass range, the relation between ZAMS mass and compactness is also more complex, as already discussed in \cite{sukhbold.woosley:2014}.
Not surprisingly, we see a correlation between the explosion energy, the ejected Ni mass, and to a lesser degree the remnant mass. The total ejected mass is dominated by the mass at collapse, so we see a similar trend of the total ejecta mass with ZAMS mass as seen in Figure~\ref{fig:prog_mass_structure} 

\begin{figure}  
\begin{center}
	\includegraphics[width=0.48\textwidth]{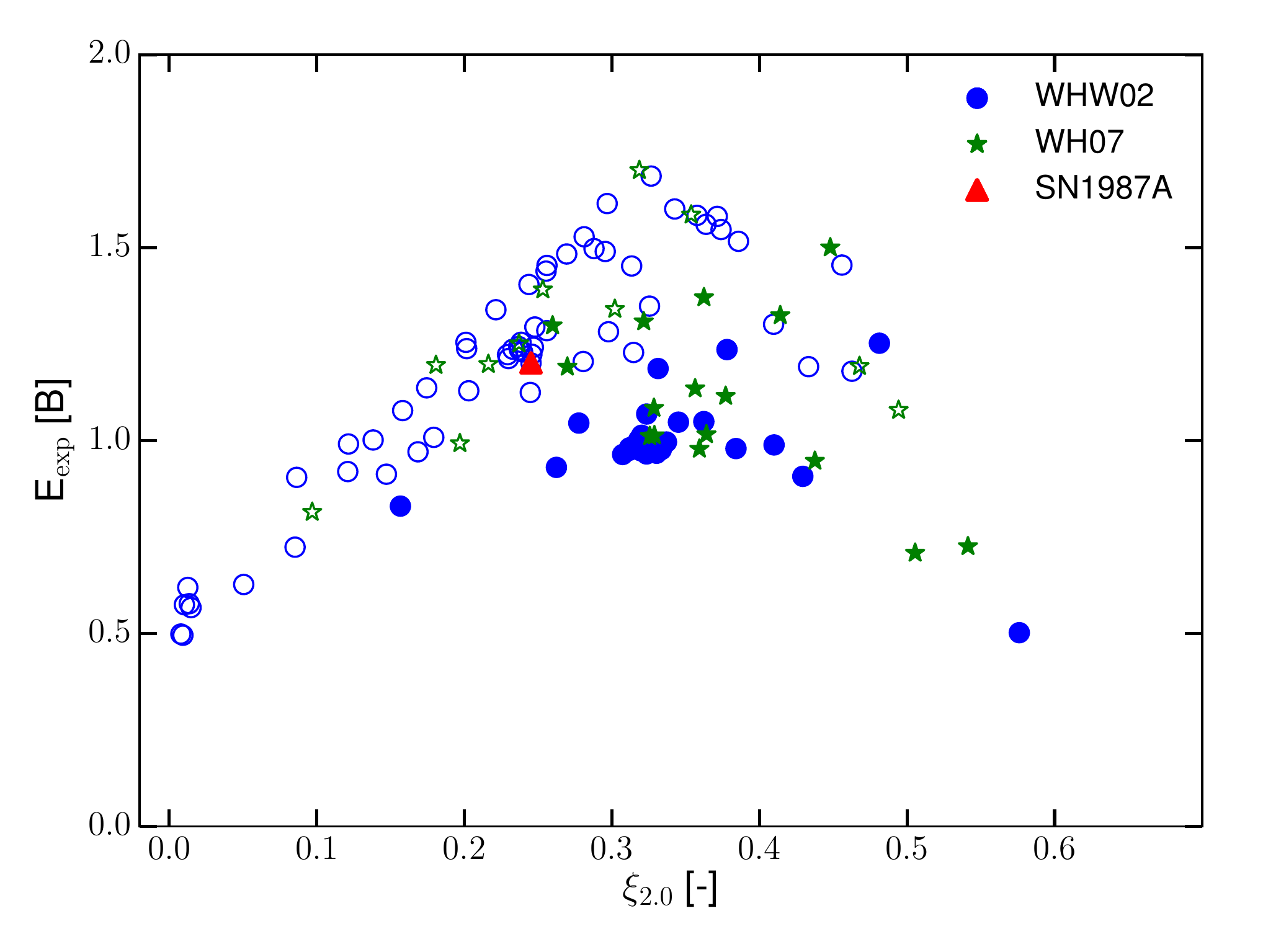} \\
    \includegraphics[width=0.48\textwidth]{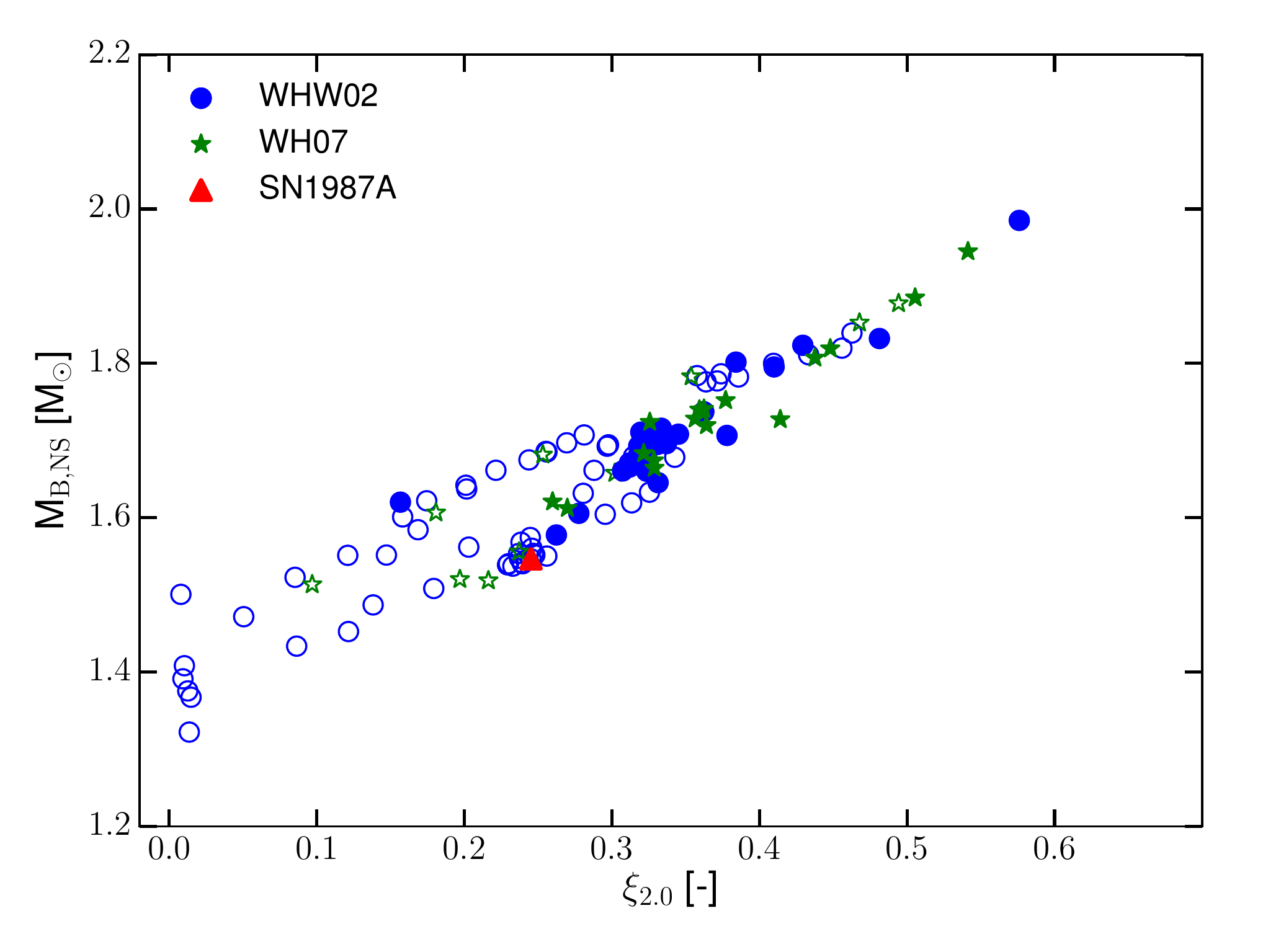} \\
   	\includegraphics[width=0.48\textwidth]{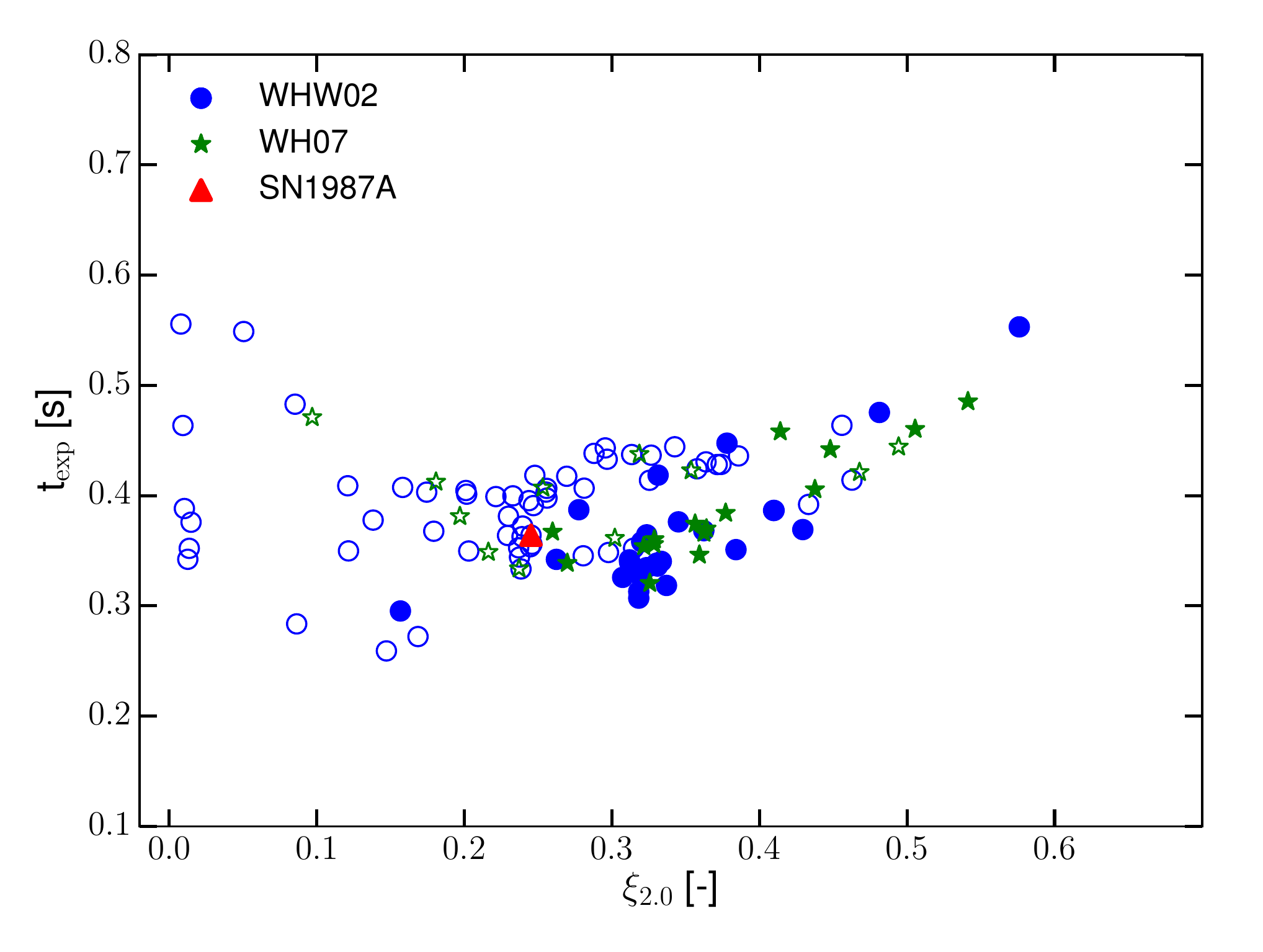}  \\
 	\caption{From top to bottom: explosion energy, remnant mass (baryonic mass), and explosion time for the WHW02 series (blue circles) and WH07 series (green stars) as function of the compactness $\xi_{2.0}$. Only exploding models are shown. The red triangle indicates the SN~1987A model s18.8. 
 Open symbols indicate models with ZAMS mass to the left of the compactness peak (see Figure~\ref{fig:prog_compactness}); filled symbols indicate models to the right of the compactness peak. 
		\label{fig:scan_properties-compactness}
    }
\end{center}
\end{figure}

Figure~\ref{fig:scan_properties-compactness} shows explosion energy, remnant mass, and explosion time against compactness $\xi_{2.0}$. 
This reveals several trends.
The strongest correlation is seen between the remnant mass (neutron stars only) and the compactness. More compact models have a higher mass accretion rate and hence more matter is accreted onto the PNS before the successful explosion \citep{push1}. This trend to higher remnant mass would have continued to higher compactness if we had applied a constant value of \kpush for all models, without forming any BHs. Instead, with our standard calibration for PUSH these models do not explode and eventually collapse to BHs which are not included in this Figure.
The explosion time shows a similar trend with compactness, however the distribution is broader and the lowest-compactness models do not follow the general trend. 
We note that for the models of intermediate compactness ($\xi_{2.0} \approx 0.3$), where the PUSH heating reaches maximum values, the explosion time becomes comparable to the value of $t_{\mathrm{rise}}$.
The behavior of the explosion energy with compactness has two distinct features. The highest explosion energies for each model set show a parabolic dependence on the compactness, as expected from our \kpush dependence.
In addition, we observe a large scatter (up to 0.5~B) in the explosion energies for models with compactness between 0.2 and 0.45. 
In this intermediate compactness region the explosion energies and the explosion times fall into two groups. 
In Figure~\ref{fig:prog_compactness}, one can see a peak in compactness value around 24.2~M$_{\odot}$ (WHW02) and 23.0~M$_{\odot}$ (WH07). Models to the left of the peak (i.e., with lower ZAMS mass) and to the right of the peak (i.e., with higher ZAMS mass), have similar compactness values of 0.2 to 0.45. 
Nevertheless, they are different in their behavior. In Figure~\ref{fig:scan_properties-compactness}, the models to the left of the peak in compactness are indicated by open markers and the models to the right of the peak in compactness by filled markers.
Lower ZAMS mass stars before the peak in compactness ultimately take longer to form explosions and have a higher explosion energy than higher ZAMS mass stars of the same compactness but different ZAMs and CO-core masses. For the remnant mass this split in two groups is less pronounced.

Similar studies of CCSN properties have been done in \citet{ugliano12,ertl16,sukhbold16} based on P-HOTB. By calibrating their spherically symmetric models to SN1987A and, in the case of \citet{ertl16,sukhbold16}, to lower explosion energies for less massive progenitors, the explosion properties and yields of CCSNe were investigated.
Another recent study on the explodability of CCSNe without the use of hydrodynamic simulations has been presented in \citet{mueller2016}. This work uses physically motivated scaling laws and differential equations to describe crucial quantities of CCSNe like the shock propagation, the neutron star contraction and the heating conditions. Note that with the exception of \citet{ugliano12} these other works use different progenitor samples and ranges from 9 to 120 M$_{\odot}$, i.e. the smallest mass range is covered in \citet{mueller2016}, which investigates progenitors between 10 and 32.5 M$_{\odot}$ ZAMS mass. 
Our explosion energies are compatible with values found in observations and similar to other studies \citep{ugliano12,ertl16,pejcha2015,mueller2016,sukhbold16}. 
Exceptions are the relatively high explosion energies that \citet{ugliano12} and \citet{pejcha2015} obtain for low mass progenitors. 
In \citet{sukhbold16}, slightly lower explosion energies and nickel ejecta masses are obtained with a calibration for ``Crab-like'' SNe for ZAMS masses below 12~M$_{\odot}$. 
The results presented in \citet{pejcha2015} and \citet{mueller2016} yield the largest range of explosion energies, spanning from ~0.2 to ~6~B, and from a few 0.01~B to above 2~B, respectively. 
A prominent feature is the region of non-explodability in the vicinity of 25~M$_{\odot}$ ZAMS mass. In our simulations we find BH forming models in similar mass regions to other works at 20 M$_{\odot}$, around 25~M$_{\odot}$ and between 30 and 100~M$_{\odot}$. Similar to \citet{mueller2016}, we do not find many BH forming models between 15 and 20~M$_{\odot}$. We want to stress, that in other studies \citep{ertl16,sukhbold16,mueller2016}
different progenitors have been used and thus a perfect agreement is not expected.

\subsection{Key isotopes}
\label{subsec:scan_4iso}

\begin{figure*}  
\begin{center}
	\begin{tabular}{cc}
	\includegraphics[width=0.48\textwidth]{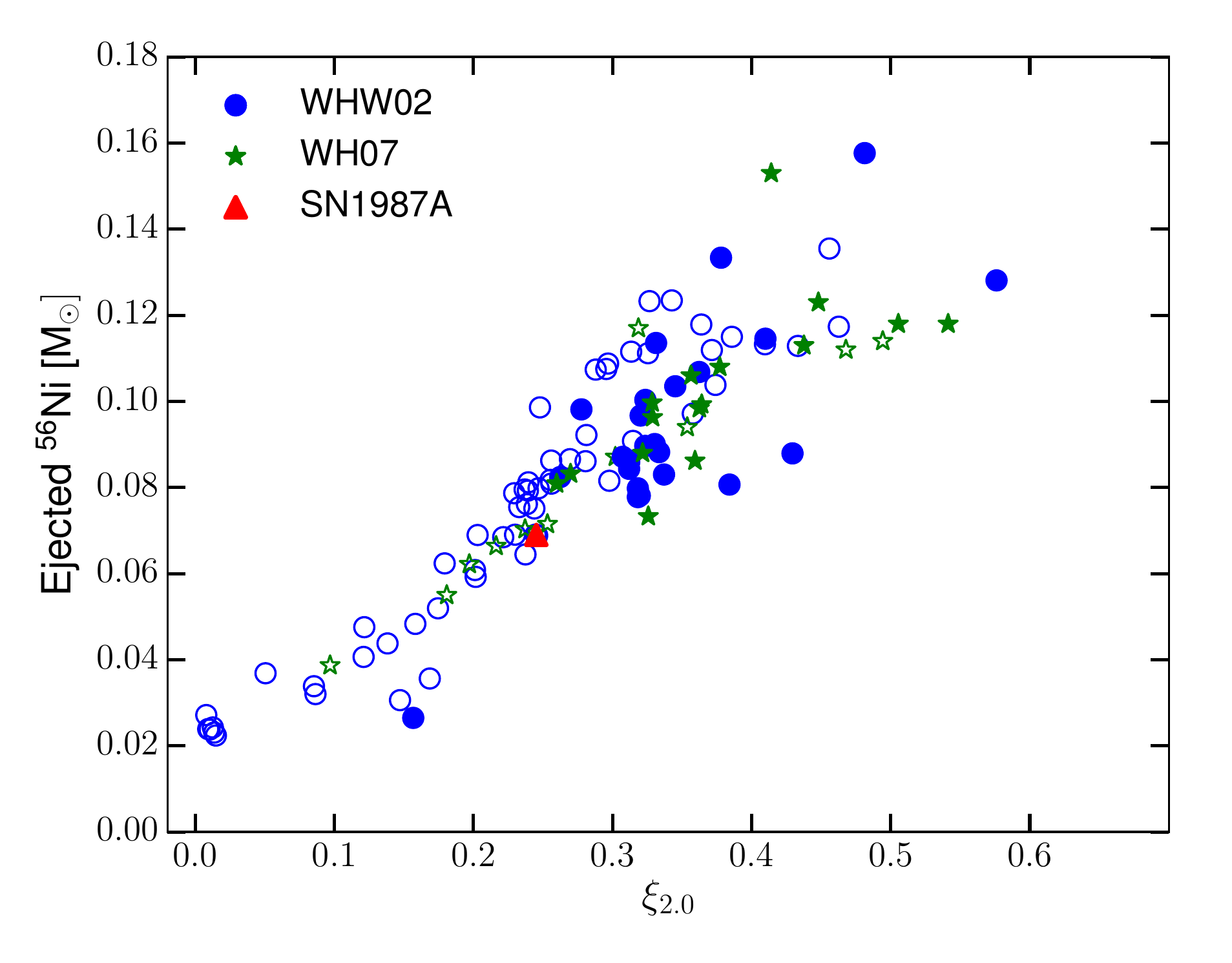} 
    & 
    \includegraphics[width=0.48\textwidth]{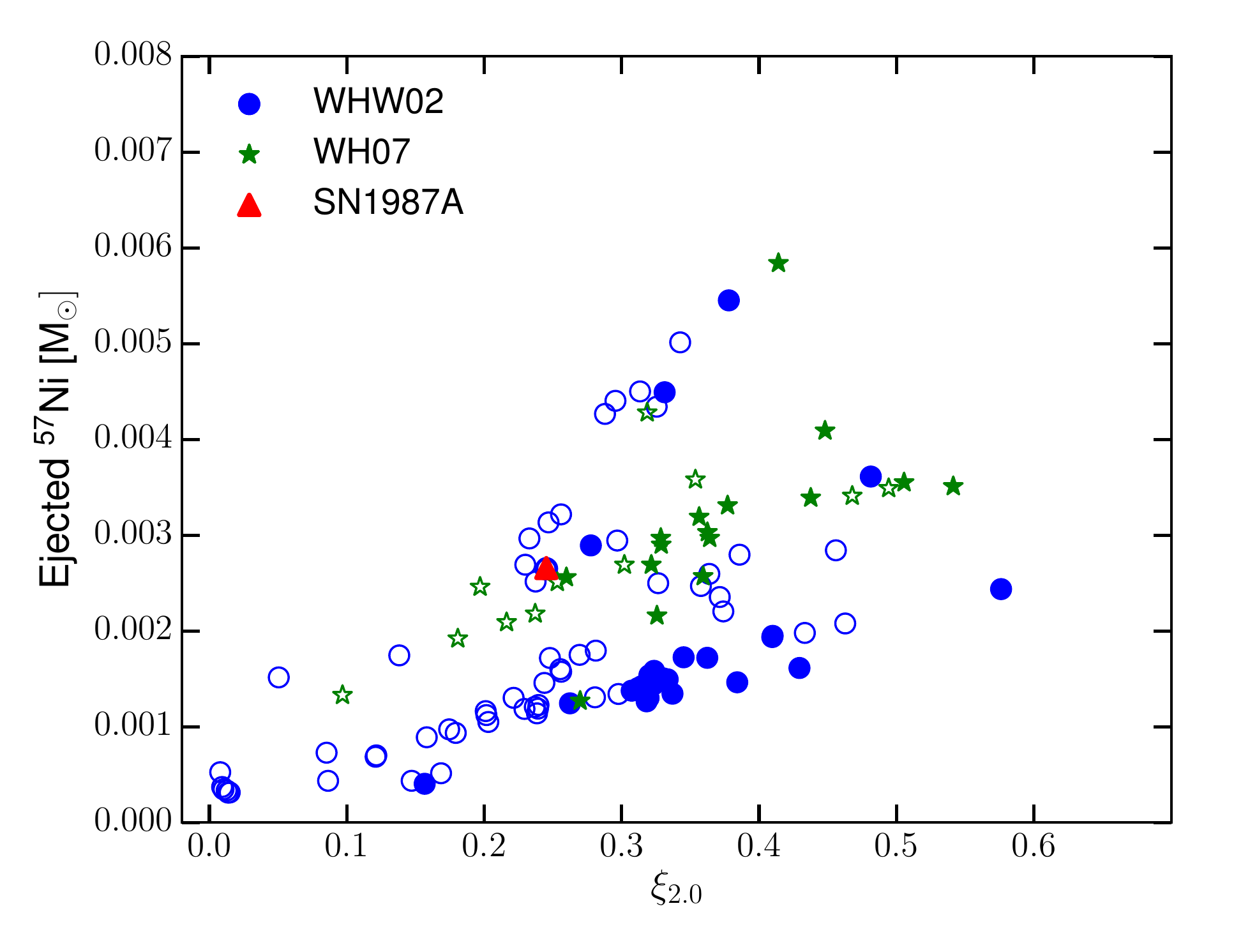} 
	\\
    \includegraphics[width=0.48\textwidth]{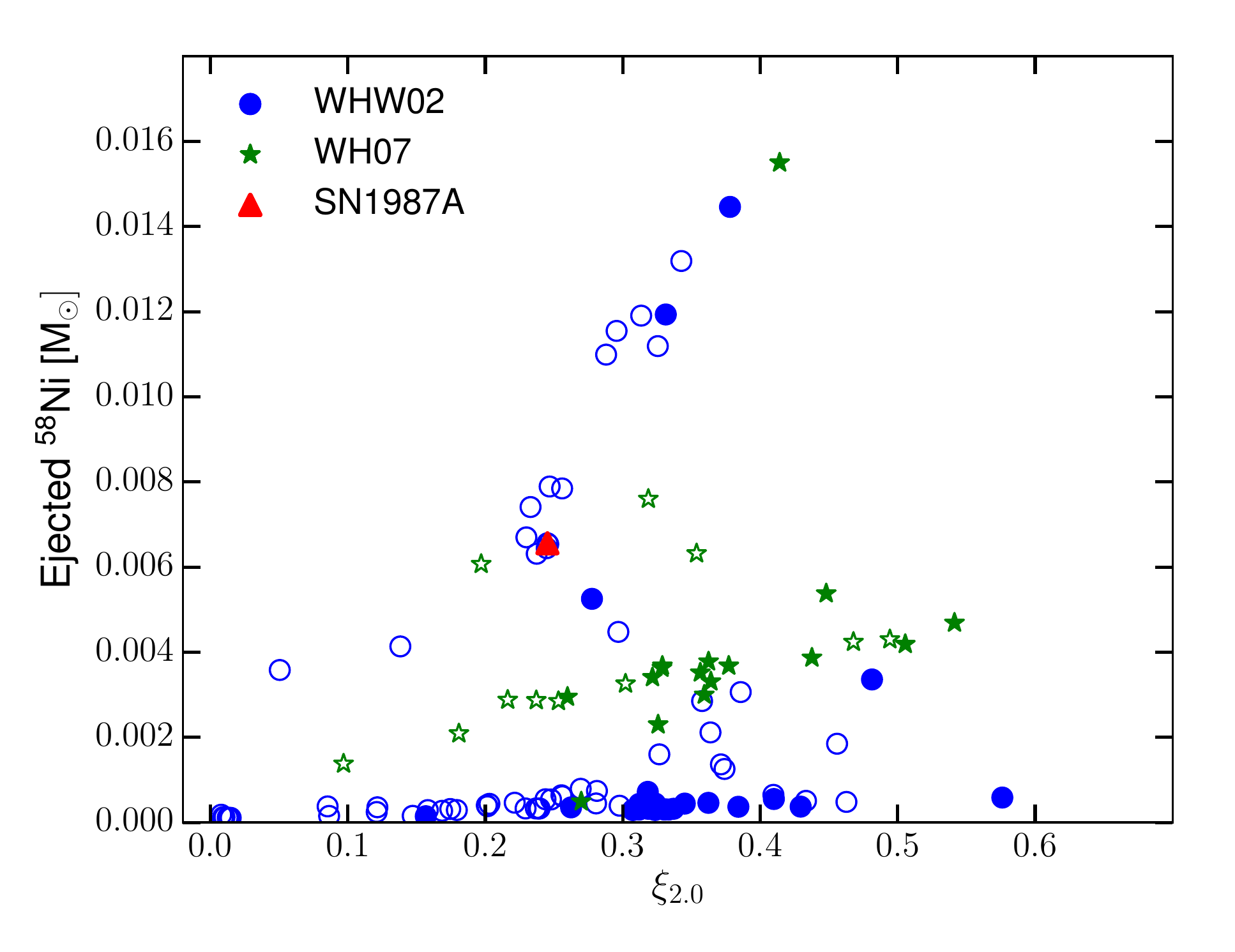} 
	& 
	\includegraphics[width=0.48\textwidth]{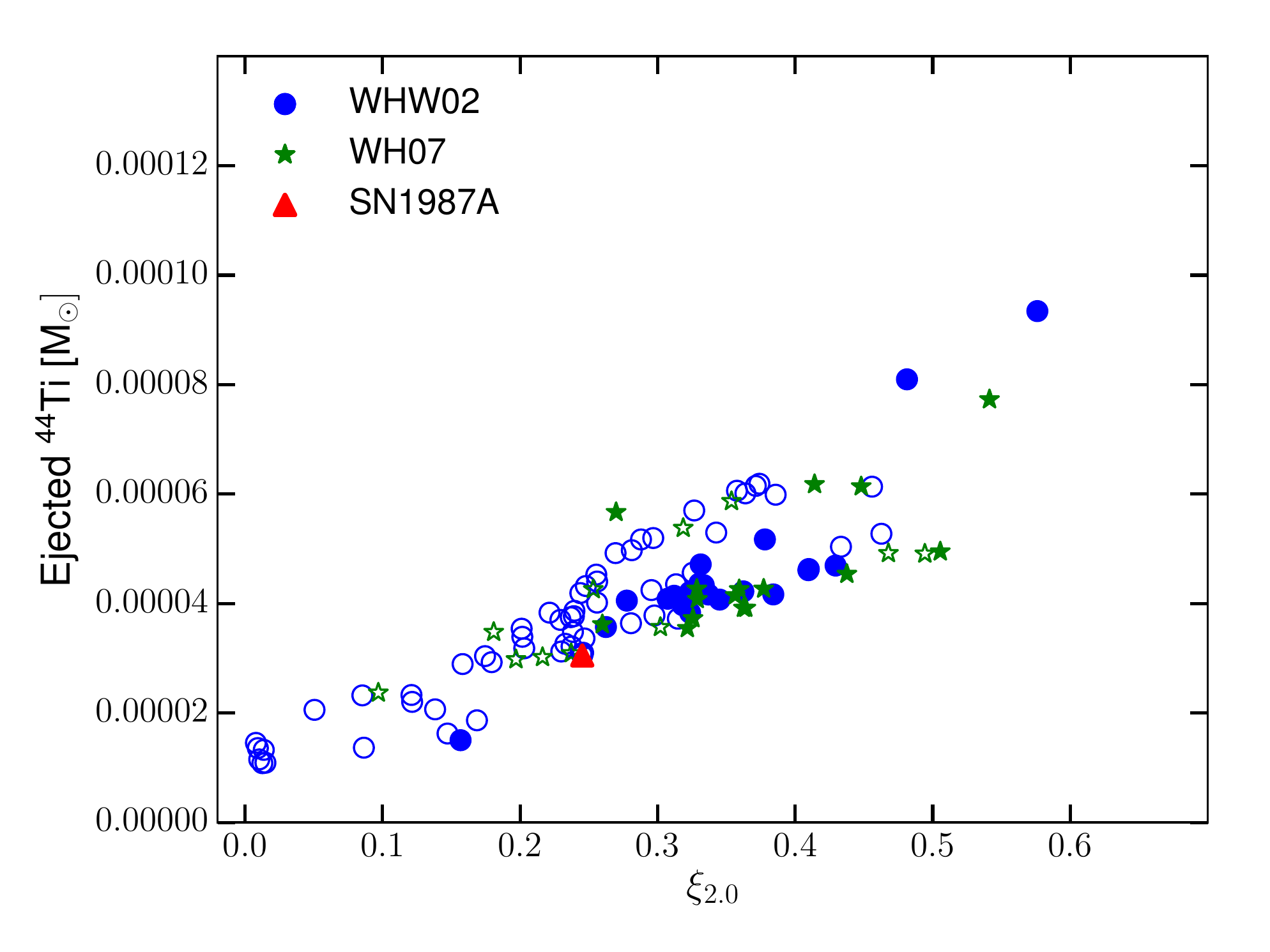} 
   	\\
	\end{tabular}
    \caption{Ejected mass of $^{56}$Ni (top left), $^{57}$Ni (top right), $^{58}$Ni (bottom left), and $^{44}$Ti (bottom right) versus compactness for WHW02 and WH07 pre-explosion models. Symbols are the same as in Figure~\ref{fig:scan_properties-compactness}.
    	\label{fig:scan_4iso-compactness}
    }
\end{center}
\end{figure*}

We now turn our attention to the behavior of four key isotopes ($^{56}$Ni, $^{57}$Ni, $^{58}$Ni, and $^{44}$Ti) with compactness. 
Figure~\ref{fig:scan_4iso-compactness} shows the ejected mass of $^{56}$Ni (top left), $^{57}$Ni (top right), $^{58}$Ni (bottom left), and $^{44}$Ti (bottom right) for all models in both series (WHW02 with blue circles; WH07 with green stars; s18.8 with a red triangle) as function of compactness. 
The amount of ejected $^{56}$Ni shows a strong linear correlation with the compactness. A similar correlation with compactness can be seen for $^{44}$Ti. Both of these isotopes are symmetric, $N=Z$ isotopes. For these isotopes, $^{56}$Ni in particular, the amount ejected depends on the explosion energy and 
how much of the Si-shell gets processed and ejected.
This trend is not shared by $^{57}$Ni and $^{58}$Ni. Instead, the correlation broadens towards high compactness (at $\xi_{2.0}$ there is almost a factor of three difference between the lowest and highest value of ejected $^{57}$Ni). A similar, but even stronger, trend is seen for $^{58}$Ni. 
While the amount of $^{56}$Ni and $^{44}$Ti mainly depends on the explosion energy, the yields of $^{57}$Ni and $^{58}$Ni depend strongly on the local electron fraction and whether regions with slightly lower $Y_e$ are ejected or not.
In general, for models of similar compactness, the models with lower ZAMS mass (open symbols in Figures~\ref{fig:scan_properties-compactness} and \ref{fig:scan_4iso-compactness}) eject more $^{57,58}$Ni and $^{44}$Ti.
An extended discussion of the nucleosynthesis yields and the trend with the local $Y_e$ of the exploding models presented here can be found in in Paper~III.

\section{Remnant properties}
\label{sec:remnants}

\begin{figure}[]
	\includegraphics[width=0.48\textwidth]{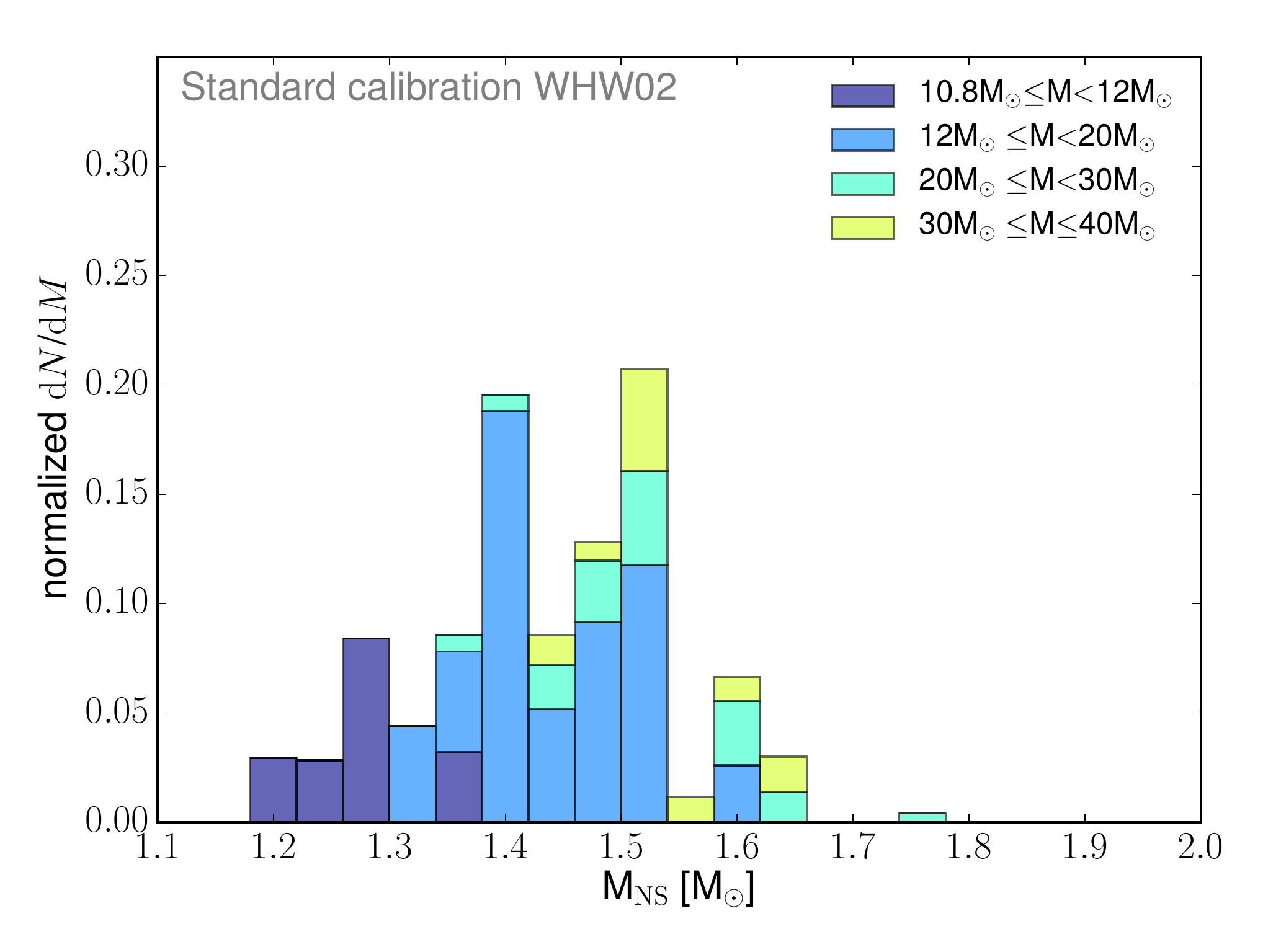} \\
    \includegraphics[width=0.48\textwidth]{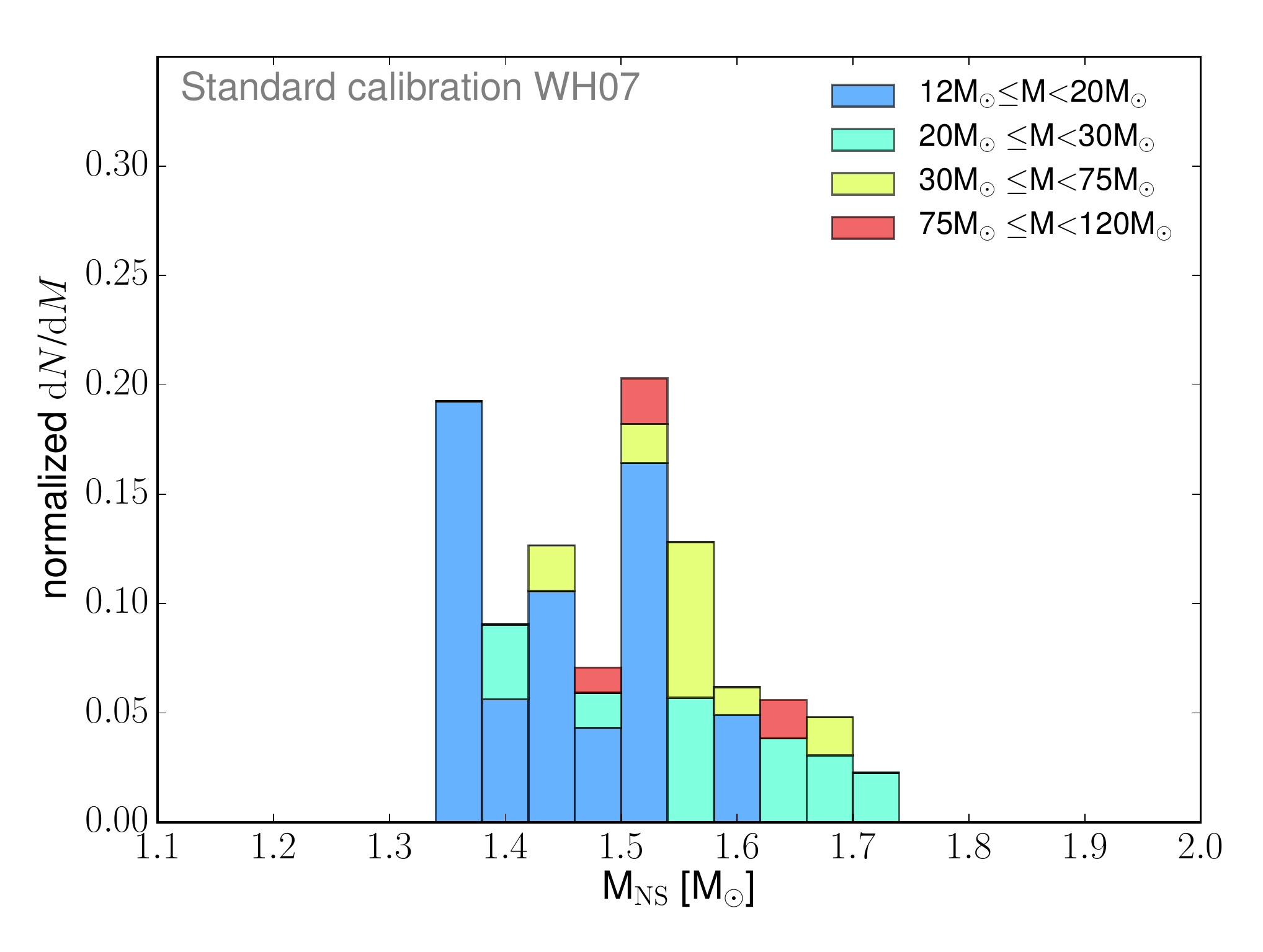}
\caption{Gravitational birth mass distributions of cold neutron stars for WHW02 (top panel) and WH07 (bottom panel) for the standard calibration. 
		\label{fig:NS_mass_distr}
    }
\end{figure}

\begin{figure}  
\begin{center}
	\includegraphics[width=0.48\textwidth]{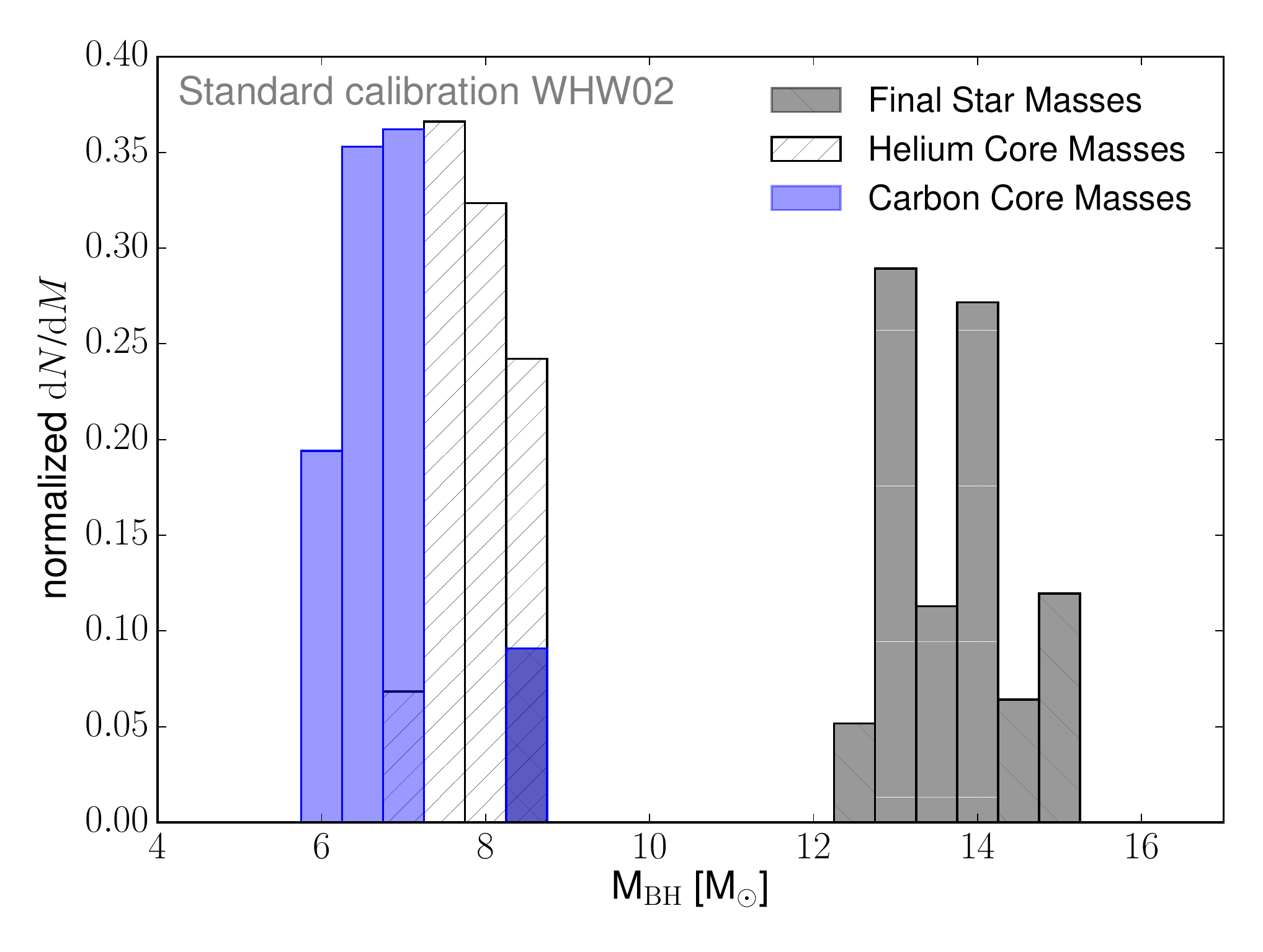} \\
    \includegraphics[width=0.48\textwidth]{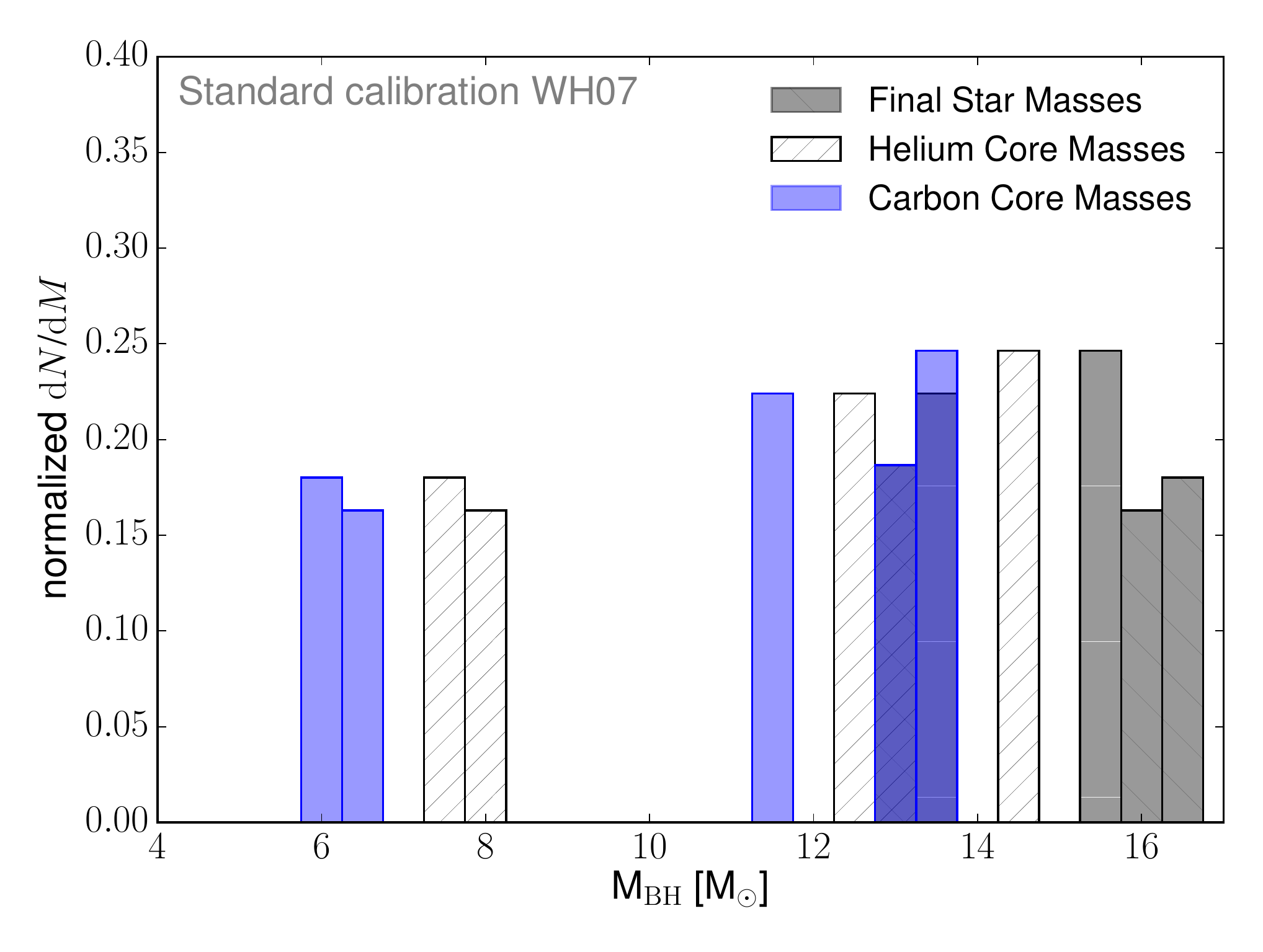} 
    \caption{BH mass distribution with and without helium envelope for WHW02 (top panel) and WH07 (bottom panel) for the standard calibration. 
    	\label{fig:bh_massdist}
    }
\end{center}
\end{figure}

\begin{figure}[]
	\includegraphics[width=0.48\textwidth]{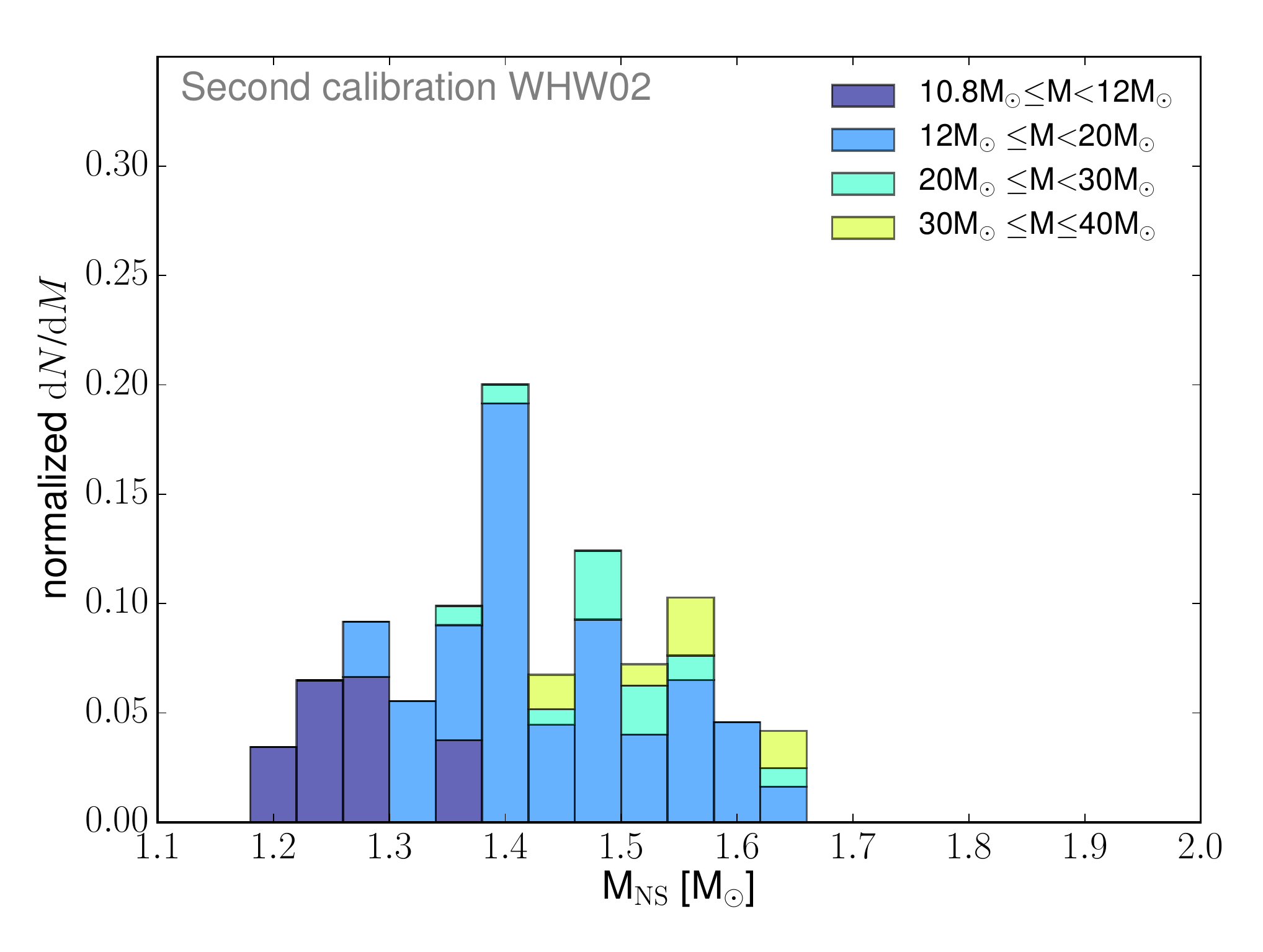}
\\
    \includegraphics[width=0.48\textwidth]{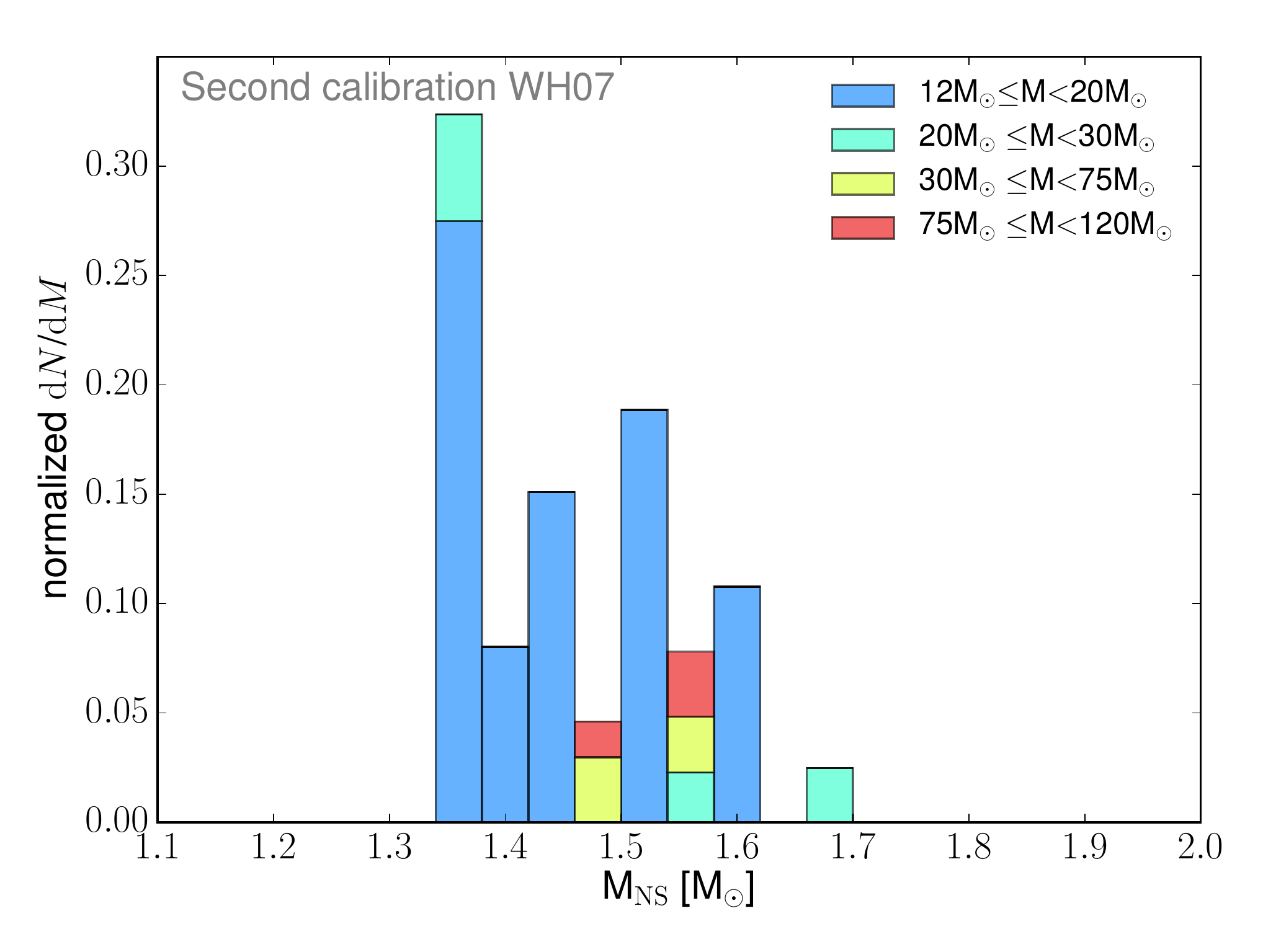}
\caption{Gravitational birth mass distributions of cold neutron stars for WHW02 and WH07 for the second calibration.
		\label{fig:NS_mass_distr2}
    }
\end{figure}

\begin{figure}  
\begin{center}
	\includegraphics[width=0.48\textwidth]{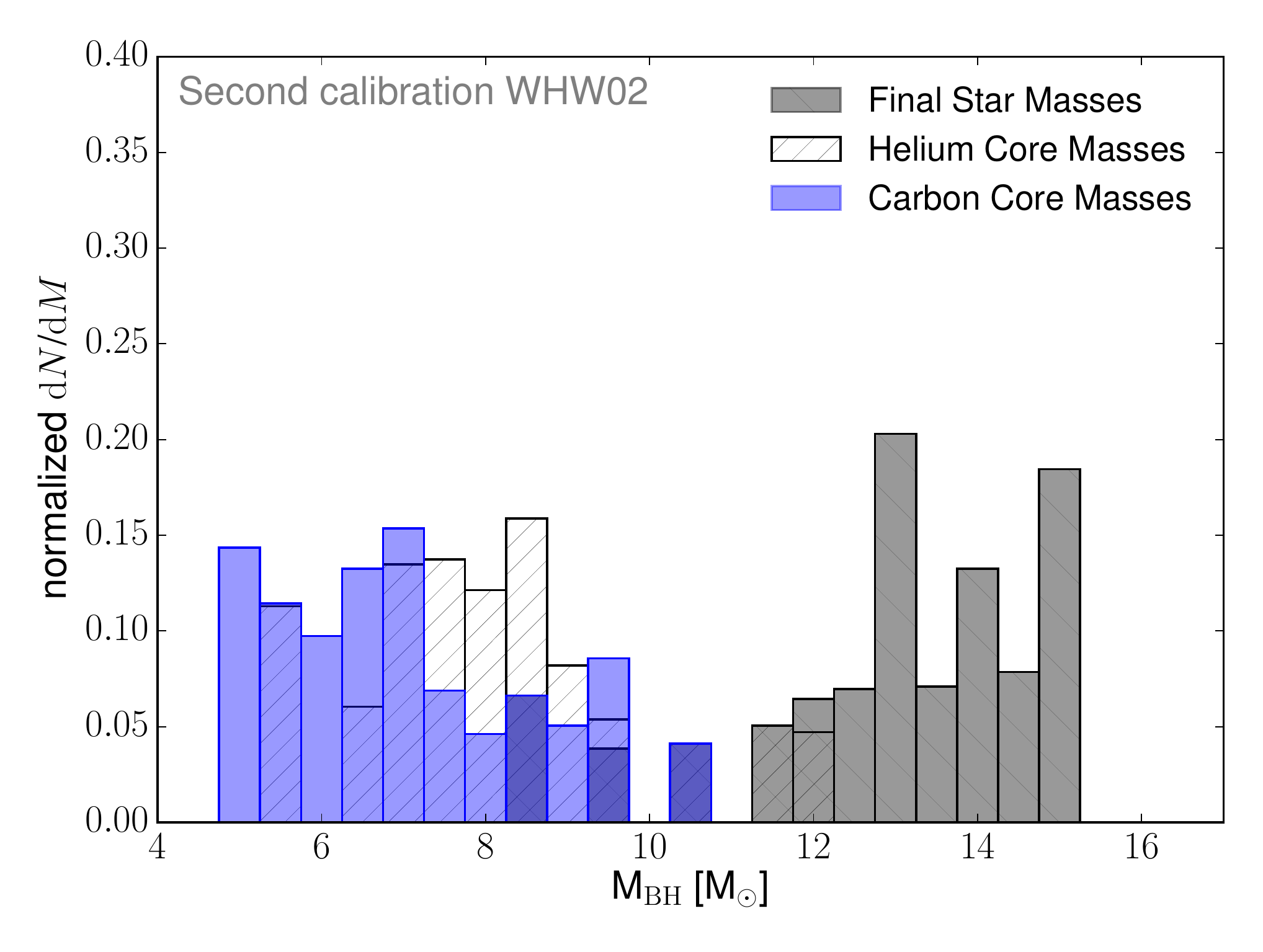} \\
    \includegraphics[width=0.48\textwidth]{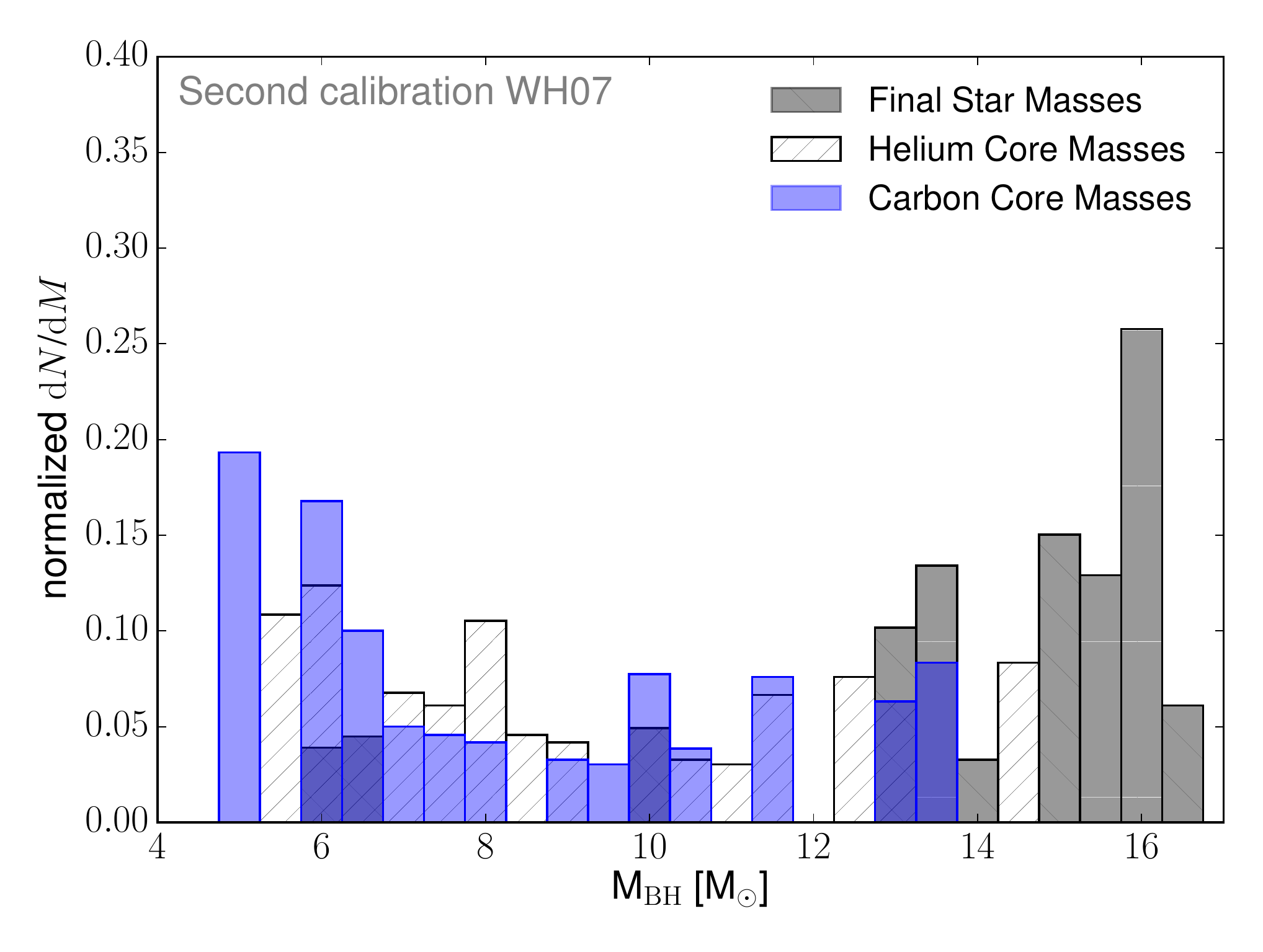} 
    \caption{BH mass distribution with and without helium envelope for the second calibration. 
    	\label{fig:bh_massdist2}
    }
\end{center}
\end{figure}

Besides explosion energies and ejected nickel masses, there are additional observables against which we can compare our results, for example the mass of the compact remnant. 
Observationally, the distribution of NS masses in slow pulsars and non-recycled high-mass eclipsing binaries has a mean of 1.28~M$_{\odot}$ (dispersion of 0.24~M$_{\odot}$), which is consistent with the formation in CCSNe \citep{ozel12}. 
With our simulation setup (no excised mass and realistic nuclear equation of state) we can follow the full evolution of the PNS and obtain the baryonic mass of the resulting newly born hot neutron star. 
The gravitational mass of the corresponding cooled neutron star (that may be observed) is smaller due to neutrino losses in the cooling phase.
We compute the zero-temperature gravitational mass of the hot neutron star formed in our simulations using the HS(DD2) nuclear equation of state. 
We arrive at a distribution of gravitational birth-masses of neutron stars by weighting the predicted neutron star masses as a function of ZAMS mass with the initial mass function (IMF) of stars according to \citet{salpeter} which is sufficiently accurate for stars with masses above 10~M$_{\odot}$.
In Figure~\ref{fig:NS_mass_distr}, we show the gravitational birth-mass distribution of cool neutron stars for the two sets of pre-explosion models from our standard calibration. 
The different colors indicate different ranges of ZAMS masses of the pre-explosion models. 
The two samples combined include stars with masses between 10.8 and 120~M$_{\odot}$ (see also Table~\ref{tab:all_prog_series}).
The resulting neutron-star masses are between 1.2 and 1.8~M$_{\odot}$. The lowest neutron-star masses between 1.2 and 1.4~M$_{\odot}$ are only noteworthily populated by the WHW02 sample.  
In Figure~\ref{fig:NS_mass_distr} we see that lower ZAMS-mass stars contribute most of lower neutron-star masses around 1.4~M$_{\odot}$ and the higher ZAMS-mass stars are the main contribution to the neutron-star masses in the vicinity of 1.6~M$_{\odot}$. 
The resulting distribution of neutron-star birth masses for the second (less energetic) calibration does not differ too much from our standard calibration, as can be seen in Figure~\ref{fig:NS_mass_distr2}.
In comparison to the reported observed NS distribution, the mass range of our predicted distribution is somewhat shifted to higher masses.
Note, that due to a lack of progenitor models below 10.8~\msun, remnants from even lighter progenitors are not included, which due to the applied Salpeter-IMF also would considerably contribute to the lightest neutron stars and would arguably reduce the lower limit of the predicted NS mass distribution. Also note that for the WHW02 sample we did not include the outlier with 75~M$_{\odot}$ ZAMS mass. 
The results are mass distributions of single-star systems and do not consider the possible effects present in binary systems, such as mass accretion or loss due to a companion, that can lead to a very different evolution of the stars. 
Therefore, a statistically significant comparison of our models with the observed NS masses of double systems is beyond the scope of this paper. 
Overall, our NS masses are 0.1 to 0.2~M$_{\odot}$ higher than those in \citet{ugliano12} and in \citet{pejcha2015} who used the same WHW02 pre-explosion models. Our results are in general agreement with \citet{ertl16}, who use different pre-explosion models below 30~M$_{\odot}$. \citet{Nakamura2014} performed 2D axisymmetric simulations up to 1~s post bounce of the same WHW02 pre-explosion models. Similar to us, they find NS masses in the range of 1.2 to 2.1~M$_{\odot}$. It should be noted that none of their simulations failed to explode on the timescale of their simulations. 
Since we do not take into account fallback, which has been found to not have a major effect in other studies \citep{sukhbold16}, all our exploding models result in neutron stars. This is a consequence of the 1D treatment, where it is not possible to have infalling and outgoing matter at the same time. Thus, we do not obtain any fallback for exploding models in our simulations, unlike in multi-D simulations where it is possible to have simultaneous in- and outflow, which allows for substantial fallback in 2D and 3D simulations (e.g.\ \cite{Marek2009,ott17,Chan2018}). 
In our framework, simulations of stellar collapse that run beyond the time on which PUSH is active and ultimately fail to explode and simulations that directly form a black hole contribute to the resulting birth mass distribution of BHs.
Our failed explosions for the (non-rotating) WHW02 and WH07 pre-explosion models
correspond to BH masses from failed neutrino-driven CCSNe of non-rotating (or weakly rotating) stars. 
The final mass that collapses to a BH depends on the amount of mass stripping and hence for solar metallicity progenitors can be as low as the CO-core mass. 
For the two samples considered here, the final stellar mass at the onset of collapse does not exceed $\sim$17~M$_{\odot}$ due to wind mass loss (see also Figures~\ref{fig:prog_mass_structure} and \ref{fig:prog_mass_structure2}).
Again, we weight the resulting BH masses with an IMF. 
 Figure~\ref{fig:bh_massdist} shows the resulting BH-mass distributions for our standard calibration. The different shaded regions correspond to different stellar cores that collapsed to a BH, to illustrate how mass loss may effect the final BH masses.
For the WHW02 and WH07 progenitor samples we find black-hole formation for stars between 20 and 30 M$_{\odot}$, centered around 25 M$_{\odot}$ ZAMS mass, resulting in BH masses centered around $\sim 14$~M$_{\odot}$. Stars above 30~M$_{\odot}$ that form BHs are mainly found in the WH07 sample, which overall consists of models that are more likely to collapse to BHs. For our second calibration of PUSH considerably more BHs are found, which shifts the resulting BH mass distribution to slightly lower mass (see Figure~\ref{fig:bh_massdist2}).
The gap between possible low and high mass BHs is not as strongly present as before.
For both calibrations we can compute the fraction of stars that ultimately form BHs. We do so by considering a mass range from 8 to 150~M$_{\odot}$ for the estimate and assuming that stars between 8~M$_{\odot}$ and the lowest ZAMS mass in each sample successfully explode and leave behind a NS as a remnant. Furthermore, the fate of the star with the highest ZAMS mass within each sample is continued up to 150~M$_{\odot}$. Again, we use the Salpeter IMF for our estimate. For the WHW02 progenitor sample \textasciitilde5\% of progenitor stars leave behind a black hole for the standard calibration and \textasciitilde16\% for the second calibration. The WH07 progenitor sample leads to more BHs: \textasciitilde8\% of stars have a black hole as a remnant for the standard calibration and \textasciitilde21\% for the second calibration.
Our results are broadly consistent with the observationally determined BH mass distribution ($7.8 \pm 1.2$~M$_{\odot}$, \citet{ozel10}), when we assume that the helium core mass sets the BH mass \citep{kochanek2014}. 
The BH masses found from the solar metallicity pre-explosion models (WHW02 and WH07) are not massive enough to explain the BH from the recent LIGO/VIRGO observations. These BHs can originate from low-metallicity stars which experience less (or no) mass loss during their evolution and hence collapse with almost their entire ZAMS mass.

\section{Summary and Conclusions}
\label{sec:discuss}
The investigation of the CCSN mechanism remains an intriguing and unsolved problem. The solution most likely requires multi-dimensional hydrodynamical simulations, including general relativity, a nuclear equation of state, sophisticated neutrino transport, magnetic fields and rotation, and asphericity of the progenitor structure. The high computational cost of such models still limits the number of models that can be investigated and still motivates the usage of effective spherically symmetric models for extended progenitor and outcome studies.

In this work, we have improved the PUSH method from \cite{push1}, which is used to artificially trigger parametrized core-collapse supernova explosions in spherical symmetry. The PUSH method provides a computationally affordable and robust framework to study aspects of CCSNe that require modeling many different progenitors for several seconds after the onset of the explosion. Here, we focused on applying it to two sets of non-rotating, solar metallicity pre-explosion models between 10.8~M$_{\odot}$ and 120~M$_{\odot}$ to obtain explosions and predict compact remnant properties.

The PUSH method has two free parameters (\kpush and \trise) that need to be determined from external constraints. We have used three general constraints from observations of core-collapse supernovae to set the parameters for any pre-explosion model \emph{a priori}. We required that the PUSH method
(i) reproduces the observed properties of SN~1987A for a suitable pre-explosion model, 
(ii) allows for the formation of black holes, and
(iii) results in lower explosion energies for the lowest-mass progenitors (``Crab-like SNe''). 
These requirements led to a compactness-dependent value for \kpush (and a fixed value for \trise).

Using this setup, we have simulated the death of 133 SN-progenitor models as either successful neutrino-driven explosion or as failed explosion or direct collapse to a black hole. This study has led to several interesting predictions and conclusions:
\begin{itemize}
\item As a whole, the resulting explosion energies from the PUSH method are in good agreement with the explosion energies of observed CCSNe. In addition, the same models also match observations simultaneously for $^{56}$Ni ejecta and explosion energy.
\item We have shown and discussed that it is possible to infer several interesting trends of explosion properties with compactness (and CO-core mass). The compactness is a better indicator for the expected outcome than the ZAMS mass, however it does not tell the entire story. For example, we found that for models of similar compactness a degeneracy exists that can be partially broken with the CO-core mass. 
This is consistent with the finding in recent multi-dimensional simulations, where the outcome of the CCSN simulation is related to the binding energy of the outer envelope exterior of a given mass, which has a strong correlation with compactness \cite{Burrows2018}.
\item We found linear trends of ejected $^{56}$Ni and $^{44}$Ti yields with compactness (and explosion energy). The yields of $^{57,58}$Ni do not follow the same simple correlation. Instead, the local electron fraction has a strong impact on the final yields of these isotopes.
\item The predicted outcome (neutron star or black hole) is in agreement with predictions from other works that employ a comparable parametrized approach based on the neutrino-driven CCSN mechanism. In particular, we also find a region of non-explodability around $\sim 25$~M$_{\odot}$ initial stellar mass. Note that some of these studies use different progenitor sets below 30~M$_{\odot}$.
\item We predict neutron-star mass and black-hole mass distributions that are broadly consistent with observations. However, we do not find any BHs with $m_{\mathrm{BH}} \gtrsim 18\;{\rm M}_{\odot}$ since non of the pre-explosion models has a final mass at collapse above this due to mass loss. We expect that lower metallicity pre-explosion models (which experience less mass loss) will result in more massive BHs.
\item This set of CCSN models is used for a detailed nucleosynthesis study in Paper~III \citep{push3}.
\end{itemize}

\acknowledgments
The work at NC State (KE, SC, CF) was supported by the Department of Energy through an Early CAREER Award (DOE grant No.SC0010263). 
CF acknowledges support from the Research Corporation for Science Advancement through a Cottrell Scholar Award. 
The effort at the Universit\"at Basel was supported by the Schweizerischer Nationalfond and by the ERC Advanced Grant ``FISH''.
KE acknowledges the support from GSI.
AP acknowledges supports from the INFN project "High Performance data Network" funded by the Italian CIPE.
\software{Agile \cite{Liebendoerfer.Agile}, CFNET \cite{cf06a}, Matplotlib \citep{matplotlib}}

\clearpage
\bibliographystyle{yahapj}
\bibliography{references_push,references_fkt}

\appendix

\section{Observations}
\label{subsec:obs-constraints}

Besides the well-observed SN~1987A, many other CCSNe are known with a range of progenitor masses and explosion energies.
One sees a rising trend in explosion energy and ejected nickel mass from $\sim 10$ to $\sim 20$~M$_{\odot}$ which corresponds to the transition from weaker ``Crab-like'' SNe to standard neutrino-driven SNe. 
Beyond 20~M$_{\odot}$ ZAMS mass, the properties of observed CCSNe are uncertain (see also Section~\ref{intro:1}). \citet{nomoto03} argue that they can be separated into two branches (``hypernova (HN) branch'' and ``faint SN branch'') with different magnitudes of explosion energy and ejected nickel mass, and arguably different mechanisms powering the explosions. We refer the reader for example to Figure~2 in \citet{nomoto03} \citep[see also][]{nomoto13,janka12} 
\textbf{
for the bifurcation into two branches. 
\sout{, where the bifurcation between the two branches can be seen clearly.}
Note that a definitive consensus on the existence of such a ``faint supernova branch'' has not yet been reached by observers.
}

The HN branch consists of very energetic $\gamma$-ray burst SNe (GRB-SNe) and of hypernovae (HNe) \citep{woosleybloom06,nomoto13,janka12}. For these explosions, rapid stellar rotation and strong magnetic fields are thought to be crucial. 
The term HNe originates from the exceptional brightness caused by a large production of nickel in these hyperenergetic explosions with explosion energies of $\gtrsim 10^{52}$~erg \citep{pac98,nomoto04,nomoto06}.
A possible central engine of these HNe and GRB-SNe are rapidly rotating BHs (collapsars), where accreting matter around the central compact object radiates energy in neutrinos, electromagnetic Poynting flux, and mass outflow.
An alternative scenario for the central engine of HNe and GRB-SNe is a rapidly-spinning neutron star with a strong magnetic field ($B\gtrsim 10^{15}$~Gauss) that is formed during stellar collapse. In this case, the HN or GRB-SN is powered by rotational energy which is converted into explosion energy by the magnetic field.
The resulting explosion strengths and amount of Ni produced are higher than what is typically expected for the neutrino-driven mechanism \citep{nomoto06}. This is another indication that the HN branch is likely caused by a different mechanism, distinct from the standard neutrino-driven mechanism.

The \textbf{existence of the} other branch, the faint SN branch, \textbf{is observationally less certain. It }represents CCSNe with low explosion energies. These SNe mark a transitional region to failed SNe, which indicates a mechanism that no longer can efficiently power explosions for stars above a certain ZAMS mass 
(the value is quite uncertain, with observations indicating 18~M$_{\odot}$ \citep{smartt.missingRSG:2015} to 25~M$_{\odot}$ \citep{Davis.Beasor:2018}).
These failed explosions lead to the formation of a BH either directly without an associated explosion or through the continued fallback of material onto the central object in very weak (and eventually failing) explosions \citep[see e.g.][]{lovegrove13}. The change in typical explosion energies within the faint SN branch could indicate a transition from efficient, strongly convective neutrino-driven SNe to an inefficient neutrino-driven mechanism. Thus, the observation of two branches (faint SN branch and HN branch), together with the constraint that HNe and GRB-SNe are likely due to a different and somewhat rarer mechanism, suggests that the faint SN branch represents neutrino-powered SNe.

In Table~\ref{tab:sncomp_bruenn}, we list the CCSNe included in our Figures~\ref{fig:parabolaI_s02_w07} and \ref{fig:parabolaII_s02_w07} together with their relevant observational properties. This selection follows \citet{bruenn16} and \citet{nomoto13} (only SNe with explosion energies $\le 10^{52}$~erg). These CCSNe are in the mass range between 9~M$_{\odot}$ and 30~M$_{\odot}$ and can be considered to be caused by the neutrino-driven mechanism. However, it is important to emphasize that the observationally derived ZAMS masses of the CCSN progenitors have relatively large uncertainties depending on the method used.
The first line for each SN indicates the values used in this Paper (taken from \citet{bruenn16} or \citet{nomoto13} as indicated in the References column). Additional lines indicate that significantly different values are available in the literature. We have added these for the reader's convenience. However, a detailed discussion of the observational uncertainties is beyond the scope of this paper.

\begin{table*}  
\begin{center}
	\caption{Observational Properties of CCSNe
    	\label{tab:sncomp_bruenn}
    }
\begin{tabular}{lcccll}
	\tableline \tableline
	Event & $M_{\mathrm{ZAMS}}$ & $E_{\mathrm{expl}}$ & \multicolumn{2}{c}{$m(^{56}\mathrm{Ni})$}  & References \\
	      & (M$_{\odot}$) & ($10^{51}$~erg) & \multicolumn{2}{c}{(M$_{\odot}$)} &  \\
	      &  &  & \multicolumn{2}{c}{BL ; S ; L ; O/V$^{**}$} &  \\ 
    \tableline 
    SN~1987A  & 18-21  & $1.1 \pm 0.3$  &  $0.071 \pm 0.003$ & L & 1, 2 \\ 
    &&&&& \\  
    SN~1993J  & 12-17 & 1-2 &  0.06-0.09 & L & 3, 4, 5, 6, 7 \\ 
     &&&  0.07-0.11$^{*}$ & L & 8\\ 
   
    SN~2004A  & $12.0 \pm 2.1$ & 0.76-1.3 & $0.046^{+0.031}_{-0.017}$ & O/V & 3, 9, 10, 11, 12\\
    SN~2004dj  & 12-15 & 0.7-0.9 &  $0.020 \pm 0.002$ & O/V & 3, 9, 10, 13, 14, 15, 16, 17 \\ 
    SN~2004et  & 12-15 & 1.1-1.8 & $0.062 \pm 0.02$ & S & 3, 9, 10, 18, 19 \\ 
    &25-29$^{*}$& 2.0-2.6$^{*}$ & & & 20 \\
    &  &  & $0.062 \pm 0.02^{*}$ & BL & 18 \\ 
    & 8$^{+5}_{-1}$$^*$ & & & & 21 \\
    SN~2005cs  & $9^{+3}_{-2}$ & 0.27-0.39 & $0.009 \pm 0.003$ & BL & 3, 9, 10, 22, 23 \\  
 & 17.2-19.2$^{*}$&0.41$\pm$0.03$^{*}$&&& 24 \\ 
    SN~2009kr & 11-20 &1.6-3 &&& 3, 9, 10, 25 \\
    SN~2012aw  & 14-18 & 1.0-1.7 & $0.06 \pm 0.01$ & O/V & 3, 9, 10, 18, 26, 27\\ 
  &&1-2$^{*}$ &0.06$\pm$0.01$^{*}$ &  & 27\\  
  &&1.5$^{*}$ &0.05$\pm$0.06$^{*}$ &  & 28\\  
   &&&&& \\ 
    SN~2012ec  & 14-22 & 0.6-1.9 & - & & 3, 9, 10, 29 \\  
    &&&&& \\  
SN~1994I  & \textasciitilde 13-15  & \textasciitilde $1.0$ &  \textasciitilde $0.07$ & O/V & 30, 31 \\  
    SN~2005bf  & \textasciitilde $25-30$  & \textasciitilde  $1.0-1-5$ &  \textasciitilde 0.32 & O/V & 30, 32, 33 \\
    & $8.3^{*}$ &\textasciitilde $2^{*}$ & &  & 34\\ 
    \tableline 
\end{tabular}
\end{center}
\tablecomments{
	With the exception of the values for SN~1987A, this Table mainly adapts values presented in \citet{bruenn16} and \citet{nomoto13}, and references therein.  $^{*}$ Values in rows following a given SN name represent alternative ZAMS mass, explosion energy and the amount of ejected nickel determinations for the readers convenience. $^{**}$ The abbreviations given in the column of the reported nickel values denote: values obtained with the BL-method (BL), values obtained with the S-method (S), values obtained with modeling of the lightcurve (L), and values obtained by other methods or with a combination of various methods (O/V).
	For the event SN~1994I the values given here are taken from \citet{nomoto13,nomoto94}.
    See \citet{bruenn16} for a discussion on the selection procedure.
}
\tablerefs{
	    (1)~\citet{Blinnikov2000}; 
	    (2)~\citet{Seitenzahl2014};
        (3)~\citet{bruenn16};
        (4)~\citet{shigeyama94};
        (5)~\citet{woosley94};
        (6)~\citet{young95};
        (7)~\citet{bartunov94};
        (8)~\citet{freedman94};        
        (9)~\citet{poznanski13};
        (10)~\citet{dessart10};
        (11)~\citet{hendry06};
        (12)~\citet{maund13};        
        (13)~\citet{chugai05};
        (14)~\citet{zhang06};
        (15)~\citet{maiz04};
        (16)~\citet{wang05};
        (17)~\citet{vinko09};        
        (18)~\citet{jerkstrand12};
        (19)~\citet{sahu06};
        (20)~\citet{utrobin09};
        (21)~\citet{crockett11};               
        (22)~\citet{maund05};
        (23)~\citet{takats06};
        (24)~\citet{utrobin08};
        (25)~\citet{fraser10};      
        (26)~\citet{jerkstrand14};
        (27)~\citet{bose13};  
        (28)~\citet{dallora14};        
        (29)~\citet{maund13b};         
        (30)~\citet{nomoto13};
        (31)~\citet{nomoto94};
        (32)~\citet{janka12};
        (33)~\citet{tominaga05}
        (34)~\citet{Folatelli2006}
}
\end{table*}

\clearpage
\pagebreak

\section{Tables of Simulation Results for Standard Calibration} \label{appx:tables}

The explosion properties obtained from the standard PUSH calibration for the solar metallicity progenitor samples WHW02 and WH07 (see Figure~\ref{fig:scan_properties-mass}) are provided as machine-readable tables. Only the exploding models are included in the Tables.
A portion of the Table is shown here for guidance regarding its form and content (see Table~\ref{tab:explproperties02}). The Table is published in its entirety in machine readable format.

\begin{center}
\begin{longtable*}{lccccc}

\multicolumn{6}{c}%
{{\bfseries \tablename\ \thetable{} -- WHW02 models (standard calibration)}} \label{tab:explproperties02}\\
	
\hline
\multicolumn{1}{l}{M$_{\rm ZAMS}$} & \multicolumn{1}{c}{M$_{\rm collapse}$} & \multicolumn{1}{c}{E$_{\rm expl}$}  & \multicolumn{1}{c}{t$_{\rm expl}$} & \multicolumn{1}{c}{M$_{\rm ^{56}Ni}$} &  \multicolumn{1}{c}{M$_{\rm remn}$} \\ 
\multicolumn{1}{l}{(M$_{\odot}$)}  & \multicolumn{1}{c}{(M$_{\odot}$)} & \multicolumn{1}{c}{(B)} & \multicolumn{1}{c}{(s)} &  \multicolumn{1}{c}{(M$_{\odot}$)} & \multicolumn{1}{c}{(M$_{\odot}$)} \\ 
\endfirsthead

\multicolumn{6}{c}%
{{\bfseries \tablename\ \thetable{} -- (continued)}} \\
\hline
\multicolumn{1}{l}{M$_{\rm ZAMS}$} & \multicolumn{1}{c}{M$_{\rm collapse}$} & \multicolumn{1}{c}{E$_{\rm expl}$}  & \multicolumn{1}{c}{t$_{\rm expl}$} & \multicolumn{1}{c}{M$_{\rm ^{56}Ni}$} &  \multicolumn{1}{c}{M$_{\rm remn}$} \\ 
\multicolumn{1}{l}{(M$_{\odot}$)}  & \multicolumn{1}{c}{(M$_{\odot}$)} & \multicolumn{1}{c}{(B)} & \multicolumn{1}{c}{(s)} &  \multicolumn{1}{c}{(M$_{\odot}$)} & \multicolumn{1}{c}{(M$_{\odot}$)} \\ 
\hline
\endhead
\hline
\endfoot

\hline
\endlastfoot

 \tableline
18.0 & 14.50 & 1.45 & 0.44 & 1.12E-01 & 1.62 \\
18.2 & 14.59 & 1.24 & 0.39 & 7.98E-02 & 1.55 \\
18.4 & 14.83 & 1.58 & 0.42 & 9.72E-02 & 1.78 \\
18.6 & 14.86 & 1.22 & 0.36 & 6.88E-02 & 1.56 \\
18.8 & 15.05 & 1.20 & 0.36 & 6.90E-02 & 1.55 \\
19.0 & 15.04 & 1.58 & 0.43 & 1.12E-01 & 1.78 \\

\end{longtable*}
\end{center}

\end{document}